\documentclass[preprint]{emulateapj}
\usepackage{CJK,amsmath,amssymb,graphicx,longtable,booktabs,lscape}
\pdfoutput=1
\newcommand{\magarcsec}{mag arcsec$^{-2}$}
\newcommand{\king}{KINGPHOT }

\slugcomment{ApJ, accepted}

\shorttitle{NGVS. X. Ultra-Compact Dwarfs in M87, M49 and M60}
\shortauthors{Liu et al.}

\begin{document}
\title{The Next Generation Virgo Cluster Survey. X. Properties of Ultra-Compact Dwarfs in the M87, M49 and M60 Regions.}

\author{Chengze Liu,\altaffilmark{1,2}
Eric W. Peng,\altaffilmark{3,4}
Patrick C\^ot\'e,\altaffilmark{5}
Laura Ferrarese,\altaffilmark{5}
Andr\'es Jord\'an,\altaffilmark{7}
J. Christopher Mihos,\altaffilmark{8}
Hong-Xin Zhang,\altaffilmark{3,4,6,9,10}
Roberto P. Mu{\~n}oz,\altaffilmark{7}
Thomas H. Puzia,\altaffilmark{7}
Ariane Lan\c{c}on,\altaffilmark{11}
Stephen Gwyn,\altaffilmark{5}
Jean-Charles Cuillandre,\altaffilmark{12}
John P. Blakeslee,\altaffilmark{5}
Alessandro Boselli,\altaffilmark{13}
Patrick R. Durrell,\altaffilmark{14}
Pierre-Alain Duc,\altaffilmark{15}
Puragra Guhathakurta,\altaffilmark{16}
Lauren A. MacArthur,\altaffilmark{5,17}
Simona Mei,\altaffilmark{18,19,20}
Rub\'en S\'anchez-Janssen,\altaffilmark{5}
and Haiguang Xu\altaffilmark{1,2}
}

\altaffiltext{1}{Center for Astronomy and Astrophysics, Department of Physics and Astronomy, Shanghai Jiao Tong University, Shanghai 200240, China; czliu@sjtu.edu.cn}
\altaffiltext{2}{Shanghai Key Lab for Particle Physics and Cosmology, Shanghai Jiao Tong University, Shanghai 200240, China}
\altaffiltext{3}{Department of Astronomy, Peking University, Beijing 100871, China}
\altaffiltext{4}{Kavli Institute for Astronomy and Astrophysics, Peking University, Beijing 100871, China}
\altaffiltext{5}{Herzberg Institute of Astrophysics, National Research Council of Canada, Victoria, BC V9E 2E7, Canada}
\altaffiltext{6}{National Astronomical Observatories, Chinese Academy of Sciences, A20 Datun Rd, Chaoyang District, Beijing 100012, China}
\altaffiltext{7}{Departamento de Astronom\'ia y Astrof\'isica, Pontificia Universidad Cat\'olica de Chile, 7820436 Macul, Santiago, Chile}
\altaffiltext{8}{Department of Astronomy, Case Western Reserve University, Cleveland, OH, USA}
\altaffiltext{9}{CAS-CONICYT Fellow}
\altaffiltext{10}{Chinese Academy of Sciences South America Center for Astronomy, Camino EI Observatorio \#1515, Las Condes, Santiago, Chile}
\altaffiltext{11}{Observatoire Astronomique de Strasbourg, Universit\'e de Strasbourg \& CNRS, UMR 7550, 11 rue de l'Universit\'e, F-67000 Strasbourg, France}
\altaffiltext{12}{CEA/IRFU/SAp, Laboratoire AIM Paris-Saclay, CNRS/INSU, Universit\'e Paris Diderot, Observatoire de Paris, PSL Research University, F-91191 Gif-sur-Yvette Cedex, France}
\altaffiltext{13}{Aix-Marseille Universit\'e, CNRS, LAM (Laboratoire d'Astrophysique de Marseille) UMR 7326, 13388, Marseille, France}
\altaffiltext{14}{Department of Physics and Astronomy, Youngstown State University, One University Plaza, Youngstown, OH 44555, USA}
\altaffiltext{15}{AIM Paris Saclay, CNRS/INSU, CEA/Irfu, Universit\'e Paris Diderot, Orme des Merisiers, F-91191 Gif sur Yvette cedex, France}
\altaffiltext{16}{UCO/Lick Observatory, Department of Astronomy and Astrophysics, University of California Santa Cruz, 1156 High Street, Santa Cruz, CA 95064, USA}
\altaffiltext{17}{Department of Astrophysical Sciences, Princeton University, Princeton, NJ 08544, USA}
\altaffiltext{18}{GEPI, Observatoire de Paris, CNRS, University of Paris Diderot, Paris Sciences et Lettres (PSL), 61, Avenue de l'Observatoire 75014, Paris  France}
\altaffiltext{19}{University of Paris Denis Diderot, University of Paris Sorbonne Cit\'e (PSC), 75205 Paris Cedex 13, France}
\altaffiltext{20}{California Institute of Technology, Pasadena, CA 91125, USA}

\begin{abstract}
We use imaging from the Next Generation Virgo cluster Survey (NGVS) to present a comparative study of ultra-compact dwarf (UCD) galaxies associated with three prominent Virgo sub-clusters: those centered on the massive, red-sequence galaxies M87, M49 and M60. We show how UCDs can be selected with high completeness using a combination of half-light radius and location in color-color diagrams ($u^*iK_s$ or $u^*gz$).  Although the central galaxies in each of these sub-clusters have nearly identical luminosities and stellar masses, we find large differences in the sizes of their UCD populations, with M87 containing $\sim$ 3.5 and 7.8 times more UCDs than M49 and M60, respectively. The relative abundance of UCDs in the three regions scales in proportion to sub-cluster mass, as traced by X-ray gas mass, total gravitating mass, number of globular clusters, and number of nearby galaxies. We find that the UCDs are predominantly blue in color, with $\sim 85$\% of the UCDs having colors similar to blue GCs and stellar nuclei of dwarf galaxies. We present evidence that UCDs surrounding M87 and M49 may follow a morphological sequence ordered by the prominence of their outer, low surface brightness envelope, ultimately merging with the sequence of nucleated low-mass galaxies, and that envelope prominence correlates with distance from either galaxy. Our analysis provides evidence that tidal stripping of nucleated galaxies is an important process in the formation of UCDs.
\end{abstract}


\setcounter{footnote}{0}

\section{Introduction}
\label{sec:intro}

For many decades, it was widely believed that galaxies and star clusters were completely unrelated populations in the manifold of stellar systems, characterized by distinct stellar populations, structural and dynamical properties, and, presumably, formation histories (e.g., \citealt{1985ApJ_295_73Kormendy, 1997AJ_114_1365Burstein}; see also the overview of \citealt{2012AJ_144_76Willman} and references therein). However, in the late 1990s, multiple spectroscopic surveys of the Fornax cluster uncovered an apparently new type of compact stellar system with properties that seemingly bridged the gap between compact, low-mass galaxies and globular clusters (GCs). First,  \citet{1999A+AS_134_75Hilker} used the 2.5 m Dupont telescope to carry out a spectroscopic survey of stellar systems in the vicinity of NGC1399 and reported the discovery of two Fornax members so compact that they were barely resolved in ground-based images. Three other compact cluster members were identified by  \citet{2000PASA_17_227Drinkwater} and \citet{2001ApJ_560_201Phillipps} in an AAT/2dF Fornax spectroscopic survey. With absolute magnitudes in the range $-13.8 \lesssim M_B \lesssim -11.6$, these objects have luminosities comparable to those of the faintest dwarf galaxies then known in the Local Group, but effective radii about an order of magnitude smaller. As a result, these rare systems came to be known as ``ultracompact dwarf" galaxies (UCDs).

Soon afterwards, UCDs were discovered in the Virgo cluster using a combination of HST imaging and Keck spectroscopy \citep{2005ApJ_627_203Hacsegan}. Additional Virgo UCDs were later identified in an AAT/2dF spectroscopic survey \citep{2006AJ_131_312Jones}. More recent studies have increased the number of known UCDs in Virgo, though mainly in the immediate vicinity of M87 (e.g., \citealt{2011AJ_142_199Brodie}).
To date, UCDs have been found in a wide range of environments, including the Centaurus
\citep{2007A+A_472_111Mieske}, Hydra \citep{2007ApJ_668_35Wehner, 2011A+A_531_4Misgeld}, Coma \citep{2009MNRAS_397_1816Price, 2010ApJ_722_1707Madrid, 2011ApJ_737_86Chiboucas}, Antlia \citep{2013MNRAS_430_1088Caso} and Perseus \citep{2014MNRAS_439_3808Penny} clusters, as well as in proximity to some group and field galaxies
\citep{2007MNRAS_378_1036Evstigneeva, 2011ApJ_737_13Madrid, 2009MNRAS_394_97Hau, 2011MNRAS_414_739Norris}.

Generally speaking, the UCD samples amassed by these surveys have continued to close the once-prominent gap between GCs and ``normal", low-mass galaxies, further blurring the distinction between these families of stellar systems.\footnote{A working definition of UCDs has recently been proposed by \citet{2012AJ_144_76Willman}, who take UCDs to be systems with sizes $10 \le r_h \le 100$~pc and absolute magnitudes $-13 \le M_V \le -9$. However, a variety of such definitions exist, and it is worth remembering that all such definitions remain subjective in nature.}
Still, the role of selection effects in shaping our perceptions of ``families", ``gaps", and ``trends" among stellar systems is a matter of concern. Existing UCD catalogues are heterogeneous in nature, having been assembled from a patchwork of surveys with different  biases and completeness limits. This is understandable because UCDs are notoriously difficult to find and study: i.e., most are too small to be resolved in typical ground-based images, rendering spectroscopic surveys the obvious route forward \citep{1999A+AS_134_75Hilker, 2000PASA_17_227Drinkwater, 2006AJ_131_312Jones, 2007A+A_472_111Mieske, 2007MNRAS_378_1036Evstigneeva, 2009A+A_507_1225Adami, 2009AJ_137_498Gregg}. Unfortunately, spectroscopy is both time consuming and inefficient: for instance, the 2dF survey of \citet{2006AJ_131_312Jones} yielded just nine UCDs from a radial velocity sample of 1501 objects, for an overall success rate of $\sim$ 0.6\%.

In principle, high-resolution Hubble Space Telescope (HST) imaging can circumvent this problem by allowing UCDs in the local universe to be resolved directly and separated from GCs on the basis of size and luminosity \citep{2004AJ_128_1529Mieske, 2005ApJ_627_203Hacsegan, 2008AJ_136_2295Blakeslee, 2009MNRAS_394_97Hau, 2010ApJ_722_1707Madrid, 2011ApJ_737_13Madrid, 2012ApJ_760_87Strader, 2013MNRAS_430_1088Caso, 2014MNRAS_443_1151Norris}. Although this approach can improve discovery efficiency, candidates must still be confirmed spectroscopically because contamination from background galaxies can be significant, particularly at faint magnitudes. The foremost concern, though, is HST's small field of view, which severely limits sky coverage and hence spatial completeness in the nearest groups or clusters: e.g., the search for UCDs in the Virgo cluster by \citet{2005ApJ_627_203Hacsegan}, undertaken as part of the ACS Virgo Cluster Survey (ACSVCS; \citealt{2004ApJS_153_223Cote}), covered one hundred sightlines but just $\sim$0.3\% of the total cluster area.

More than 15 years after their discovery, the origin of UCDs remains surprisingly obscure. A number of scenarios --- with varying degrees of theoretical/numerical underpinning --- have been proposed: UCDs could be: (1) otherwise normal objects within the high-luminosity tail of the GC luminosity function; \citep{2002A+A_383_823Mieske, 2012A+A_537_3Mieske}; (2) end-products of the aggregation of young massive star clusters formed during the interaction of gas-rich galaxies \citep{2002MNRAS_330_642Fellhauer, 2005MNRAS_359_223Fellhauer, 2012A+A_547_65Bruns}; (3) the remnants of nucleated dwarf galaxies that have been stripped, or ``threshed", during tidal interactions \citep{2001ApJL_552_105Bekki, 2003Natur_423_519Drinkwater, 2010ApJ_724_64Paudel, 2013MNRAS_433_1997Pfeffer, 2014Natur_513_398Seth,2015MNRAS_449_1716Janz}.

Of course, these scenarios need not be mutually exclusive, and it is entirely possible that multiple formation pathways exist, as noted by \citet{2005ApJ_627_203Hacsegan} and others \citep{2006AJ_131_2442Mieske, 2011AJ_142_199Brodie, 2011MNRAS_412_1627Chilingarian, 2011A+A_525_86DaRocha, 2011MNRAS_414_739Norris}. For instance, \citet{2012A+A_537_3Mieske} examined the specific frequency of UCDs in different environments and showed that they seem to match those of GCs very well, suggestive of a direct GC-UCD link. Other studies found some UCDs to be surrounded by faint halos \citep{2005ApJ_627_203Hacsegan, 2008AJ_136_2295Blakeslee} or even asymmetric extensions \citep{2005A+A_439_533Richtler, 2011AJ_142_199Brodie} that could be explained by nucleated dwarf progenitors. There is also evidence that mergers of super star clusters has played a role in the formation of some young massive objects that could be analogs for some UCD progenitors (e.g., \citealt{2004A+A_416_467Maraston}). Thus, the immediate tasks are to decide which mechanisms have been involved in UCD formation, and to assess their relative importance as a function of time and environment.

A large and carefully selected sample of UCDs --- preferably identified in a survey with a well characterized selection function --- is needed to better understand the origin of these puzzling systems. As the nearest rich cluster of galaxies \citep{2005ApJS_156_113Mei, 2009ApJ_694_556Blakeslee}, Virgo is an obvious choice for a comprehensive UCD survey. At a distance of $\sim$16.5~Mpc, a UCD in Virgo with a typical half-light radius of $\sim 20$ pc (corresponding to $\sim0\farcs25$) can, as we will show below, be resolved in high-quality ground-based images. Moreover, the cluster itself is quite rich, containing $\sim$ 2000 cataloged member galaxies of virtually all morphological types \citep{1987AJ_94_251Binggeli}, tens of thousands of GCs, and large numbers of nucleated galaxies \citep{1987AJ_94_251Binggeli, 2006ApJS_165_57Cote, 2005ApJ_634_1002Jordan, 2009ApJ_703_939Harris}, making it possible to compare the properties of these stellar systems within a single, homogenous survey.

The Next Generation Virgo cluster Survey (NGVS) is a large program carried out with the 3.6 m Canada-France-Hawaii Telescope (CFHT) between 2008 and 2013. The project has been described in detail in \citealt{2012ApJS_200_4Ferrarese} (Paper~I). Briefly, the survey used the MegaCam instrument to perform panoramic imaging of the Virgo cluster, from its center to virial radius, in the $u^*giz$ filters ($\approx$ 100 deg$^{2}$). For a subset of the survey area, $r$- and $K_s$-band imaging is also available (see, e.g., \citealt{2014ApJS_210_4Munoz}). Especially important for the study of UCDs, the NGVS image quality is uniformly excellent, with a median $i$-band seeing of FWHM $\simeq0\farcs54$ (the first and third quartiles are $\simeq0\farcs52$ and $\simeq0\farcs56$, see also Figure 8 in \citealt{2012ApJS_200_4Ferrarese}). Overall, the depth, areal coverage, and image quality of the survey, as well as the availability of extensive color information, make it an ideal resource for the study of compact stellar systems like UCDs. We will identify UCD candidates using NGVS multi-wavebands data and show that this photometry-based sample is very clean. Because the survey spans a wide range of local density, its also provides an opportunity to investigate the role played by environment in the formation and evolution of UCDs.

In this paper, we use NGVS imaging to carry out a systematic study of UCDs in the Virgo cluster. We focus on three important, high-density regions within the cluster: the sub-clusters centered on the massive, red-sequence galaxies M87, M49 and M60. In future papers, we shall extend our analysis to include the entire cluster, including regions of low density. Other papers in the NGVS series related to the topics considered here include studies of the cluster-wide GC populations in Virgo \citep{2014ApJ_794_103Durrell}, the abundance and dynamical properties of star clusters, UCDs and galaxies in the cluster core \citep{2014ApJ_792_59Zhu,2015ApJ_802_30Zhang,2015ApJ_807_88Grossauer}, the structure and dynamics of compact galaxies in Virgo \citep{2015ApJ_804_70Guerou} and the physical classification of stellar and galactic sources using optical and NIR imaging \citep{2014ApJS_210_4Munoz, 2014ApJ_797_102Raichoor}. \citet{2015ApJ_802_30Zhang} is especially relevant to this paper, in that it presents a detailed kinematic analysis of UCDs and GCs in the core of the cluster.

This paper is structured as follows. We give a brief description of the NGVS imaging and data reduction procedures in \S 2.1 and \S 2.2, while a discussion of our UCD selection methods is given in the remainder of \S2. In \S 3, we examine and compare the properties of UCDs in the M87, M49 and M60 regions, paying particularly close attention to the M87 region where we have the benefit of both $K_s$-band imaging and extensive spectroscopic coverage. The implications of our findings are given in \S 4, while \S 5 presents our conclusions and outlines some possible directions for future work.

\section{Observations}
\label{sec:obs}

\subsection{Survey Overview and Data Reduction}

The NGVS is a deep optical imaging survey in the $u^*$, $g$, $i$ and $z$ bands, with additional partial coverage in $r$, carried out with MegaCam, the 340 megapixel camera on CFHT \citep{2003SPIE_4841_72Boulade}.\footnote{Outside of the M87 region, $r$-band imaging is available at only a fraction of the full survey depth, ranging from 1374 to 4461~sec.} The survey covers an area of $\Omega = 104$ deg$^2$ inscribed within the virial radii of the Virgo A ($R_{200}=1.55$ Mpc) and Virgo B ($R_{200}=0.96$ Mpc) subclusters \citep{2012ApJS_200_4Ferrarese}, which are centered on M87 and M49, respectively. Beginning with the NGVS ``pilot program" (see \citealt{2012ApJS_200_4Ferrarese}), images were collected during six consecutive observing seasons, 2008 through 2013. Each MegaCam pointing (i.e., NGVS field) measures roughly $1^{\circ} \times1^{\circ}$ on the sky.  There is some overlap between fields, so that the 104 deg$^2$ survey area is covered by 117 distinct NGVS pointings. Each NGVS field is assigned a pair of ordered numbers according to its location within the survey grid. Field ($+0,+0$) is in the cluster center and includes M87; the field numbers grow with increasing right ascension and declination. In addition, four background fields were also observed (see Table~1 of \citealt{2012ApJS_200_4Ferrarese}). The background fields are located well beyond the survey boundaries (offset from M87 by $\approx16^{\circ}$), at about three virial radii from the center of the A subcluster and at Galactic latitudes corresponding to the highest and lowest values spanned by the main survey.

Exposure times in seconds for the five  filters are:
\begin{equation}
\begin{array}{lcrcl}
T_{u^*} & = & 11\times582 & = & 6402 \\
T_{g} & = & 5\times634 & = & 3170 \\
T_{r} & = & 7\times687 & = & 4809 \\
T_{i} & = & 5\times411 & = & 2055 \\
T_{z} & = & 8\times550 & = & 4400 \\
\end{array}
\end{equation}
Different dithering patterns were used for the different filters with the exception of the $g$ and $i$ bands which shared an identical strategy. In the $g$-band, the limiting magnitude for point sources is $25.9$ mag (10$\sigma$) while the surface brightness limit for extended sources is $\mu_g\sim29.0$~\magarcsec~at 2$\sigma$ above the mean sky level. This means that the NGVS is roughly three magnitudes deeper than either the Sloan Digital Sky Survey (SDSS; \citealt{2000AJ_120_1579York}) or the Virgo Cluster Catalog (VCC; \citealt{1985AJ_90_1681Binggeli}). In addition, the NGVS image quality is excellent: the FWHM for every image, in all filters, is always better than 1$^{\prime\prime}$. The best seeing conditions were reserved for the $i$ band, where the FWHM never exceeded $0\farcs6$, with a median of $0\farcs54$. The $g$ band images have longer exposure time and the larger FWHM with a median of $0\farcs80$ \citep{2012ApJS_200_4Ferrarese}.

The pre-processing for all  NGVS images includes bad pixel masking, overscan and bias subtraction, flat fielding, fringing correction (which is necessary only for the $i$ and $z$ bands) and scattered light correction. The individual images are then corrected for a global background map and stacked using the {\it MegaPipe} pipeline \citep{2008PASP_120_212Gwyn} with the adoption of an artificial skepticism method (see \citealt{2012ApJS_200_4Ferrarese} for details). The final stacks are then photometrically and astrometrically calibrated to SDSS images and scaled to have photometric zero point of $m_{\rm AB}$~=~30.

For the Virgo core region centered on M87 (a 3.62 deg$^2$ survey that is comprised of the {\tt NGVS+0+0}, {\tt NGVS+0+1}, {\tt NGVS-1+0} and {\tt NGVS-1+1} fields), a deep $K_s$ band imaging survey (Next Generation Virgo cluster Survey - InfraRed, hereafter NGVS-IR) was also carried out with CFHT, using the WIRCam instrument (see \citealt{2014ApJS_210_4Munoz} for more details). The NGVS-IR has a limiting magnitude of $K_s\sim24.4$ mag ($5\sigma$) for point sources and a surface brightness limit of $\mu_{K_s} \simeq 24.4$ \magarcsec. The seeing was always better than $0\farcs7$, with the median seeing of the stacked mosaic being $0\farcs54$. The NGVS-IR raw images were processed using the {\tt $`$I$`$iwi} processing pipeline v2.0 \footnote{see http://www.cfht.hawaii.edu/Instruments/ \\ Imaging/WIRCam/IiwiVersion2Doc.html}, with cosmic rays removed and saturated pixels corrected. After the sky removal, the images were stacked together using the SWARP software package \citep{2002ASPC_281_228Bertin}. Full details on NGVS-IR reductions can be found in \citet{2014ApJS_210_4Munoz}.

\subsection{Catalog Generation and SExtractor Photometry}
\label{sec:cata}

To construct an NGVS compact-source catalog,  SExtractor \citep{1996A+AS_117_393Bertin} was first run on each NGVS field in double image mode. Objects were detected in the $g$-band images using SExtractor parameters: {\tt DETECT\_THRESH=1.5}, {\tt ANALYSIS\_THRESH=5.0}, {\tt DETECT\_MINAREA=2.0}, {\tt DEBLEND\_NTHRESH=32.0} and {\tt DEBLEND\_MINCONT=0.002}, and then measured in each of the $u^*griz$ images. More than 25 million objects were detected and measured across the 117 NGVS fields. All the objects were detected on the {\it MegaPipe} global background stacked images, which have the highest photometric accuracy for point-like sources such as stars and GCs. Note, however, that this catalogue is not suitable for investigations of extended objects, especially low surface brightness galaxies. Future papers in this series will discuss the identification and characterization of extended sources in the survey.

By contrast, UCDs at the distance of Virgo are both bright and compact (see, e.g., \citealt{2005ApJ_627_203Hacsegan} for an early demonstration of this based on HST/ACS imaging) so we expect to detect virtually every UCD in the cluster. As discussed below, the main obstacle facing a systematic study of UCDs in Virgo is the high level of contamination from stars, compact background galaxies and spurious features such as asterisms or blends.

\subsection{Measurement of Half-Light Radii: Methodology, Simulations and Consistency Checks}
\label{sec:size}

While UCDs are often taken to have half-light radii, $r_h$, larger than 10~pc, the exact dividing point with GCs is somewhat arbitrary. Nevertheless, the great majority of GCs are smaller than $r_h \sim 10$~pc; in the Virgo cluster, which lies at a distance of 16.5 Mpc \citep{2005ApJS_156_113Mei, 2009ApJ_694_556Blakeslee}, the average GC half-light radius is $2.7\pm0.35$ pc $\sim 0\farcs035$ \citep[similar results can be found in \citealt{2013ApJ_779_94Webb}] {2005ApJ_634_1002Jordan}. By contrast, the half-light radii of known UCDs span a range, from 11 pc to 93 pc with a median of $\sim 20$~pc or $0\farcs25$ \citep{2005ApJ_627_203Hacsegan, 2008AJ_136_461Evstigneeva}. Virgo galaxies are much larger than GCs or UCDs: i.e., the effective radii of even the most compact galaxies in Virgo are larger than 150 pc $\sim 2^{\prime\prime}$ \citep{2006ApJS_164_334Ferrarese}. Thus, a typical GC in Virgo can only be resolved at HST resolution, but a typical UCD can be resolved in high-quality ground-based images, like those from the NGVS. Note that the MegaCam pixel size is 0\farcs187, which corresponds to $\sim15$~pc at this distance.

In order to measure structural parameters --- including half-light radii and total magnitudes --- for the many thousands of GC candidates in the ACSVCS, \citet{2005ApJ_634_1002Jordan} developed the \king software package.  Briefly, \king fits --- inside a fitting radius, $r_{\textrm{fit}}$ --- PSF-convolved \citet{1966AJ_71_64King} models to each GC candidate and then determines structural parameters via $\chi^2$ minimization. Needless to say, reliable PSF determination is essential for the successful application of this code. In the case of the ACSVCS, the \king measurement limit on $r_h$ was found to be $\sim1$ pc, which is about $1/8$th of the instrumental PSF for ACS/WFC (FWHM~$\sim$~0\farcs1). The uncertainty on $r_h$ was also found to grow when $r_h$ became comparable to $r_{\textrm{fit}}$.

In the present study, \king has been adapted to measure the half-light radii for UCDs and bright GCs in the NGVS images. Not surprisingly, the main challenge in an application of \king to ground-based images lies in the estimation of the PSF. Although the NGVS image quality is superb for a ground-based survey,  the PSF is nevertheless much broader than that of HST (by about a factor of five in the $i$ band). In addition, ground-based PSFs can change from field to field, and may vary significantly across an individual MegaCam pointing.

Complete details on the construction of PSFs in the NGVS will be presented in Gwyn et al. (2015). Briefly, candidate PSF stars were first identified using SExtractor and DAOphot (only stars identified by {\it both} SExtractor and DAOphot were used). Additionally, stars within a distance of 40 pixels from other bright objects were discarded. The PSF was then generated using the DAOphot routine \texttt{psf} with a bivariate Gaussian and allowed to vary across the field with second order variations. In each NGVS field, the residual images for the PSF stars were checked to ensure there were no systematic trends with magnitude or position, and  \king was run only on bright point sources with high signal-to-noise (S/N) ratios. Measurements in the $g$ and $i$ bands (which have the highest S/N and the best image quality) were made independently. Our final $r_h$ values represent weighted averages of the measurements in the $g$ and $i$ bands.

For the catalog described in \S\ref{sec:cata}, a set of aperture magnitudes (within diameters of 3, 4, 5, 6, 7, 8, 16 and 32 pixels)\footnote{These are equivalent to aperture radii of 0\farcs28, 0\farcs37, 0\farcs47, 0\farcs56, 0\farcs65, 0\farcs75, 1\farcs5 and 3\farcs0. In physical units, these radii are 22, 30, 37, 45, 52, 60, 120 and 240~pc at the distance of Virgo.} were measured for all objects. Because the PSFs differ from field to field, the magnitude within a given aperture does not represent the same percentage of total flux. Accordingly, we corrected the aperture magnitudes to {\it infinite} aperture in a two-stage process. First, a sample of bright (but unsaturated) stars were selected in each field and their 16-pixel aperture magnitudes corrected to infinity (calibrated to SDSS PSF magnitudes). Then, the other aperture magnitudes (3, 4, 5, 6, 7 and 8 pixels) were corrected to a 16-pixel aperture.

\begin{figure}
\epsscale{1.15}
\plotone{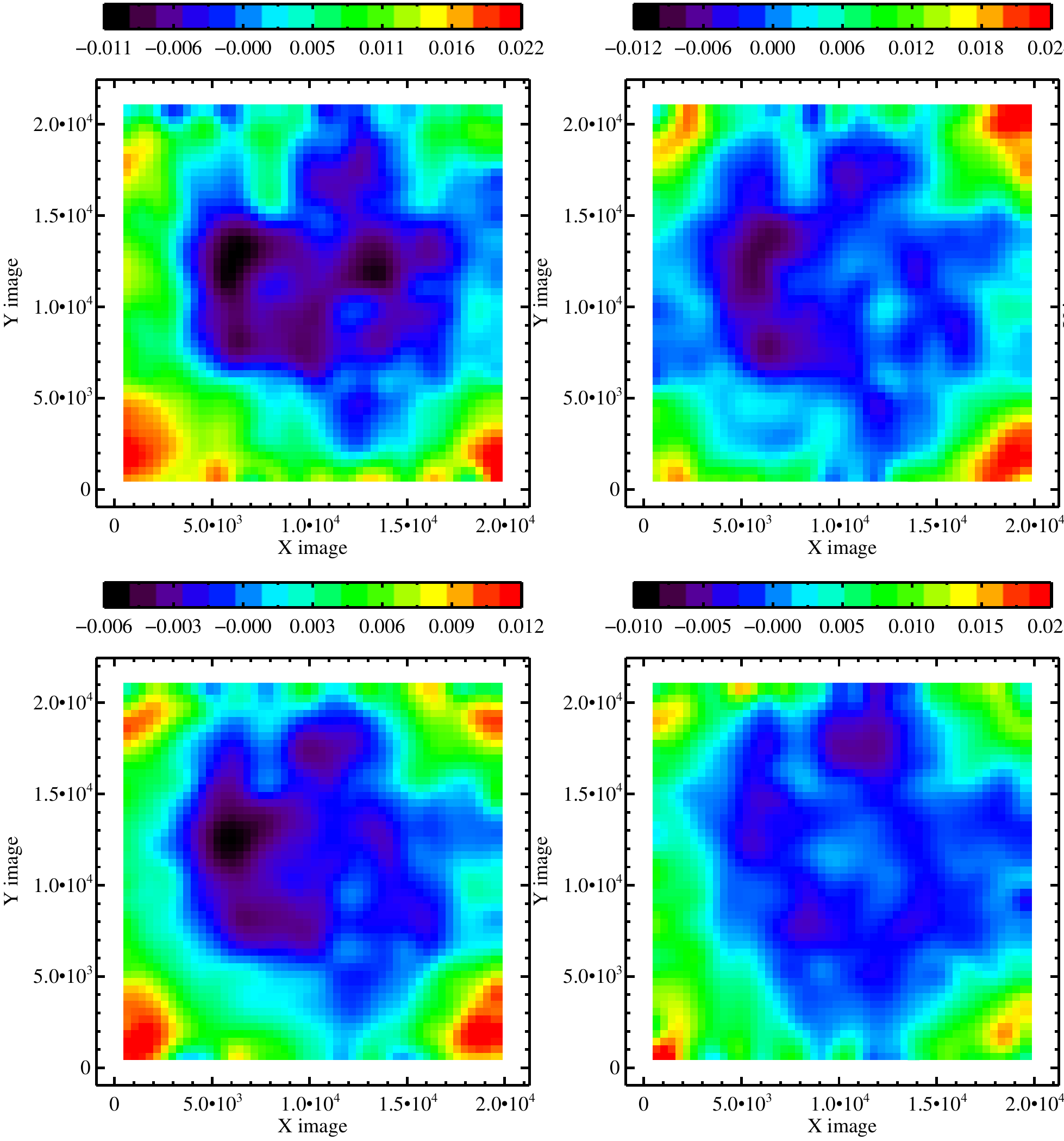}
\caption{Aperture correction map (from 8-pixel diameter apertures to 16-pixel diameter apertures)  made by stacking all individual NGVS fields. The four panels show results for the $u^*$ (upper-left panel), $g$ (upper-right panel), $i$ (lower-left panel) and $z$ (lower-right panel) bands. Color coding shows the deviations in magnitudes from the mean correction across each MegaCam pointing.}
\label{fig:ap8toap16}
\end{figure}

After the application of these corrections, we have a homogeneous database of point-source photometry across the entire NGVS survey area, all calibrated against the SDSS. This catalog also accounts for the small but measurable spatial variations in the PSF within individual fields, thereby tightening the stellar color-color and color-magnitude diagrams and aiding in the selection of targets (see, e.g., Figure~1 of \citealt{2014ApJ_794_103Durrell}). Figure~\ref{fig:ap8toap16} shows the aperture correction map made by stacking individual MegaCam pointings. The color coding highlights regions of negative and positive deviations in blue and red, respectively.

\begin{figure}
\epsscale{1.15}
\plotone{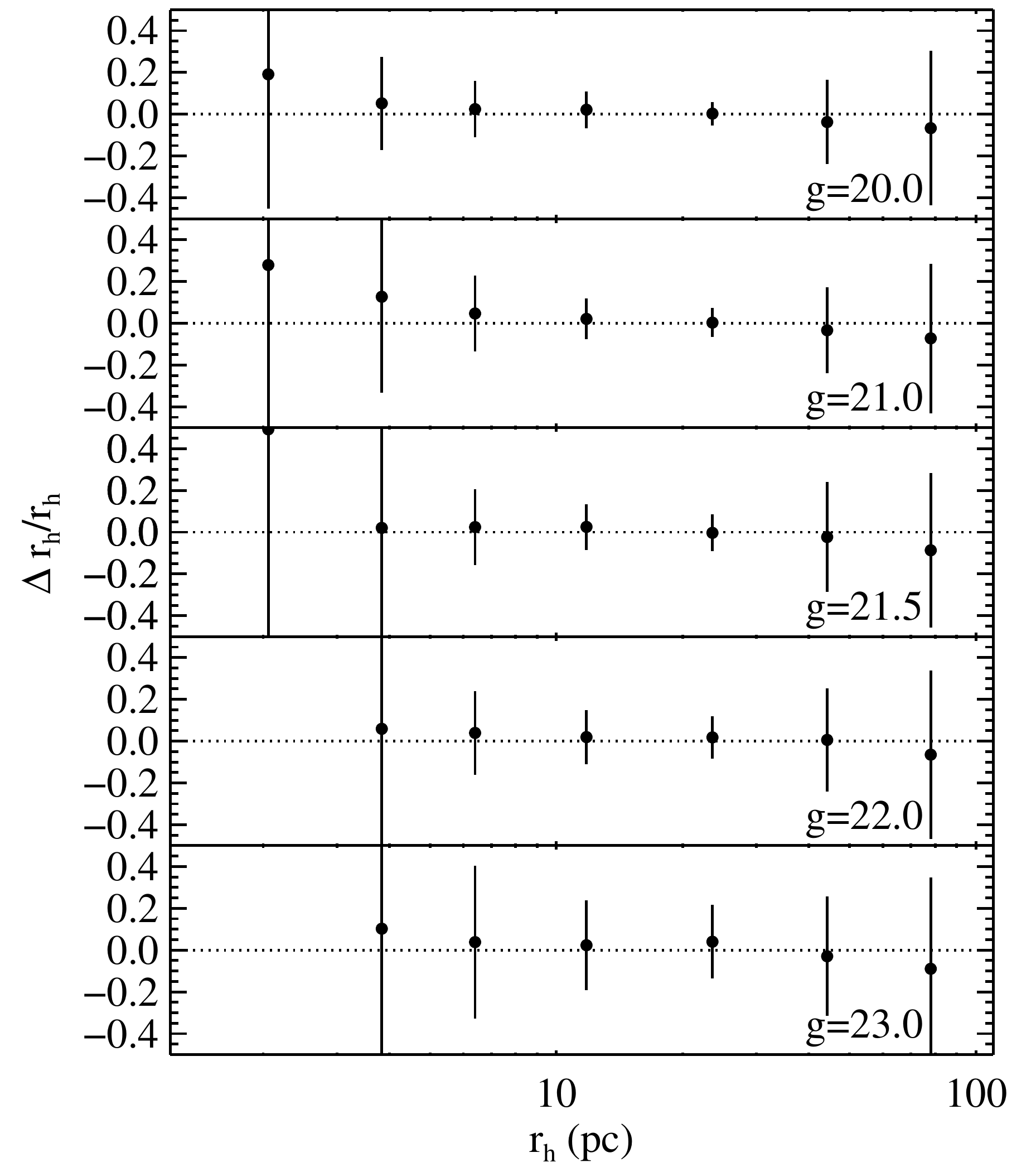}
\caption{Fractional difference between the measured and input half-light radius, $\Delta r_h \equiv r_{\textrm{h,detected}} - r_{\textrm{h,real}}$, plotted as a function of $r_h$. The points show the mean $\Delta r_h/r_h$ in bins of $r_h$. Errorbars show the standard deviation of sizes measured by KINGPHOT. The mean magnitude of the simulated objects is reported in the lower right corner of each panel.}
\label{fig:rh_sim_king}
\end{figure}

It is important to determine the ranges in effective radius and magnitude over which the NGVS \king measurements  are reliable. We have addressed this question in several ways. First, we carried out a large number of simulations in which artificial sources were generated using PSF-convolved King models plus a realistic amount of noise. The simulated objects were then added randomly to NGVS images and their sizes measured using KINGPHOT. The artificial objects had $g$-band magnitudes in the range 20 to 23 mag, half-light radii between 1 and 100~pc, and concentration parameters, $c \equiv \log_{10}{r_t/r_c}$, between 0.5 and 2.3\footnote{Here $r_t$ and $r_c$ are tidal and core radius, respectively.}. Figure~\ref{fig:rh_sim_king} compares the actual and measured cluster sizes from these simulations; the symbols show mean values of $\Delta r_h/r_h$ while the errorbars show the standard deviation at each $r_h$. Results are shown, from top to bottom, in bins of decreasing magnitude ($g$ = 20 to 23 mag). The \king measurements are found to be in good agreement with the actual sizes over a wide range in radius, with the exception of the most compact objects which have input sizes of $r_{\rm h} \sim 2$~pc.

The simulations show that there is a tendency to overestimate the sizes of the smallest objects and underestimate the sizes of the largest objects. The most reliable \king measurements are obtained for objects brighter than $g\sim21.5$ mag and having half-light radii of $r_h\sim30$~pc. Of course, there are good reasons to expect that the measurements for the larger ($r_h\gtrsim50$~pc) objects would show increased scatter. First, at fixed magnitude, larger objects always have lower mean surface brightness, and the lower S/N ratio will naturally lead to an enhanced scatter. Second, in our analysis the \king fitting radius for the NGVS is 7 pixels. This corresponds to $\simeq105$~pc at the distance of Virgo. \citet{2005ApJ_634_1002Jordan} have shown that the \king measurements begin to show a bias when $r_h \gtrsim r_{\textrm{fit}}/2$, which translates to $\gtrsim$~50~pc for the NGVS. As noted above, UCDs are often defined to have sizes between 10 and 100 pc, although the exact limits are arbitrary and several recent studies have found that most UCDs have $10\le r_h \le 30$~pc (\citealt{2011AJ_142_199Brodie, 2011ApJ_737_86Chiboucas, 2011ApJS_197_33Strader, 2012MNRAS_422_885Penny}). Our choice of fitting radius thus seems appropriate for the majority of known UCDs. We conclude from the simulations that the \king measurements should be reliable for UCDs and GCs brighter than $g \simeq 21.5$ and having half-light radii in the range $\sim$10--100~pc.

\begin{figure}
\epsscale{1.0}
\epsscale{1.10}
\plotone{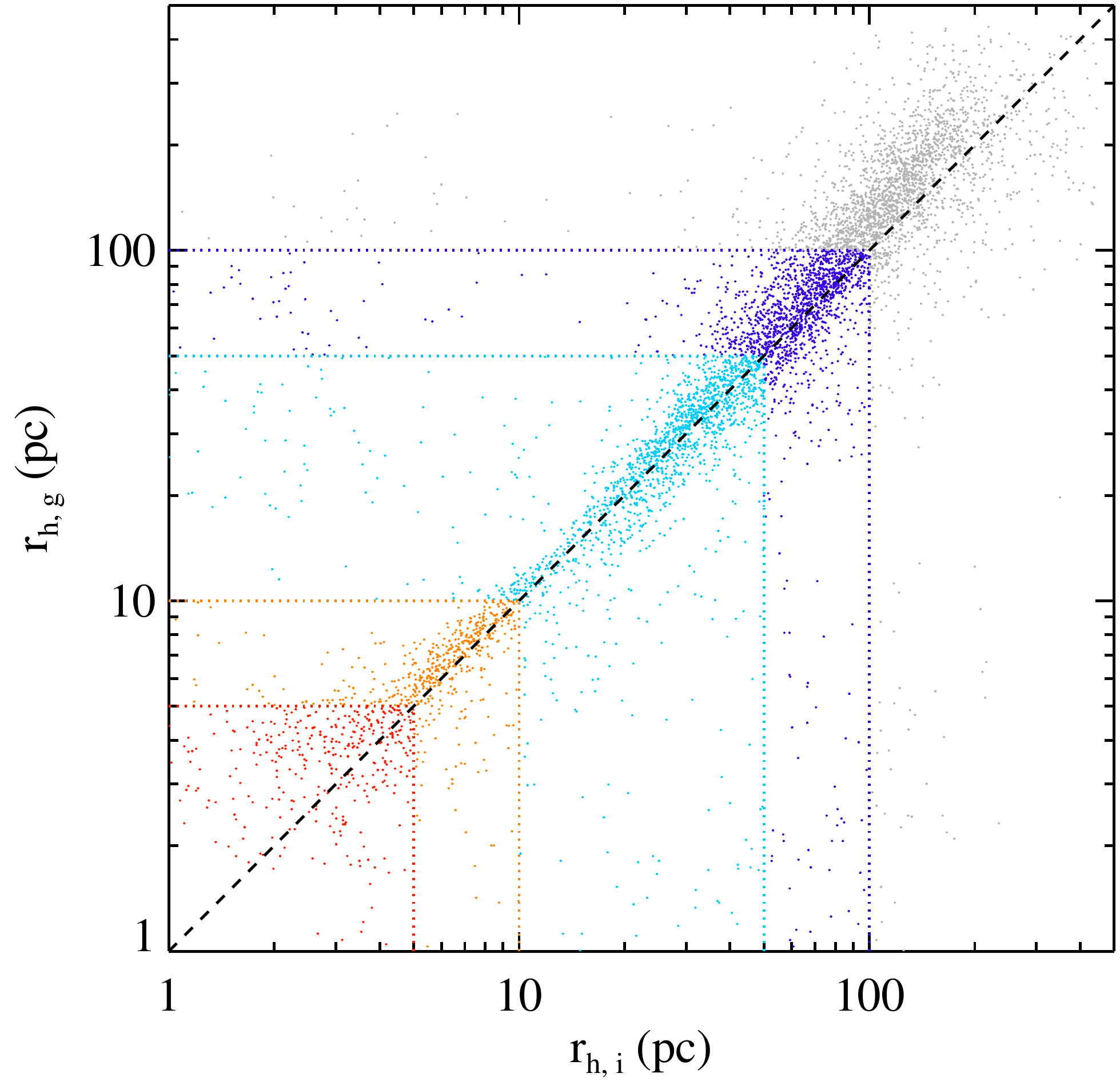}
\caption{A comparison of half-light radii measured in the $g$ and $i$ bands for objects brighter than $g=21.5$. The dashed line shows the one-to-one relation, while the dotted lines show half-light radii of 5, 10, 50 and 100 pc.}
\label{fig:rg_ri}
\end{figure}

\begin{figure}
\epsscale{1.10}
\plotone{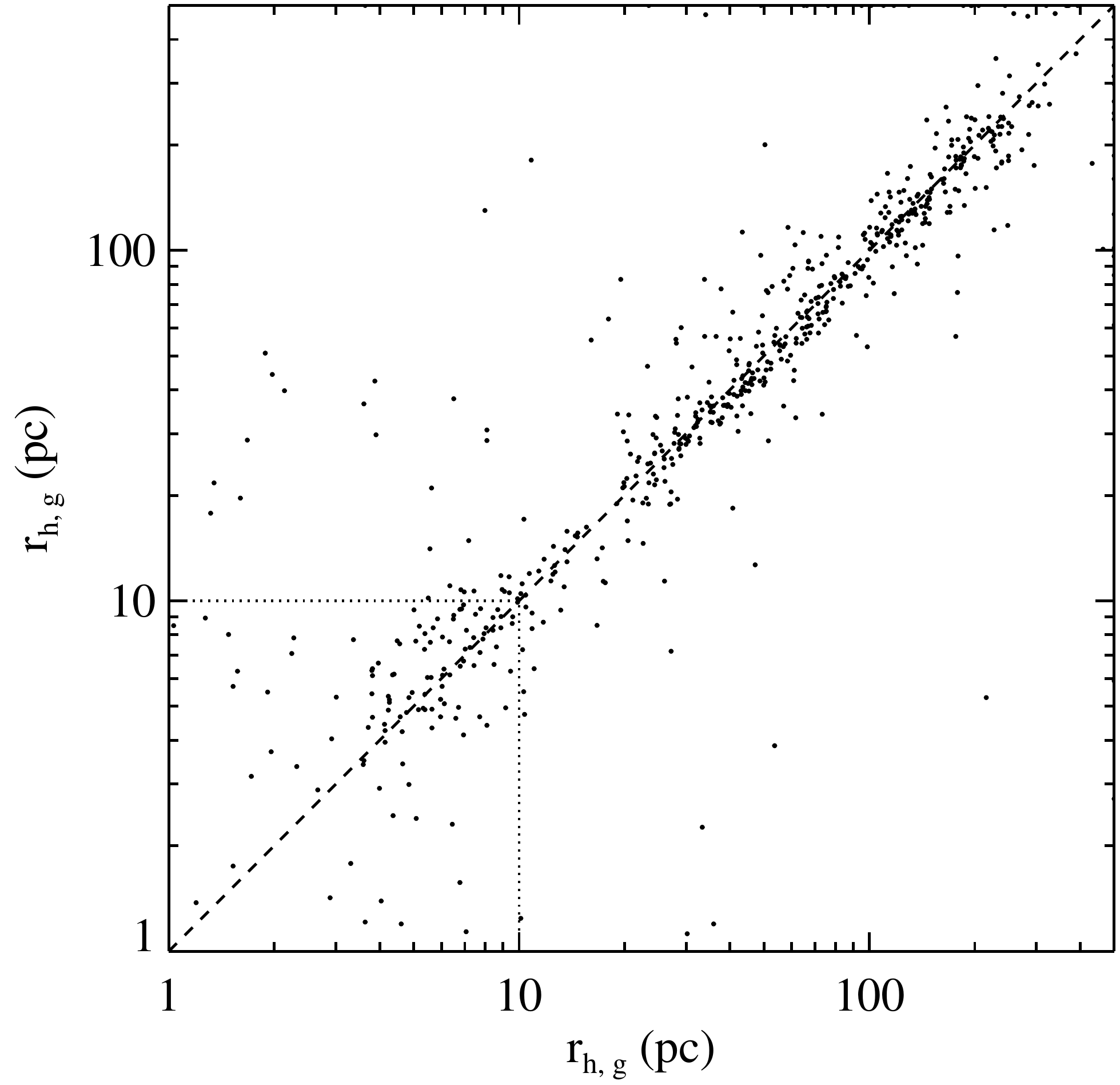}
\caption{Comparison of $g$-band half-light radii measured in the overlap regions of the four NGVS fields surrounding M87. All objects shown in this plot have $g \le 21.5$. The dashed line shows the one-to-one relation while the dotted lines shows a half-light radius of $r_h=10$~pc.}\label{fig:rg_overlap}
\end{figure}

A second check on the reliability of the \king sizes is possible using the independent $g$- and $i$-band measurements. We compare the measurements in these two bands in Figure \ref{fig:rg_ri}, now showing only those objects with $g \le 21.5$ mag (see above). The dotted lines show sizes of 5, 10, 50 and 100~pc. The half-light radii are in good agreement over a wide range in radius --- roughly from 5~pc to 100~pc. A third check on the NGVS size measurements is possible using regions of overlap between adjacent NGVS fields, where some sources can have their sizes measured in completely independent stacks. Figure \ref{fig:rg_overlap} shows this comparison for the overlap regions in the four M87 fields. The agreement is once again very good, apart from the most compact objects. In fact, this comparison is likely to be overly pessimistic since the measurements in this case are made entirely at the edges of individual MegaCam fields where the image quality is poorest.

\begin{figure}
\epsscale{1.10}
\plotone{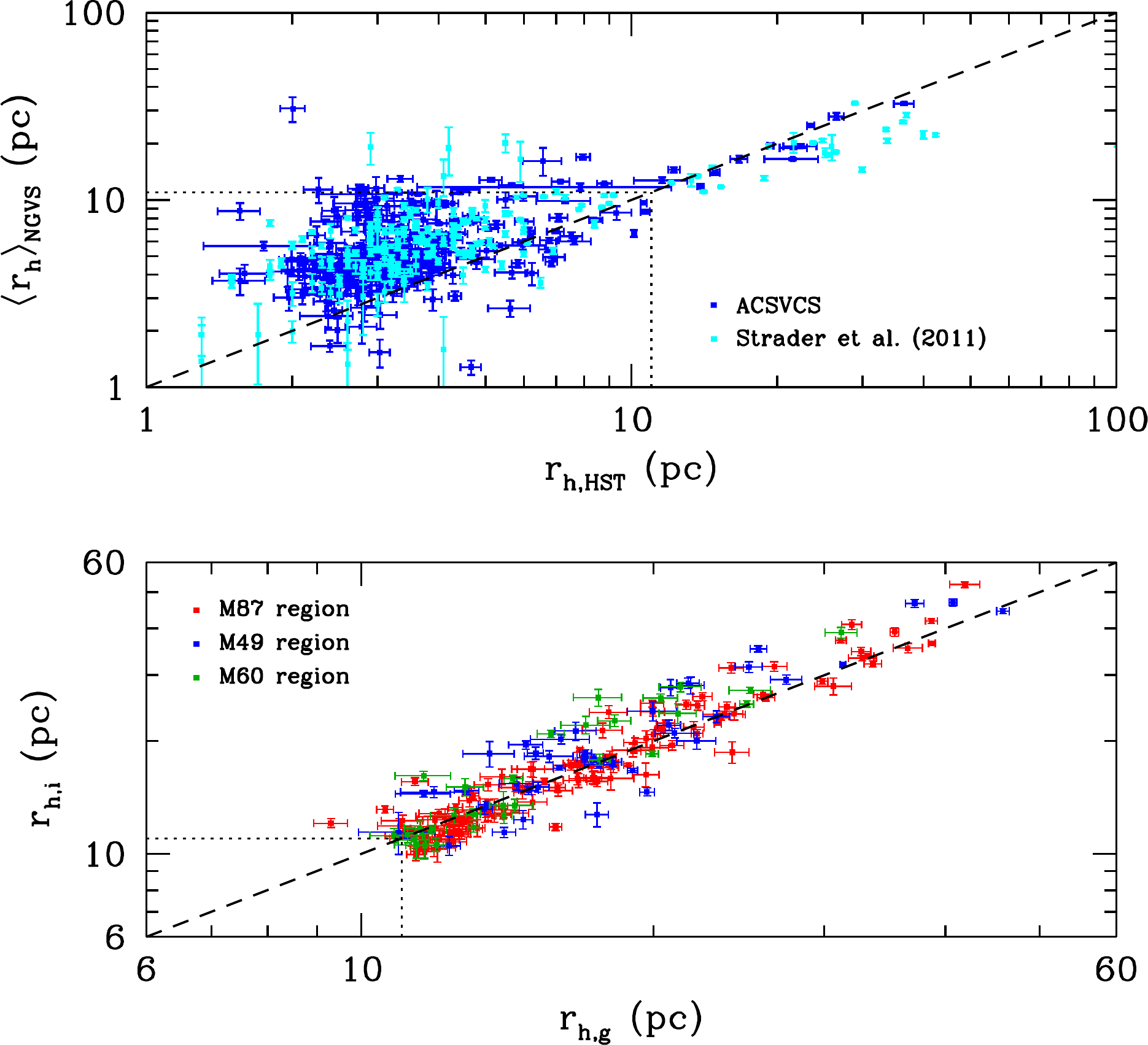}
\caption{(Upper Panel) Comparison of GC and UCD half-light radii measured with the NGVS to those measured from HST-based imaging. The NGVS errorbars show the weighted mean \king uncertainties in the $g$ and $i$ bands. The blue symbols show objects with published sizes from the ACSVCS \citep{2005ApJ_634_1002Jordan, 2005ApJ_627_203Hacsegan, 2009ApJS_180_54Jordan} while the cyan symbols show HST measurements from  \citet{2011ApJS_197_33Strader} for GCs and UCDs in the M87 region.
(Lower Panel) Comparison of NGVS half-light radii measured in the $g$ and $i$ bands for our final sample of UCD candidates in the M87, M49 and M60 regions. See \S\ref{sec:size} and \S\ref{sec:selection} for details on the selection procedure.
}
\label{fig:rh_ngvs_hst}
\end{figure}

Finally, we can compare our sizes with those measured from HST images. The blue squares in the upper panel of Figure \ref{fig:rh_ngvs_hst} show the results of this comparison for GCs and UCDs from the ACSVCS \citep{2005ApJ_634_1002Jordan, 2005ApJ_627_203Hacsegan}. Note that the NGVS sizes shown here refer to weighted mean sizes measured in the $g$ and $i$ bands. The agreement is quite good for objects with $r_h \gtrsim$~10~pc, although it is worth bearing in mind that the ACSVCS sizes were also made with KINGPHOT. The cyan squares in this figure show objects from \citet{2011ApJS_197_33Strader}, who collected HST images and measured half-light radii for spectroscopically confirmed GCs and UCDs using the ISHAPE program of \cite{1999A+AS_139_393Larsen}. For the smallest objects ($r_h \lesssim$ 5~pc), the NGVS sizes are clearly overestimated, as expected from the simulations shown in Figure~\ref{fig:rh_sim_king}. For larger objects, the NGVS measurements generally compare favorably with the HST values, although there is a $\sim$ 25\% downward offset for $r_h \gtrsim$~10~pc relative to the values of \citet{2011ApJS_197_33Strader}. The explanation of this offset is unclear, and while it may be tempting to attribute the offset to the use of different algorithms (i.e., KINGPHOT vs. ISHAPE), previous studies have shown the codes to yield fully compatible results (e.g., \citealt{2009ApJ_703_42Peng}), at least for GC-sized sources. In any case, our aim in this study is not to measure UCD sizes at a level of precision that rivals HST, but simply to identify any UCDs with sizes in the range $\sim$ 10 to 100~pc.

\begin{figure}
\epsscale{1.10}
\plotone{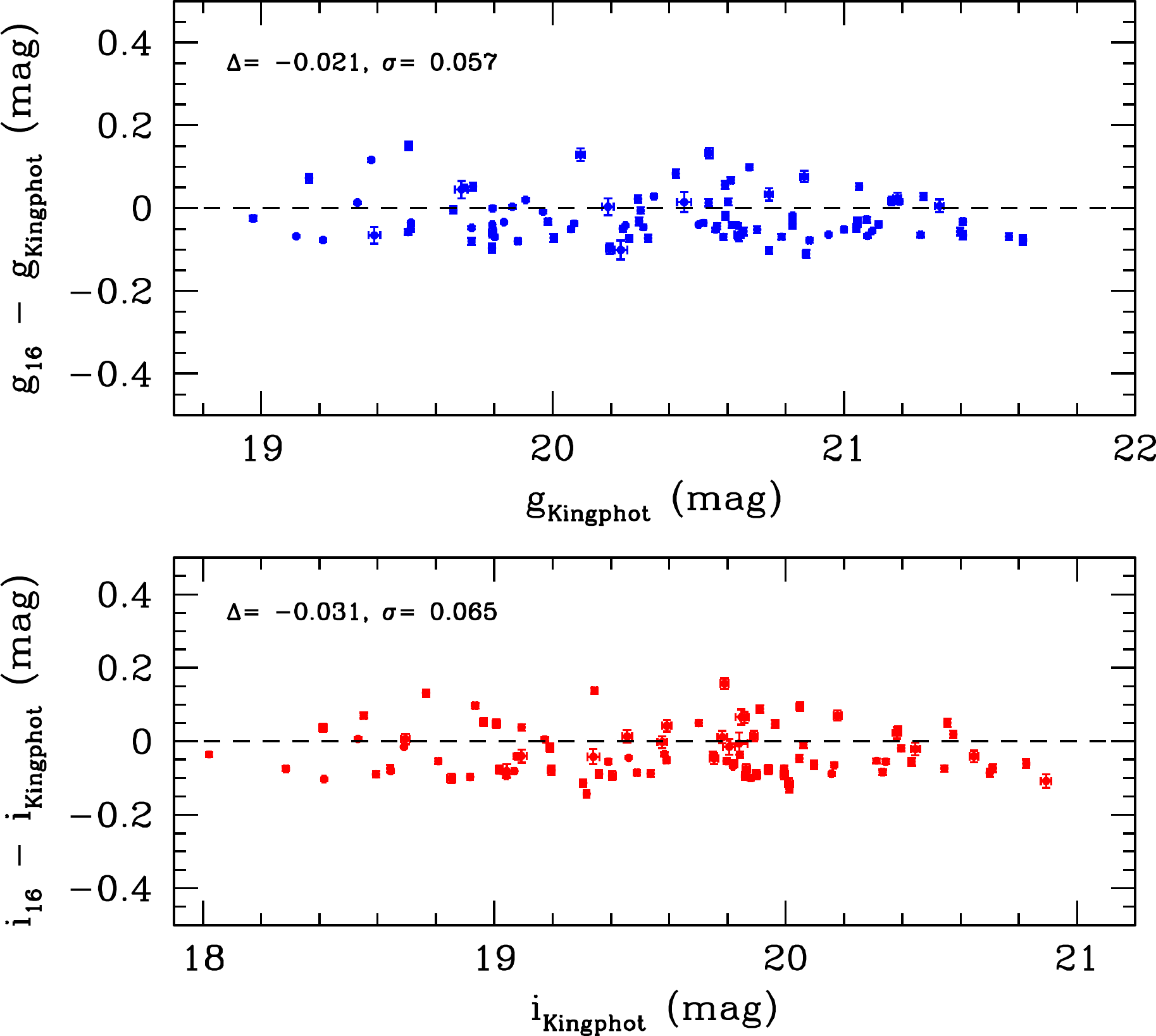}
\caption{Comparison of the total UCD magnitude derived from \king versus that from aperture-corrected, 16-pixel-diameter aperture photometry. The upper and lower panels show results for the $g$ and $i$ bands, respectively. The objects shown here are the final sample of UCDs identified in the M87 region (see \S\ref{sec:cat} for details).}
\label{fig:phot}
\end{figure}

After some experimentation, candidate UCDs identified on the basis of their magnitudes and colors (\S\ref{sec:selection}) were considered to have sizes appropriate for bonafide UCDs if three conditions were satisfied: (1) the measurement errors on $r_h$ were less than 1.5~pc, in both the $g$ and $i$ bands; (2) the fractional difference in the size measurements was $|r_{h,g} - r_{h,i}|$/$\langle{r_h}\rangle \le 0.5$; and (3) the weighted mean half-light radius was $\langle{r_h}\rangle \ge 11$~pc. Additionally, we limit our selection to sources brighter than $g = 21.5$ mag. Taken together, these criteria ensure that objects with uncertain sizes, spurious sources, and compact objects that may suffer from biased size measurements, are eliminated from our analysis.  The lower panel of Figure~\ref{fig:rh_ngvs_hst} compares the $g$ and $i$-band sizes for candidate UCDs identified in our three program regions.

Before proceeding, we pause to explain the photometric measurements used in this analysis. For each UCD candidate, we have $u^*giz$  aperture magnitudes (and, in the M87 region, $u^*grizK_s$), measured in a series of ever larger apertures as described above. We also have PSF-convolved King-model fits from \king in the $g$ and $i$ bands. In principle, these latter values should provide the best photometry for the UCDs. However, there are two limitations with their use: (1) the \king measurements are available only in two of the six bands; and (2) analogous measurements exist only for the brightest ($g \lesssim 21.5$ mag) objects. For these reasons, in this paper we generally rely on the corrected, 16-pixel diameter aperture magnitudes when comparing the UCD and GC properties. In addition, aperture {\it colors} are most reliable when using relatively small apertures, so colors are based on measurements within an 8-pixel diameter aperture. Figure~\ref{fig:phot} compares the 16-pixel aperture and \king magnitudes for our final sample of UCDs (see \S\ref{sec:cat}) in the M87 region, demonstrating that there is good agreement between the two sets of measurements.

\begin{center}
\begin{deluxetable*}{lccccccccccccccc}
 \tablewidth{0pt}
 \tablecaption{Basic Data for M87, M49 and M60.
 \label{tab:3galaxies}}
 \tablehead{
  \colhead{Parameter} &
  \colhead{Units} &
  \colhead{M87} &
  \colhead{M49} &
  \colhead{M60} &
  \colhead{Reference\tablenotemark{1}} \\
  \colhead{(1)} &
  \colhead{(2)} &
  \colhead{(3)} &
  \colhead{(4)} &
  \colhead{(5)} &
  \colhead{(6)} }
 \startdata
    $\alpha({\rm J2000})$                 & deg                      & 187.7059365     & 187.4448543    & 190.9165323   & 1 \\
    $\delta({\rm J2000})$                  & deg                      & +12.3911224     & +8.0004889     & +11.5527000   & 1 \\
    $g$                                             & mag                      &  9.03           &  8.71          &  9.11         & 1 \\
    $\langle(g-i)\rangle$                    & mag                      &  1.15           &  1.12          &  1.10         & 1 \\
    $R_e$                                    & arcsec                   & 96.3            & 105.7          & 76.0          & 1 \\
    ${\langle{\mu}_g\rangle}_e$              & mag arcsec$^{-2}$        & 20.95           & 19.67          & 19.22         & 1 \\
    $c_{82}$                                 &                          & 10.06           &  9.85          &  9.71         & 1 \\
    $\log_{10}[{\it L}_{g,\star}/L_{g,\odot}]$ &                        & 10.92           & 11.04          & 10.89         & 1 \\
    $\log_{10}[{\it M}_{\star}/M_{\odot}]$   &                          & 11.55           & 11.63          & 11.46         & 1 \\
    $\log_{10}[{\it M}_{\rm gas}/M_{\odot}]$ &                          & 13.28           & 11.64          & $\sim$11      & 3,4 \\
    $\log_{10}[{\it M}_{\rm DM}/M_{\odot}]$  &                          & 14.15           & 13.94          & 13.54         & 3,4 \\
    VCC Morphology                           &                          & E2/S0$_1$(2)    & E0             & S0$_1$(2)     & 2 \\
    NGVS Classification Type                 &                          & E-Q             & E(sh$|$st)-Q   & E(sh:)-Q      & 1 \\
    NGC                                      &                          & 4486            & 4472           & 4649          &   \\
    VCC                                      &                          & 1316            & 1226           & 1978          & ~2
\enddata
\tablenotetext{$\dagger$}{Key to references: (1) NGVS ;
(2) \citet{1985AJ_90_1681Binggeli}; (3) \citet{1999A+A_343_420Schindler}; (4) \citet{2006ApJ_646_899Humphrey}. All magnitudes and surface brightnesses have been extinction corrected and transposed to the SDSS photometric system. Luminosities and masses are calculated assuming the distance moduli listed in Mei et al. (2007).}
\end{deluxetable*}
\end{center}

\subsection{Sample Selection}
\label{sec:selection}

To select a sample of UCDs, we must rely on a combination of parameters: magnitudes, half-light radii, colors (specifically, their location within the color-color diagram or diagrams), and mean effective surface brightness. The magnitude range adopted for the selection of both UCDs and GCs in our study is $18.5 \le g \le 21.5$ mag ($-12.7 \lesssim M_g \lesssim -9.7$). The bright limit is set by saturation in the NGVS long exposures while the faint limit reflects the minimum S/N needed to measure accurate sizes using KINGPHOT (see \S\ref{sec:size}). Note that we select both UCD and GC candidates from their location in color-color diagrams (\S\ref{sec:uik} and \ref{sec:ugz}) and identify UCDs on the basis of their larger sizes ($11 \le r_h \le 100$~pc, \S\ref{sec:size}) and lower mean surface brightness at fixed luminosity~(\S\ref{sec:ugz}).

\begin{figure}
\epsscale{1.10}
\plotone{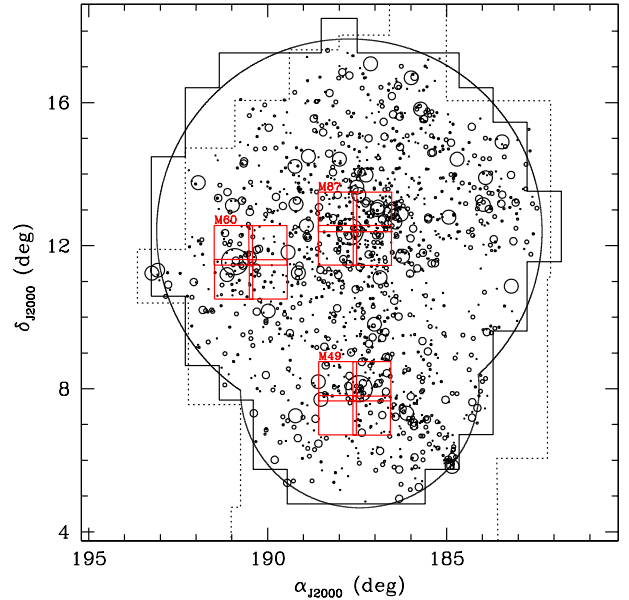}
\caption{Location within the Virgo cluster of the three regions (M87, M49 and M60) whose UCD populations are examined in this paper. Red squares show individual NGVS pointings. VCC galaxies that are confirmed or likely members of the cluster are shown as black circles, with symbol size proportional to luminosity. The solid and dotted lines show the respective boundaries of the NGVS and VCC surveys. The large circles denote the virial radii of the Virgo A and B subclusters (see \citealt{2012ApJS_200_4Ferrarese}).}
\label{fig:ngvs}
\end{figure}

In this work, we aim to study the properties of UCDs around three luminous early-type galaxies in the Virgo cluster: M87 (=NGC4486, VCC1316), M49 (=NGC4472, VCC1226) and M60 (=NGC4649, VCC1978). Some basic properties of these three galaxies are listed in Table~\ref{tab:3galaxies}, including coordinates, magnitudes, colors, effective radii, mean effective surface brightness, concentration $c_{\rm 82}$ (defined as the ratio of the radii containing 80\% and 20\% of the luminosity derived from a curve-of-growth analysis), stellar luminosity, stellar mass, X-ray gas mass, dark matter mass, VCC morphology, NGVS classification type, and other names.

Although these are the three brightest galaxies in the cluster\footnote{The blue luminosities of the three galaxies are quite similar, differing from their mean luminosity by only $\lesssim$~25\%.}, they occupy very different environments, as shown in Figure~\ref{fig:ngvs}. As is well known, Virgo is not a dynamically relaxed system but consists of several distinct subclusters (e.g., \citealt{1987AJ_94_251Binggeli, 1993A+AS_98_275Binggeli, 1999A+A_343_420Schindler, 2007ApJ_655_144Mei}). Subcluster A, which is centered on M87, is the dominant component of Virgo. Subcluster B, which is centered on M49 and located $\sim$4$^{\circ}$ to the south of M87, is considerably smaller and more compact than subcluster A \citep{1993A+AS_98_275Binggeli}. The third, less conspicuous, subcluster (C) is centered on a small group of galaxies containing M59 and M60 located $\sim$3$^{\circ}$ to the east of M87 \citep{1987AJ_94_251Binggeli}.\footnote{A fourth subcluster is probably associated with M86 (VCC763), located $\simeq 1\fdg5$  WNW of M87 (e.g., \citealt{1999A+A_343_420Schindler, 1999LNP_530_9Binggeli, 2007ApJ_655_144Mei}).}
Similarly, X-ray imaging of the cluster shows some dramatic differences in the properties of the hot gas in the vicinity of these galaxies \citep{1994Natur_368_828Bohringer}. Located at the center of subcluster A, M87 is unsurprisingly embedded in a vast, massive reservoir of
X-ray-emitting gas. M49 contains much less ($\sim$ 1/40) hot gas than M87, but substantially more than M60 (4-5$\times$).

These three galaxies also reside in quite different environments in terms of their surrounding diffuse light. Located at the cluster center, M87 possesses an extended diffuse envelope and is embedded in a web of intracluster light \citep{2009ApJ_698_1879Mihos, 2009A+A_507_621Castro-Rodriguez, 2010ApJ_720_569Rudick}. However, this web is highly concentrated in the central regions of Virgo (\citealt{2009A+A_507_621Castro-Rodriguez, 2014A+A_570_69Boselli}, Mihos et al. 2015, in preparation). Rather than being embedded in an extended envelope of diffuse light, M49 shows tidal features scattered througout its halo \citep{2010ApJ_715_972Janowiecki,2012A+A_543_112ArrigoniBattaia,2012ApJS_200_4Ferrarese}, while M60 shows diffuse light associated with M59 \citep{2008ApJ_675_136Yan}.  These differences in the prominence and morphology the diffuse light surrounding each galaxy are likely related to the different dynamical histories of the three regions.

\subsubsection{$u^*iK_s$ Selection}
\label{sec:uik}

\begin{figure}
\epsscale{1.10}
\plotone{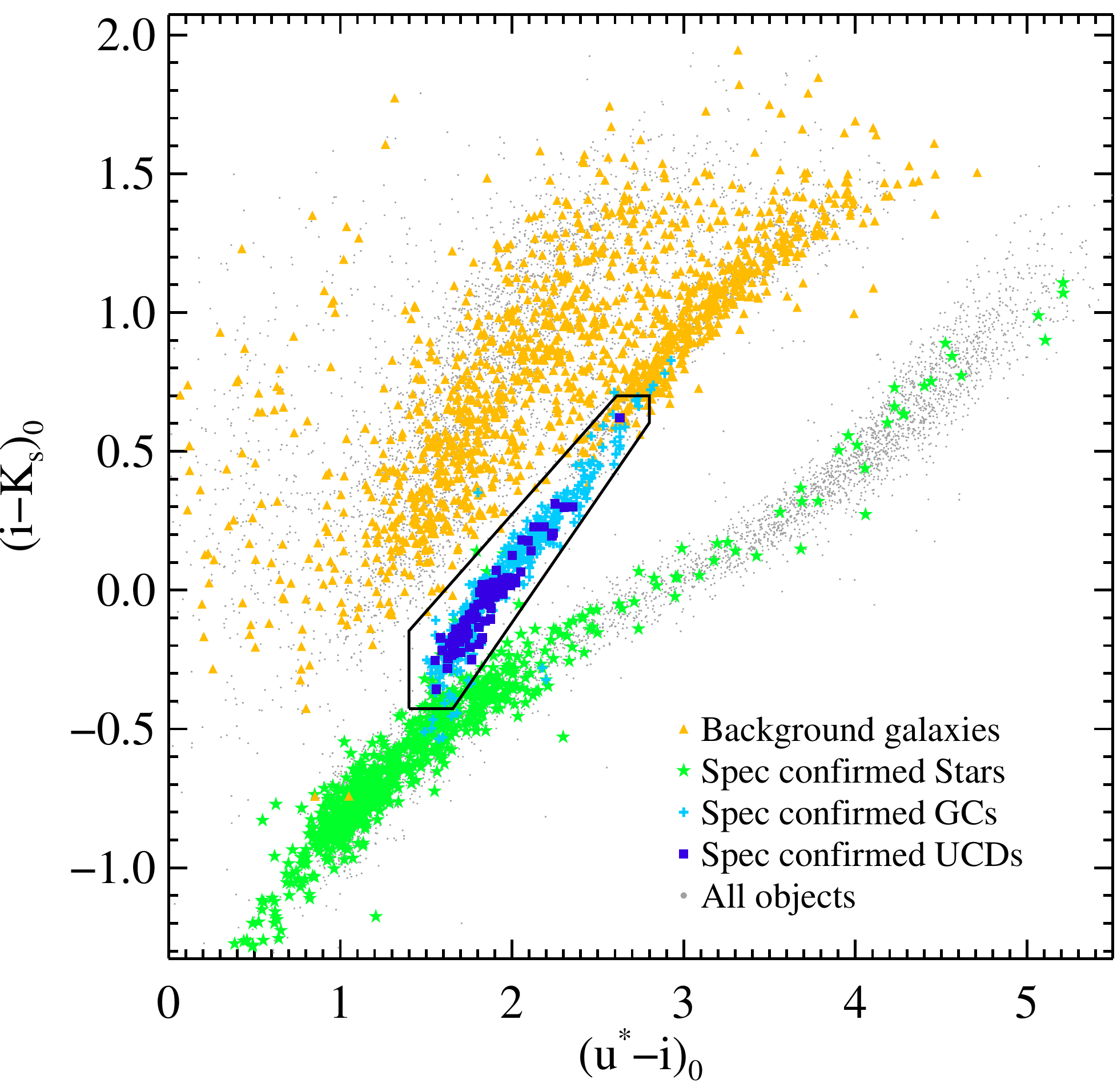}
\caption{Distribution of sources in the M87 region in the ($u^*-i$)-($i-K_s$) diagram (gray symbols).  Blue squares, cyan crosses, green stars and orange triangles show the location of spectroscopically confirmed UCDs, GCs, stars, and background galaxies, respectively. }
\label{fig:ui_ik_pilot}
\end{figure}

In the NGVS pilot program region (the four fields in the immediate vicinity of M87: {\tt NGVS+0+0}, {\tt NGVS+0+1}, {\tt NGVS-1+0} and {\tt NGVS-1+1}), we have imaging in all five optical bands, $u^*griz$ \citep{2012ApJS_200_4Ferrarese}, plus $K_s$-band imaging from the NGVS-IR \citep{2014ApJS_210_4Munoz}. In this region, we are thus able to select UCD and GC candidates directly from the ($u^*-i$)-($i-K_s$) color-color diagram (= $u^*iK_s$) which was shown by \cite{2014ApJS_210_4Munoz} to be remarkably effective in separating UCDs and GCs from foreground stars and background galaxies. Indeed, Figure~\ref{fig:ui_ik_pilot} shows that the sources in the M87 region --- many of which have been observed spectroscopically (see \citealt{2014ApJ_792_59Zhu,2015ApJ_802_30Zhang} --- can be neatly separated into three main groups in the $u^*iK_s$ diagram: GCs and UCDs, background galaxies, and foreground stars. The polygon in this figure shows our adopted selection region for UCDs and GCs, a region that includes all of the spectroscopically confirmed UCDs and most of the confirmed GCs in the vicinity of M87 \citep{2001ApJ_559_812Hanes, 2011ApJS_197_33Strader, 2015ApJ_802_30Zhang}.

\begin{figure}
\epsscale{1.10}
\plotone{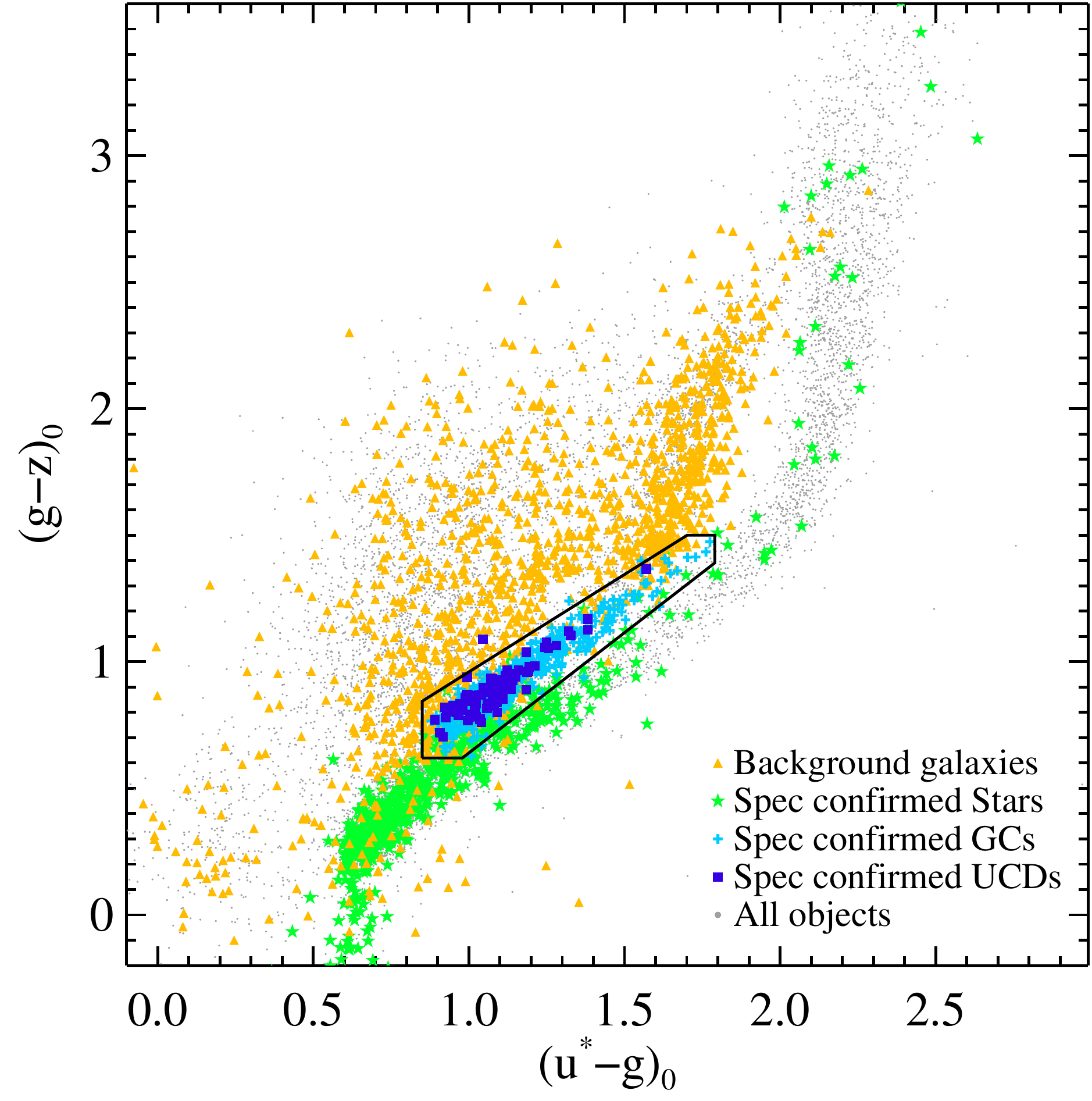}
\caption{Distribution of sources in the M87 region in the ($u^*-g$)-($g-z$) diagram (gray symbols). Blue squares, cyan crosses, green stars and orange triangles show the location of spectroscopically confirmed UCDs, GCs, stars, and background galaxies, respectively. }
\label{fig:ug_gz_pilot}
\end{figure}

\begin{figure}
\epsscale{1.10}
\plotone{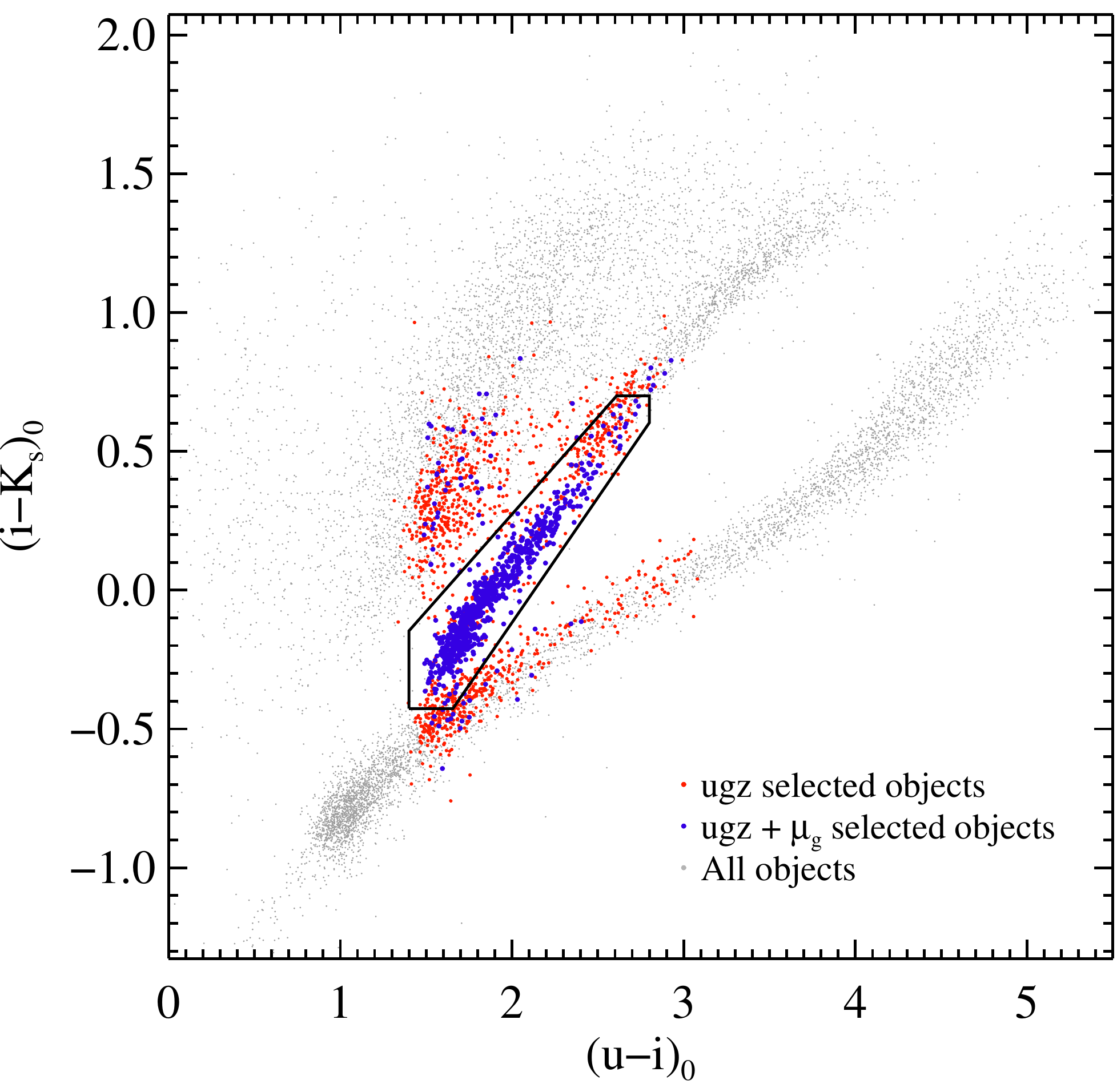}
\caption{Distribution of $u^*gz$-selected sources from Figure~\ref{fig:ug_gz_pilot} in the $u^*iK_s$ diagram. Red and blue symbols denote the combined sample of UCD and GC candidates before and after a second selection on mean effective surface brightness. }
\label{fig:ugz_in_uik}
\end{figure}

\begin{figure}
\epsscale{1.15}
\plotone{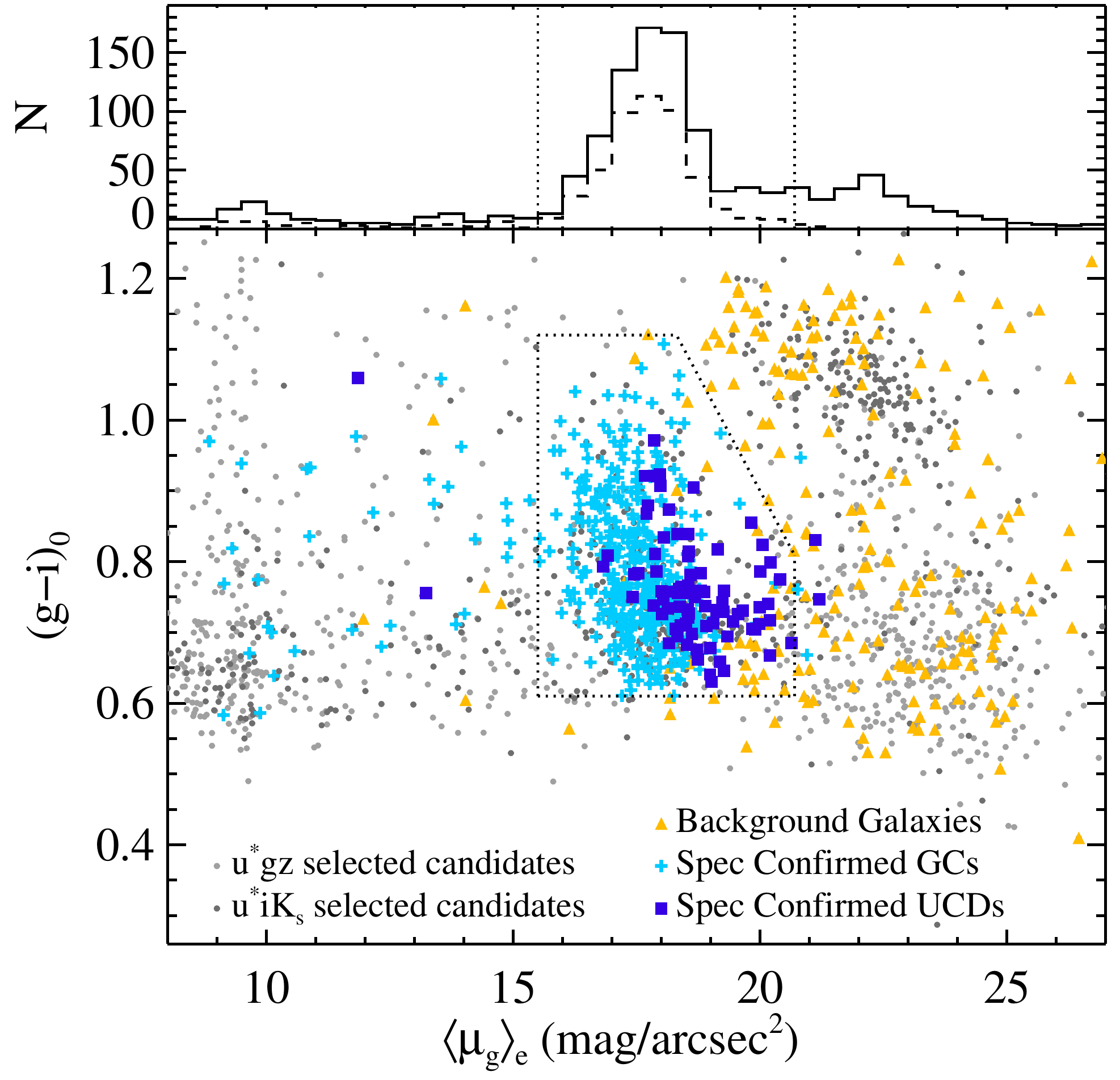}
\caption{(Upper Panel). Histogram of mean effective surface brightness, $\langle{\mu_g}\rangle_e$, for sources in the M87 region. The solid histogram shows the distribution for color-selected (both $u^*iK_s$- and $u^*gz$-selected) UCDs and GC candidates, while the dashed histogram shows the distribution for the subset of spectroscopically confirmed UCD and GCs. The dotted lines show our adopted surface brightness limits for the selection of UCDs. (Lower Panel). Dependence of ($g-i$) color on mean effective surface brightness. Gray dots show all $u^*gz$-selected UCD and GC candidates. Spectroscopically confirmed UCDs, GCs and galaxies are shown by blue squares, cyan crosses and orange triangles, respectively. }
\label{fig:mu_selection}
\end{figure}

\subsubsection{$u^*gz$ Selection}
\label{sec:ugz}

Unfortunately, of the 117 MegaCam fields that make up the NGVS, a $K_s$-band photometric catalog is available at this time in only the four pilot program fields. For the remaining fields, including the M49 and M60 regions, we must rely on the ($u^*-g$)-($g-z$) color-color diagram (= $u^*gz$) to identify UCD and GC candidates. While there is certainly a separation of UCDs and GCs from foreground stars and background galaxies in this color-color space, it is not as dramatic as in the case of $u^*iK_s$ \citep{2014ApJS_210_4Munoz}. As Figure~\ref{fig:ug_gz_pilot} demonstrates, the sample of $u^*gz$-selected UCD and GC candidates (blue and cyan dots) shows some overlap with the regions occupied by spectroscopically confirmed stars and galaxies (green and orange dots). As the previous figure, the polygon in Figure~\ref{fig:ug_gz_pilot} shows our adopted selection region for UCDs and GCs.

To better understand contamination arising from the selection of UCDs and GCs in the $u^*gz$ diagram, we show in Figure~\ref{fig:ugz_in_uik} the distribution of $u^*gz$-selected candidates from the M87 region in the $u^*iK_s$ diagram (red dots). For comparison, the complete sample of sources in this region is shown by the faint gray dots. Clearly, the level of contamination is non-negligible, with many of the $u^*gz$-selected UCD and GCs candidates located in the regions occupied by stars or galaxies.

To reduce the contamination in this $u^*gz$-selected sample, we impose an additional selection on mean effective surface brightness, $\langle{\mu}_g\rangle_e$. In the upper panel of Figure~\ref{fig:mu_selection}, the spectroscopically confirmed UCDs and GCs are found to occupy a relatively narrow range in surface brightness (dashed histogram). The lower panel of Figure~\ref{fig:mu_selection} shows mean effective surface brightness plotted against ($g-i$) color for all $u^*gz$- and $u^*iK_s$-selected sources from the M87 region (gray and black dots, respectively). Spectroscopically confirmed UCDs, GCs and galaxies are shown by the blue, cyan and orange symbols, as indicated in the legend. The irregular pentagon (denoted by the dotted lines) shows the region of $\langle{\mu}_g\rangle_e$-($g-i$) space used to identify UCDs and GCs; the diagonal boundary in the upper right corner was used to minimize contamination from background galaxies, which tend to have red colors and low surface brightness. While a selection on surface brightness does mean that a small number of (high surface brightness) GCs will be missed, all but four of the spectroscopically confirmed UCDs are identified using this approach.

Figure~\ref{fig:ugz_in_uik} plots the $u^*gz$ and surface brightness-selected objects in the $u^*iK_s$ diagram as blue dots. Although a handful of objects are still found in the regions of the diagram occupied by stars or galaxies, the sample is clearly much cleaner than before. We notice that the $u^*iK_s$-selection does introduce many background galaxies, as seen in the lower panel of Figure~\ref{fig:mu_selection} (the cloud of black points in the upper right corner). Therefore, we proceed by selecting UCD candidates relying on a combination of $u^*iK_s$ and $\langle{\mu}_g\rangle_e$-($g-i$) diagram for the M87 region, and a combination of $u^*gz$ and $\langle{\mu}_g\rangle_e$-($g-i$) diagrams for M49 and M60 regions.

\subsubsection{Visual Inspection of UCD Candidates}
\label{sec:visual}

After the color-color selection, the UCD candidates in all three regions (M87, M49 and M60) were separated from GCs on the basis of their half-light radii. Operationally, the UCDs were defined as those objects with $r_h > 11$~pc (\S\ref{sec:size}; see also \citealt{2004PASA_21_375Drinkwater, 2012AJ_144_76Willman, 2012A+A_547_65Bruns, 2012MNRAS_422_885Penny}) and $M_g<-9.7$~mag. Up to this point, UCD selection was performed in a completely objective manner. However, a final, somewhat subjective step --- a visual inspection of each candidate --- is required to eliminate obvious contaminants or artifacts. Such contaminants were usually found to be nucleated dwarf galaxies or compact background galaxies. After a visual inspection, each candidate was assigned a classification code to identify it as a UCD or an another type of object: (1) = {\tt probable UCD}; (2) = {\tt dwarf nucleus}; (3) = {\tt background galaxy}; (4) = {\tt other type of contaminant} (usually a candidate UCD with uncertain properties due to its being located close to a bright star, stellar halo or CCD bleed trail).

\begin{figure}
\epsscale{1.10}
\plotone{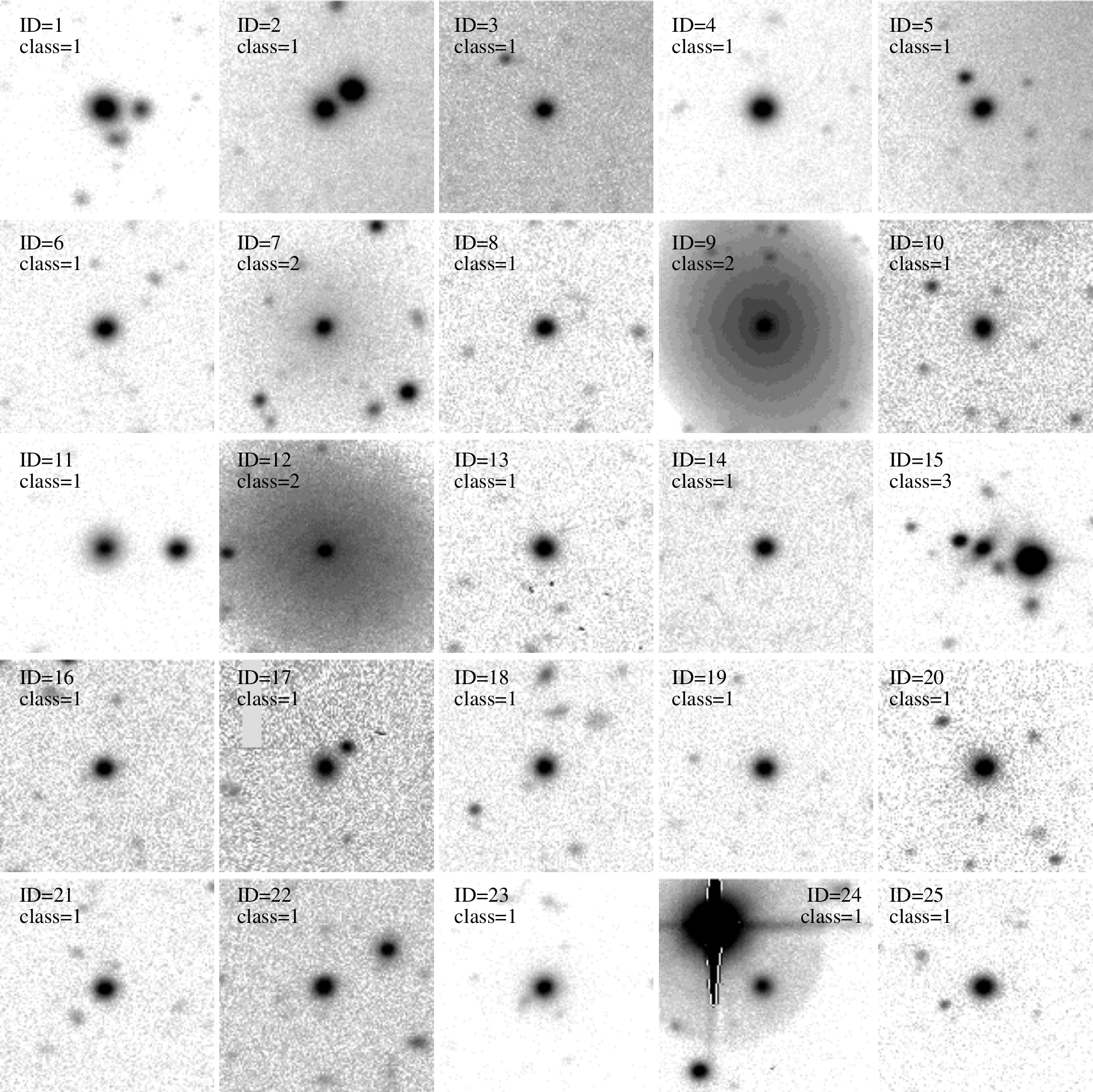}
\caption{UCD candidates 1--25 in the M87 region selected on the basis of magnitude, color and half-light radius (see Tables~\ref{tab:ucd1} and \ref{tab:ucd2}). Most of these candidates are {\tt class 1} objects, although some objects are known nucleated dwarf galaxies (7, 9, 12) or likely background galaxies (15). Each panel measures 60$\times$60 pixels (11\farcs2$\times$11\farcs2). North is up and east to the left.
}
\label{fig:finder1}
\end{figure}

\begin{figure}
\epsscale{1.10}
\plotone{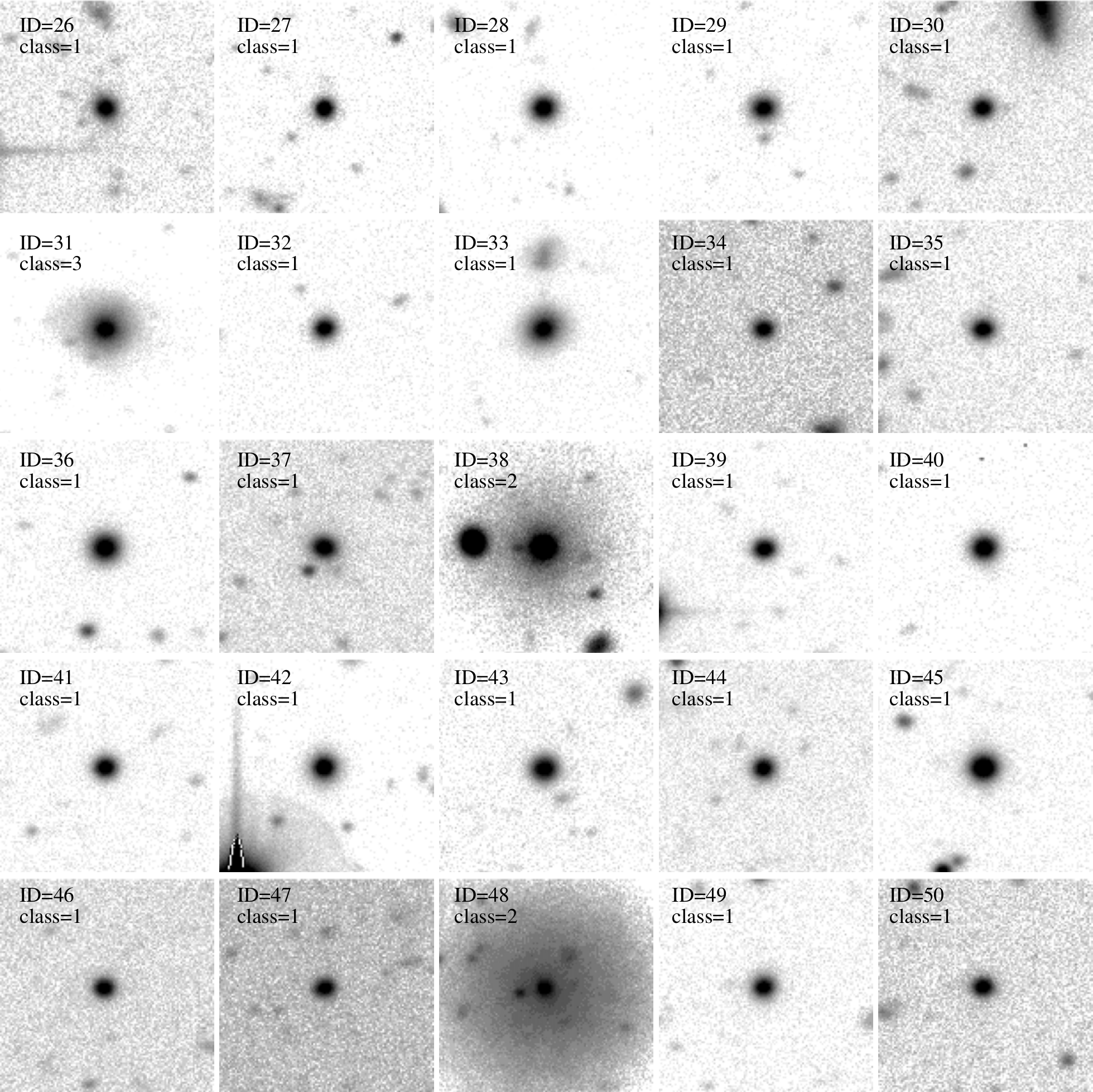}
\caption{Same as Figure~\ref{fig:finder1} except for objects 26--50.
}
\label{fig:finder2}
\end{figure}

\begin{figure}
\epsscale{1.10}
\plotone{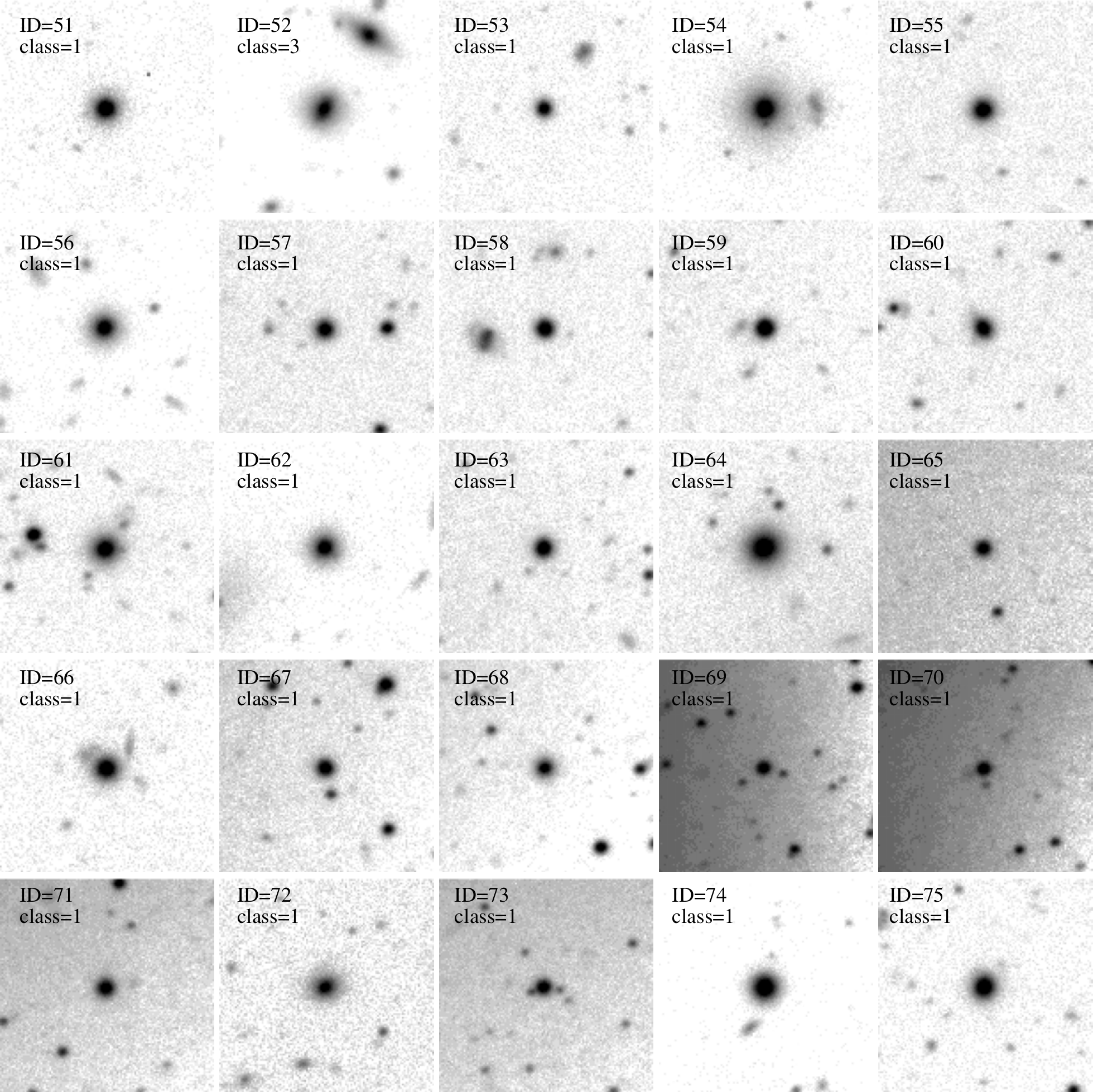}
\caption{Same as Figure~\ref{fig:finder1} except for objects 51--75.
}
\label{fig:finder3}
\end{figure}

\begin{figure}
\epsscale{1.10}
\plotone{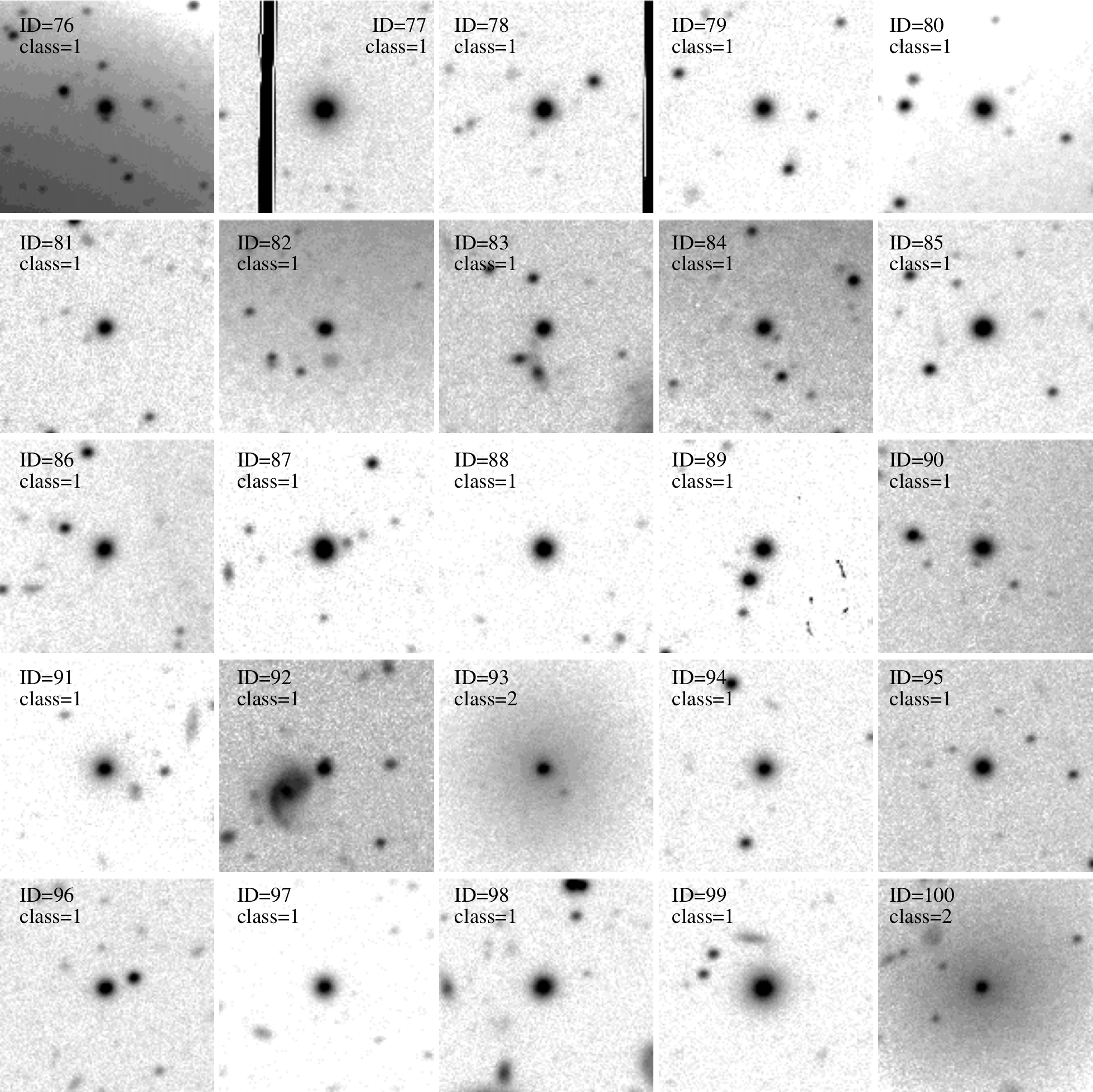}
\caption{Same as Figure~\ref{fig:finder1} except for objects 76--100.
}
\label{fig:finder4}
\end{figure}

\begin{figure}
\epsscale{1.10}
\plotone{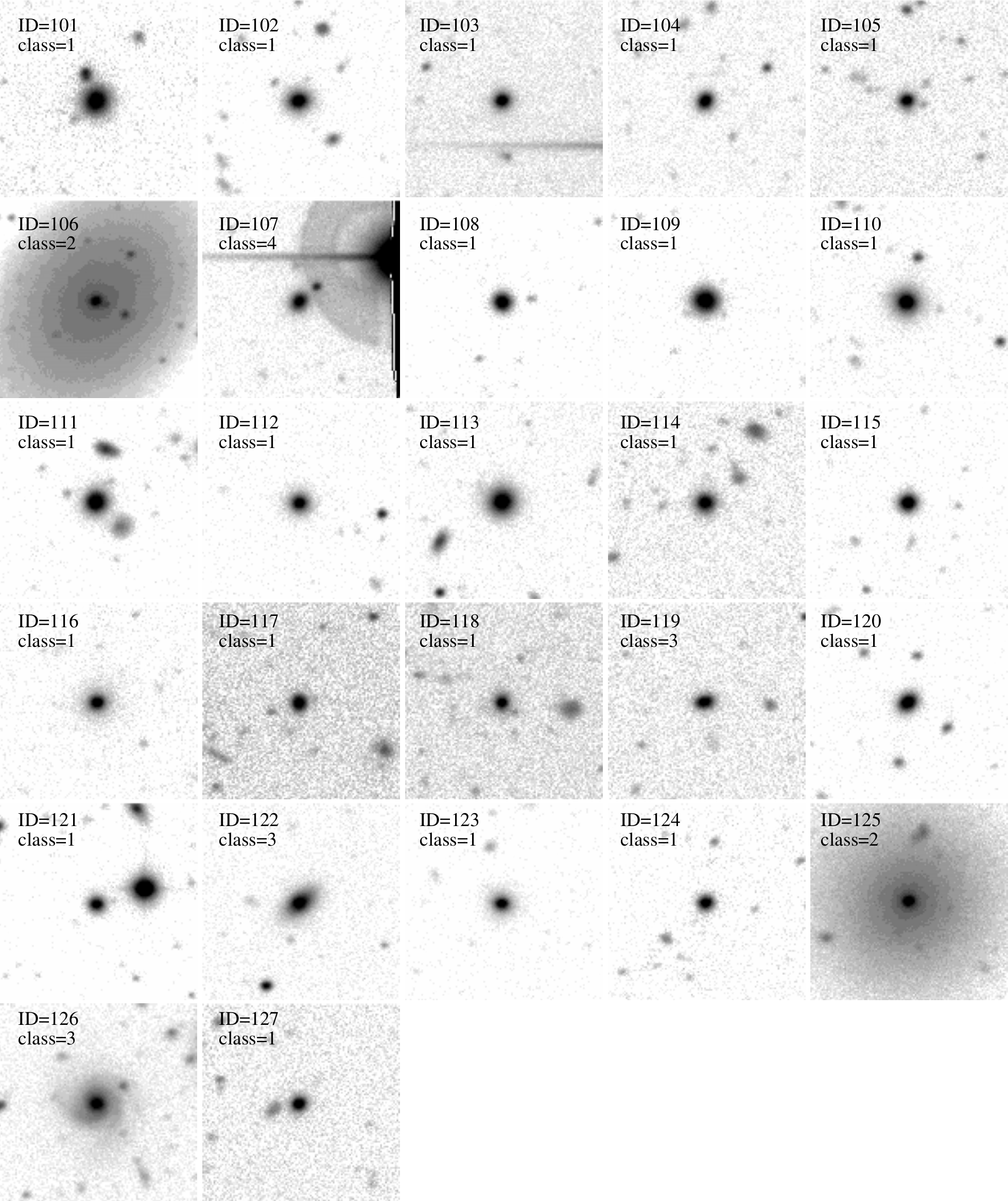}
\caption{Same as Figure~\ref{fig:finder1} except for objects 101--127.
}
\label{fig:finder5}
\end{figure}

Figures \ref{fig:finder1}--\ref{fig:finder5} present $g$-band mosaics for our complete, objectively selected sample of candidate UCDs in the M87 region. Each panel has been labelled with both our {\tt class} parameter and the object identification number from Tables~\ref{tab:ucd1} and \ref{tab:ucd2}. These 127 objects make up our sample of UCD candidates that satisfy the joint selection criteria on magnitude, color and size described above.

\subsubsection{Constraints on Membership and Contamination from Spectroscopy}
\label{sec:spectroscopy}

While the selection criteria described above can provide a very clean sample of likely UCDs from NGVS imaging alone, there is no substitute for spectroscopic redshifts in distinguishing UCDs from foreground stars or background galaxies.

Of course, there is no single spectroscopic survey of the Virgo cluster that is complete to the depth of the UCD samples considered here, but there nevertheless exists a huge amount of spectroscopic data  for bright, compact sources in the direction of Virgo, especially in the cluster core. Such studies include early surveys carried out with MMT \citep{1987AJ_93_779Huchra} and Palomar \citep{1990AJ_99_1823Mould}, as well as subsequent programs with
Keck \citep{1997ApJ_486_230Cohen, 2000AJ_119_162Cohen, 2011ApJS_197_33Strader},
CFHT \citep{2001ApJ_559_812Hanes} and AAT \citep{2006AJ_131_312Jones}.
Beyond the immediate vicinity of M87, the Sloan Digital Sky Survey (SDSS; \citealt{2000AJ_120_1579York}) observed many targets over the full extent of the cluster, although these were mainly Milky Way stars, QSOs and background galaxies. Most recently, a number of coordinated NGVS spectroscopic programs have been undertaken with the MMT, AAT and other facilities (see, e.g., \citealt{2015ApJ_802_30Zhang}). Thus, there exists a rather extensive spectroscopic database that can be compared to our UCD catalog.

Among the sample of 127 UCD candidates found in the M87 region (before visual inspection), 85 have measured velocities  which are listed in Table~\ref{tab:ucd2}. Using these 85 objects, we will estimate the contaminants in both the $u^*iK_s$- and $u^*gz$-selected UCD samples. Of these, 70 and 73 candidates, respectively, were deemed to be likely UCDs on the basis of $u^*iK_s$ and  $u^*gz$ diagrams, {\it and} were classified as {\tt class~=~1} systems. Four of 70 objects in the $u^*iK_s$-selected sample and eight of 73 objects in the $u^*gz$-selected sample have velocities ($v_r>3500$~km/s) that show them to be background galaxies. We would expect $\sim6\%$ of $u^*iK_s$-selected candidates and $\sim11\%$ of $u^*gz$-selected candidates remaining after visual inspection to be misclassified background galaxies.

As noted in \S\ref{sec:intro}, a companion paper in the NGVS series presents a kinematic study of UCDs in the M87 region \citep{2015ApJ_802_30Zhang}. Since their sample is a little different from ours, we pause to compare the slightly different selection strategies used in the two studies. The \citet{2015ApJ_802_30Zhang} analysis is based on radial velocities for 97 UCDs which were identified using a combination of NGVS and HST imaging. Among this sample, 86 UCDs with radial velocity measurements were selected from the NGVS to have $r_h \ge 11$~pc and $g \leq 21.5$. Eleven more UCDs\footnote{Three of these 11 objects were not detected in the NGVS due to their proximity to bright stars.} with measured radial velocities were also included in the \citealt{2015ApJ_802_30Zhang} analysis. Based on earlier, accurate HST size measurements, some of these objects have sizes in the range $9.5 \le r_h \le 11$~pc, which is just below the adopted NGVS cutoff in $r_h$, while the others are larger than $r_h=11$~pc but fainter than $g=21.5$ mag. In short, \citet{2015ApJ_802_30Zhang} adopted slightly more flexible selection criteria in order to explore the kinematics of the M87 UCD system using the largest possible radial velocity sample. Our sample selection relies entirely on NGVS imaging in an effort to maximize spatial completeness and sample homogeneity.

\subsection{Control Fields and Estimation of Contamination}
\label{sec:control_fields}

The NGVS includes four background (BG) fields located far from the cluster center. Each field is offset from M87 by $\sim$16$^{\circ}$, or roughly three virial
radii.\footnote{We note that none of these fields contains a known galaxy cluster with $z\lesssim 0.1$, although a galaxy group, located at $z\sim0.02$ \citep{2007ApJ_655_790Crook}, does fall within the BG4 field.} Because these four fields were observed in the same filters as the other survey fields ($u^*griz$), and under typical NGVS observing conditions with identical exposure times and dithering strategies, we can use these fields to estimate the level of contamination in our UCD samples selected from $u^*gz$ imaging.

In all, the background fields contain a total of 10\,322 objects between our UCD magnitude selection limits of $18.5\leq g \leq 21.5$. Of these, 1145 objects are additionally located in the region of the $u^*gz$ diagram occupied by UCDs and GCs. The selection on surface brightness described in \S\ref{sec:ugz} is quite effective at removing galaxies, however, and leaves a total of only 174 sources. Among these 174 sources, 19 were found to satisfy the three size conditions described in \S\ref{sec:size}. After a visual inspection, only 9 ``UCDs" were left in the sample, a slightly higher number than the four spurious sources identified based on spectroscopy (\S\ref{sec:spectroscopy}) of $u^*iK_s$-selected sources. While the field-to-field variance may be significant (there were 2, 1, 4 and 2 contaminants found in BG1, BG2, BG3 and BG4 fields, respectively), the number density of contaminants in the $u^*gz$-selected sample is $\Sigma(\rm UCD)_{u^*gz} = 2.2\pm0.7$~deg$^{-2}$.

Unfortunately, there is no $K_s$-band imaging for these four background fields. To estimate the contamination in our $u^*iK_s$-selected sample, we use the CFHTLS-D2 field as a proxy background field for the NGVS. CFHTLS-D2 field is one of the four CFHTLS Deep Survey pointings observed in $u^*griz$ filters. In this region, there is also a deep $K_s$ imaging from UltraVISTA \citep{2012A+A_544_156McCracken} which is an ultra-deep, near-infrared survey using VISTA telescope. We find 2731 objects in this field in the magnitude range $18.5 \leq g \leq 21.5$ mag; reassuringly, this is quite similar to the mean number of 2580 objects per NGVS background field in this range. A total of 36 CFHTLS objects fall within the UCD/GC region of the $u^*iK_s$ diagram, only one of which remains after selecting on size and surface brightness. Thus, the $u^*iK_s$-selected UCD sample is found to be exceptionally clean, with a contaminant surface density of $\Sigma(\rm UCD)_{u^*iK_s} \simeq 0.9\pm0.9$~deg$^{-2}$.

\subsection{Catalog of UCD Candidates}
\label{sec:cat}

The complete sample of 127 UCD candidates in the M87 region is presented in Tables~\ref{tab:ucd1} and \ref{tab:ucd2}. Recall that this sample was selected entirely on the basis of magnitude, color (specifically, location in the $u^*iK_s$ and/or $u^*gz$ diagrams), half-light radius and surface brightness, as described in the preceding sections.  For completeness, we list all candidates that were selected {\it before} visual inspection and including even those objects which are known from spectroscopy to be background galaxies. A total of 16 of these 127 UCD candidates were subsequently rejected following a careful examination of the NGVS images (i.e., the objects with {\tt class} parameters of {\tt 2, 3} or {\tt 4} in Table~\ref{tab:ucd2}).

Information on the M87 UCD candidates is divided into two tables. Table~\ref{tab:ucd1} gives, from left to right, each object's identification number, NGVS identifier, right ascension, declination, extinction \citep{1998ApJ_500_525Schlegel}, $g$-band magnitude measured in an aperture with a diameter of 16 pixels and corrected to the SDSS PSF magnitude system, and a variety of 8-pixel diameter aperture colors. Note that these are the observed magnitudes, prior to any correction for extinction. The final two columns of Table~\ref{tab:ucd1} give the parameter ${\Delta}_{\rm env}$, defined in \S\ref{sec:discussion}, followed by our final UCD classification: ``1'' if the candidate is a probable UCD after considering all possible diagnostics (magnitude, color, size, morphology and/or spectroscopy) and ``0'' otherwise. A total of 92 objects have UCD classifications of 1.

Table~\ref{tab:ucd2} records the object identification number, NGVS identifier, $g$- and $i$-band \king magnitudes and radii, followed by previous $r_h$ measurements from the ACSVCS \citep{2005ApJ_634_1002Jordan, 2005ApJ_627_203Hacsegan} or from \citet{2011ApJS_197_33Strader}. The next four columns give the measured radial velocity, if available, from \cite{2011ApJS_197_33Strader}, MMT or AAT \citep{2015ApJ_802_30Zhang}, and SDSS. The NGVS {\tt class} parameter is given in the next column, followed by two flags to indicate whether the object falls inside (1) or outside (0) the GC/UCD selection regions in the $u^*iK_s$ and $u^*gz$ diagrams, respectively, and the UCD classification parameter defined above. The final column gives alternative names from the literature \citep{1985AJ_90_1681Binggeli, 2001ApJ_559_812Hanes, 2006AJ_131_312Jones, 2008MNRAS_389_1539Firth, 2009ApJ_703_939Harris, 2015ApJ_802_30Zhang}. One other object (ID = 7 = NGVS-J122643.32+121743.9) was identified as a faint, nucleated galaxy in our search for low-mass galaxies in the Virgo cluster core.

In the M87 region, 106 of the 127 candidates listed in Tables~\ref{tab:ucd1} and \ref{tab:ucd2} were selected based on magnitude, location in $u^*iK_s$ diagram, size and surface brightness ($uiK_s=1$ in Table~\ref{tab:ucd2}). Among the 106 $u^*iK_s$-selected candidates, 96 were classified as {\tt class = 1} objects. Radial velocities reveal four of these objects to be background galaxies (see \S\ref{sec:spectroscopy}), bringing our cleaned sample of UCDs in the M87 region to 92. Our cleaned sample of 92 UCD candidates is much larger than any previous studied UCD samples in this region (e.g. 34 UCDs in \citealt{2011AJ_142_199Brodie} and \citealt{2011ApJS_197_33Strader}). In addition, the homogeneity is another advantage of our sample. In addition, we also have a homogeneous sample GC and dE,N (see \S\ref{sec:gcs} and \S\ref{sec:nuc}), which allows us to compare the properties between UCDs, GCs and dwarf nuclei.

As described in  \S\ref{sec:ugz}, we also select UCD candidates based on the $u^*gz$ diagram in the M87, M49 and M60 regions. Based on magnitude, location in $u^*gz$ diagram, size and surface brightness, we find 125, 50 and 29 UCD candidates, and 110, 40 and 23 are classified as {\tt class~=~1} in these three regions. 8 and 12 candidates in the M87 and M49 regions have velocity $v_r>3500$~km/s, which we deem to be compact, background galaxies. For the M60 region, only two candidates have velocity measurements and both of them are Virgo members ($v_r<3500$~km/s). Finally, the cleaned $u^*gz$-selected samples have 102, 28 and 23 confirmed or candidate UCDs in these three regions. Table~\ref{tab:ucd_m49_1} and \ref{tab:ucd_m49_2} show the photometric and structure properties of M49 UCD candidates, while Table~\ref{tab:ucd_m60_1} and \ref{tab:ucd_m60_2} show the properties of M60 UCD candidates.

The properties of our final UCD samples are summarized in Table~\ref{tab:sample}, including the region geometry, coordinates, total area, and the number and surface density of the objects within each region. Note that the numbers given in this table refer to the total numbers of candidates identified using the different color-color diagrams, with no selection on redshift.

It is difficult to find all the UCDs within specific magnitude, color, size and surface brightness range. Based on our selection criteria, we cannot find bright UCDs ($g<18.5$) that are saturated in NGVS long exposure images. We may also miss UCDs that are located close to saturated stars or galaxy centers. In addition, the half-light radii of UCDs might increase with galactocentric distance \citep{2013MNRAS_433_1997Pfeffer}, which can introduce a location-dependent incompleteness via our adapted constant size cut $r_h\gtrsim11$~pc. Finally, uncertainties on the size and surface brightness measurements could influence the sample completeness. We know, for instance, that we miss four UCDs due to our surface brightness cut (see Figure~\ref{fig:mu_selection}). A fixed concentration $c=1.5$ is assumed (the same as in \citealt{2005ApJ_634_1002Jordan}) when we measure the half-light radius using \king. Thus, the radii of UCDs with larger concentrated ($c>1.5$) are underestimated, and vice versa. In this case, more compact UCDs ($c>1.5$) may be missed via our size criteria. The PSF variation across the survey area could also influence our size measurements. However, the variation is small (see Figure 8 of \citealt{2012ApJS_200_4Ferrarese}) and we measured the half-light radii by fitting PSF-convolved King (1996) models. So the effect of PSF variations on our UCD sample selection is small.

\subsection{Globular Cluster Sample}
\label{sec:gcs}

As seen in Figure~\ref{fig:ui_ik_pilot}, the GCs and UCDs are clearly separated from foreground stars and background galaxies. In the M87 region, it is reasonable easy to select GC candidates brighter than $g=24$ mag in the $u^*iK_s$ diagram \citep{2014ApJS_210_4Munoz}. We did not adopt other criteria like surface brightness and size since it is difficult to measure the sizes of objects down to $g=24$~mag, and most GCs are unresolved. We eliminate the UCD candidates described above and any other known background galaxies and foreground stars. Finally, we select 9497 GC candidates which are brighter than $g=24$~mag.

Around M49 and M60, where we do not have $K_s$ photometry, the GCs are selected based on $u^*gz$ color-color diagram. The $u^*gz$ diagram is not as optimal as $u^*iK_s$ to separate GCs from stars and background galaxies, especially for the faint objects. Therefore, we only select bright GCs in the M49 and M60 regions. The adopted magnitude criteria is the same as UCD selection, $18.5\leq g\leq21.5$~mag. Finally, 307 and 273 bright GCs are found in the M49 and M60 regions respectively. In this paper, we only use the M49 and M60 bright GCs to compare with UCDs, and do $not$ study the properties of GCs that need a complete sample, e.g. luminosity function. In this case, the magnitude completeness of our GC sample is not very important.

\subsection{Nucleated Galaxies in the M87  and M49 Region}
\label{sec:nuc}

A leading theory for the origin of UCDs is tidal stripping, or ``threshing", of nucleated dwarf galaxies. In this scenario, the outer envelope of a low-mass satellite is removed by the strong tidal field of a cluster or galaxy, leaving behind a nucleus that survives because of its compact nature  (e.g., \citealt{2001ApJL_552_105Bekki, 2003MNRAS_344_399Bekki, 2003Natur_423_519Drinkwater, 2013MNRAS_433_1997Pfeffer}; see also  \citealt{1994ApJ_431_634Bassino} for an early exploration of this same basic idea). It is therefore of interest to compare the properties of  UCDs and nuclei in the NGVS. Accordingly, we pause to describe the construction of a sample of nuclei for the M87 region that will be used for such comparisons.

The M87 region contains a total of 89 certain or possible nucleated galaxies from either  \citet{1985AJ_90_1681Binggeli} or the NGVS (Ferrarese et~al., in preparation). Nucleated galaxies are identified based on visual inspection of the images as well as parametric fits to the surface brightness profiles. Galaxies are deemed to be  ``certainly nucleated" when the images show clear evidence of a central luminosity excess on the scale of the PSF, and the measured surface brightness profiles lie above over the inward extrapolation of the Sersic law best fitting the profile beyond a few arcseconds. Galaxies for which either the visual inspection or the profile fits are inconclusive are classified as ``possibly nucleated"; all others are classified as non nucleated (see also \citealt{2006ApJS_165_57Cote} and \citealt{2012ApJS_203_5Turner}). Among the 89 certain or possible nucleated galaxies, nine objects were identified as UCD candidates in our automated selection process. While this provides direct evidence that at least {\it some} nuclei are robustly similar to the UCDs studied here, it is important to understand why the remaining 80 objects do not appear in our UCD sample. First, nine of the 89 galaxies were not detected by SExtractor (due either to low surface brightness or projection close to a bright source), while 16 are brighter than $g = 18.5$~mag and 23 more are fainter than $g = 21.5$~mag, meaning that 41 nuclei remain after our selection on magnitude. Among these 41 nuclei, 6 of them fall outside the selection region in the $u^*iK_s$ diagram, 18 were rejected according to their surface brightness (most of them have lower surface brightness than UCDs) and another 8 were excluded when applying our rather stringent selection on size, leaving an objectively-selected sample of 9 nuclei finally.

In what follows, we shall use both the objectively-selected sample of nine galaxy nuclei and the full sample of 64 nuclei\footnote{These 64 objects refer to the sample of 80 nuclei that appear in the SExtractor catalog after discarding the 16 bright ($g<18.5$ mag) nuclei that are saturated in the deep NGVS images.} when comparing the UCDs and nuclei in the M87 region. In the M49 region, we find 22 galactic nuclei that have magnitude $18.5\leq g\leq21.5$~mag. Two of them recovered as UCDs by our selection criteria.

\section{Results}
\label{sec:results}

The goals of our study are an improved understanding of the origin of UCDs and a comparison of their properties to those of other stellar systems in the cluster, such as galaxies and GCs. We therefore begin by examining the bulk properties of the UCDs in the M87, M49 and M60 regions in order to understand the role environment may play in their formation and evolution. In a similar vein, we compare the properties of UCDs in these fields to those of GCs to determine the extent to which GCs and UCDs may share common formation and evolutionary histories. Since the sample of UCDs in the M60 region is smaller, some results in this section focus on just the M87 and M49 regions, e.g., color distribution, color gradient, color magnitude relation, and the connection to the Intracluster light.

\subsection{Population Richness}
\label{sec:pop}

Because the UCD samples identified in the three regions were selected using different approaches (i.e., the primary diagnostic is position in the $u^*iK_S$ diagram for M87, versus location in  $u^*gz$ for M49 and M60), we may expect some differences in the level of background contamination (see Table~\ref{tab:ucd2}). The number of contaminants can be estimated robustly using either approach (\S\ref{sec:control_fields}), so we can straightforwardly compare the most basic parameter for the UCD systems in these regions: the total number of UCDs, $N_{\rm UCD}$. To ensure the most homogenous estimates for  $N_{\rm UCD}$, we focus on the results obtained when the $u^*gz$-selection is used in all three regions.

As described in \S\ref{sec:cat}, we have a total of 102, 28 and 23  cleaned $u^*gz$-selected UCD candidates in the M87, M49 and M60 regions, respectively. To compare $N_{\rm UCD}$ in the different regions, we count the number of UCD candidates within 25$R_e$ of each galaxy\footnote{We choose 25$R_e$ to make sure that the $N_{\rm UCD}$ count areas are in our survey regions.}. Finally, we find $N_{\rm UCD} \simeq 78\pm9$, $22\pm6$ and $10\pm4$ within 25$R_e$ of M87, M49 and M60. Thus, it is immediately apparent that there are large region-to-region differences in population size, despite the fact that the central galaxies have similar luminosities and stellar masses (Table~\ref{tab:3galaxies}). Indeed, in the M60 region, the detection of a UCD system is only significant at the $\sim$2.5$\sigma$ level. However, in the case of M49, and especially M87, the detection of a UCD system is unambiguous.

\begin{figure}
\epsscale{1.1}
\plotone{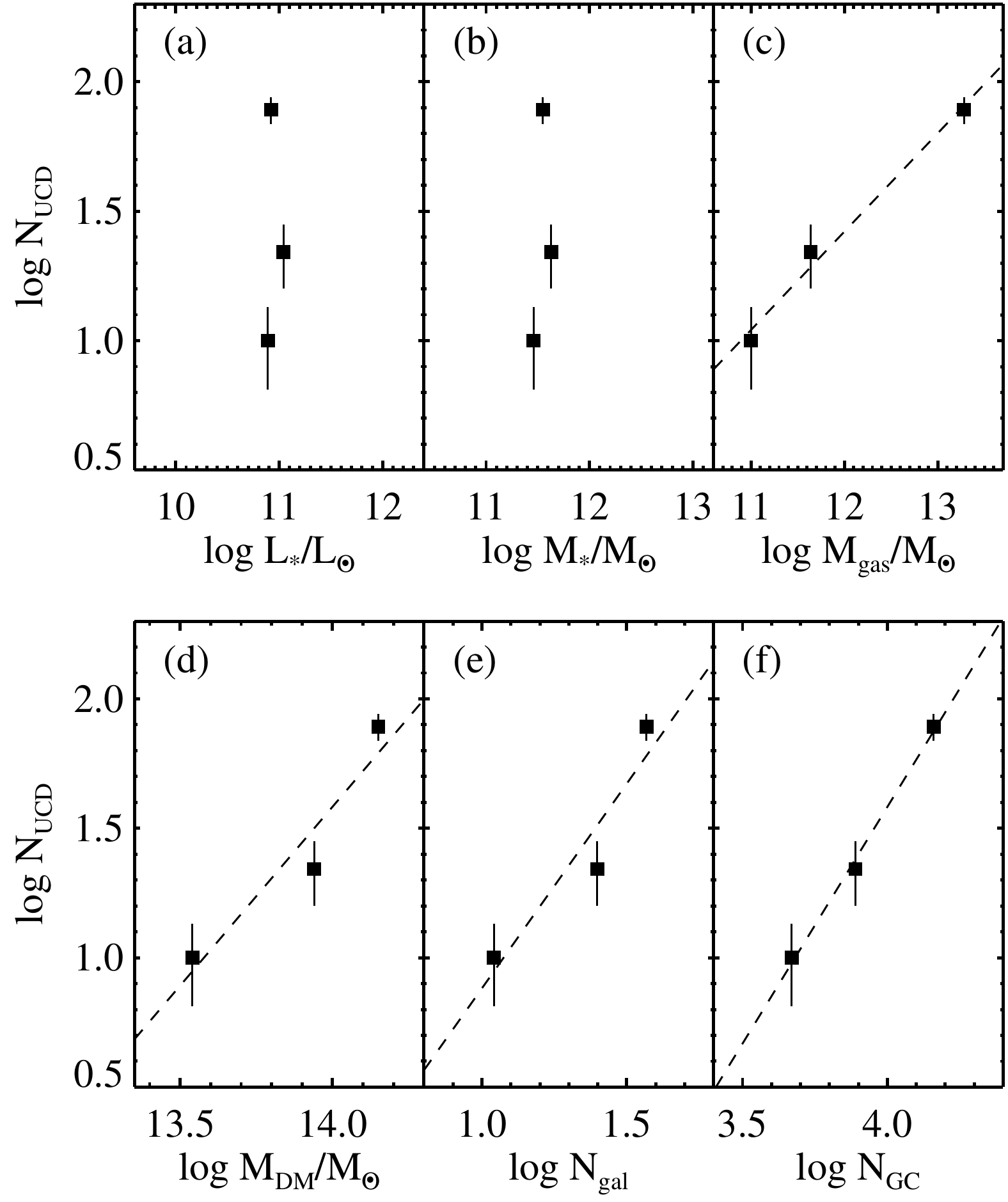}
\caption{Scaling relations for the number of UCD candidates, $N_{\rm UCD}$, within 25$R_e$ of each galaxies. Panels (a) to (f) show the variation in $N_{\rm UCD}$ as a function of galaxy $g$-band luminosity, $L_*$, stellar mass, $M_*$, X-ray gas mass, $M_{\rm gas}$, dark matter mass, $M_{\rm DM}$, number of galaxies brighter than $B=18$ within a radius of 25 $R_e$, and the total number of  globular clusters. The dashed lines show the best fit linear relations.
}
\label{fig:sn}
\end{figure}

Figure~\ref{fig:sn} shows the variation in $N_{\rm UCD}$ with several fundamental parameters for the three regions of the cluster. Panels (a) through (f) show the dependence on galaxy $g$-band luminosity, stellar mass, $M_*$, X-ray gas mass, $M_{\rm gas}$ \citep{1999A+A_343_420Schindler, 2006ApJ_646_899Humphrey}, dark matter mass, $M_{\rm DM}$ \citep{1999A+A_343_420Schindler, 2006ApJ_646_899Humphrey}, number of Virgo member galaxies located within a radius of 25$R_e$, $N_{\rm gal}$, and the total number of GCs belonging to each galaxy, $N_{\rm GC}$ \citep{2008ApJ_681_197Peng}. As already noted, there is no correlation between the luminosity or stellar mass of the galaxy and the number of surrounding UCDs: i.e., M87 and M60 have nearly identical luminosities and stellar masses, but M87 contains a rich UCD system of $\sim$80 objects, while M60 shows just a 2.5$\sigma$ enhancement in the number of UCD-like objects expected in a random field. Thus, it is clear that the M60 region is markedly deficient in UCDs compared to M87, or even M49, which again has a comparable luminosity and stellar mass.

On the other hand, the remaining panels in this figure show surprisingly tight correlations between $N_{\rm UCD}$ and the parameters describing mass of the host sub-cluster. The dashed lines show the best-fit linear relation in each case:
\begin{equation}
\begin{array}{rcl}
\log N_{\rm UCD} & = &  (0.380\pm0.031)\log M_{\rm gas}  - 3.14\pm0.11\\
                 & = &  (1.382\pm0.321)\log M_{\rm DM}  - 17.77\pm1.20\\
                 & = &  (1.581\pm0.395)\log N_{\rm gal}  - 0.70\pm0.46\\
                 & = &  (1.829\pm0.092)\log N_{\rm GC}  - 5.74\pm0.18\\
\end{array}
\end{equation}
Evidently, the large differences in UCD population size and formation efficiency disappear when $N_{\rm UCD}$ is compared to tracers of sub-cluster mass and/or richness. Normalized to the parameters such as dark matter mass and the number of neighboring galaxies, all three regions seem to have formed UCDs with comparable efficiency.

\begin{figure}
\epsscale{1.15}
\plotone{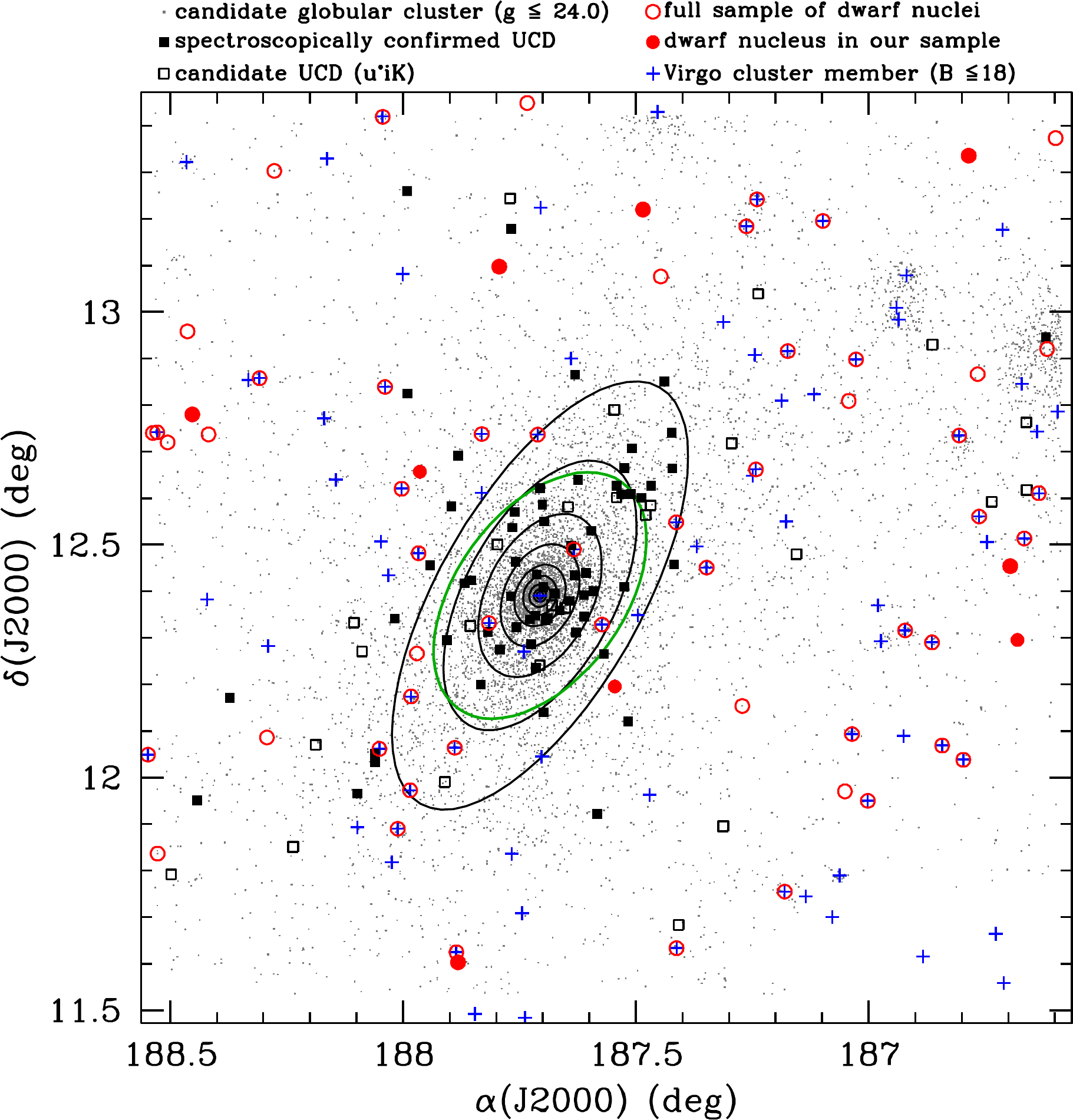}
\caption{Spatial distribution of candidate UCDs (open squares) in the M87 region. Filled squares indicate UCDs that are spectroscopically confirmed cluster members. GCs brighter than $g = 24$ as shown as gray dots. Blue crosses indicate probable member galaxies brighter than $B=18$ from \citet{1985AJ_90_1681Binggeli}. Red circles show nucleated dwarf galaxies in this region identified from the NGVS (see \S\ref{sec:nuc}). Thin black lines show the ellipses that best fit the galaxy isophotes at $\mu_g = 20, 21,~...~28$ \magarcsec. The green curve shows the ellipse that best fit the global UCD distribution (see \S\ref{sec:spatial} for details).
}
\label{fig:sd_m87}
\end{figure}

\subsection{Spatial Distribution}
\label{sec:spatial}

\begin{figure}
\epsscale{1.11}
\plotone{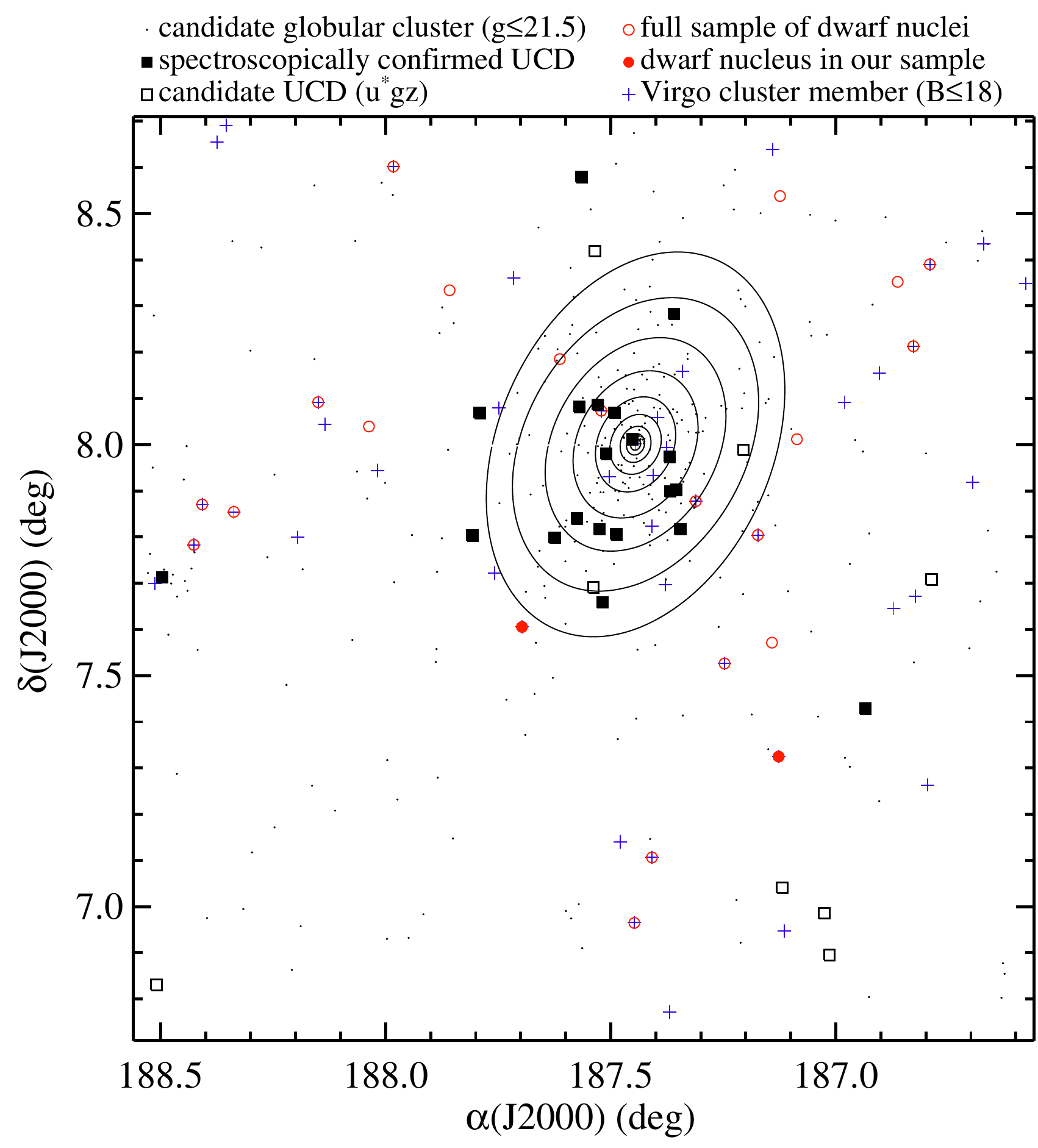}
\caption{Same as Figure~\ref{fig:sd_m87} except for the M49 region.
}
\label{fig:sd_m49}
\end{figure}

\begin{figure}
\epsscale{1.15}
\plotone{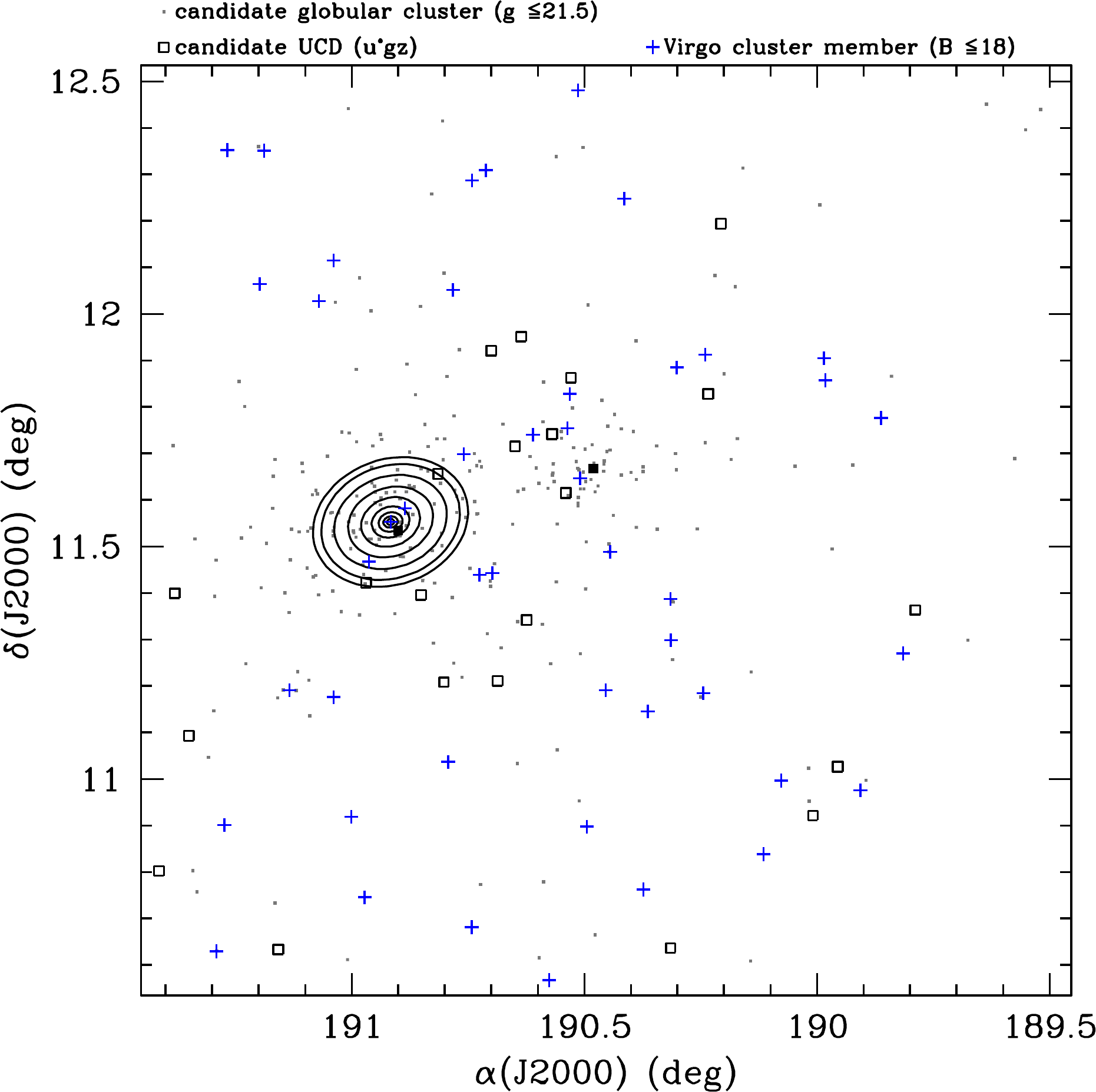}
\caption{Same as Figure~\ref{fig:sd_m87} except for the M60 region. The other galaxy whose GC system is visible at $\alpha~\simeq190\fdg5$ and $\delta~\simeq11\fdg6$ is the massive, early-type galaxy M59. The spectroscopically confirmed UCDs in this field are the bright objects discussed in \citet{2013ApJL_775_6Strader}, \citet{2014Natur_513_398Seth} and \citet{2008MNRAS_385_83Chilingarian}.
}
\label{fig:sd_m60}
\end{figure}

The spatial distribution of the confirmed and candidate UCDs in each of these three regions are shown in Figures~\ref{fig:sd_m87}, \ref{fig:sd_m49} and \ref{fig:sd_m60}. In each case, the distribution of UCDs (filled or open squares, depending on whether the object has been confirmed spectroscopically) is shown along with that of the GCs (gray dots) and Virgo member galaxies brighter than $B = 18$ (blue crosses). For comparison, the low-mass galaxies whose nuclei were identified as possible UCDs in our automated selection pipeline and full sample of nucleated dwarf galaxies in this region (as described in \S\ref{sec:nuc}) are also shown in Figure~\ref{fig:sd_m87} with filled and open red circles, respectively.

\begin{figure}
\epsscale{1.15}
\plotone{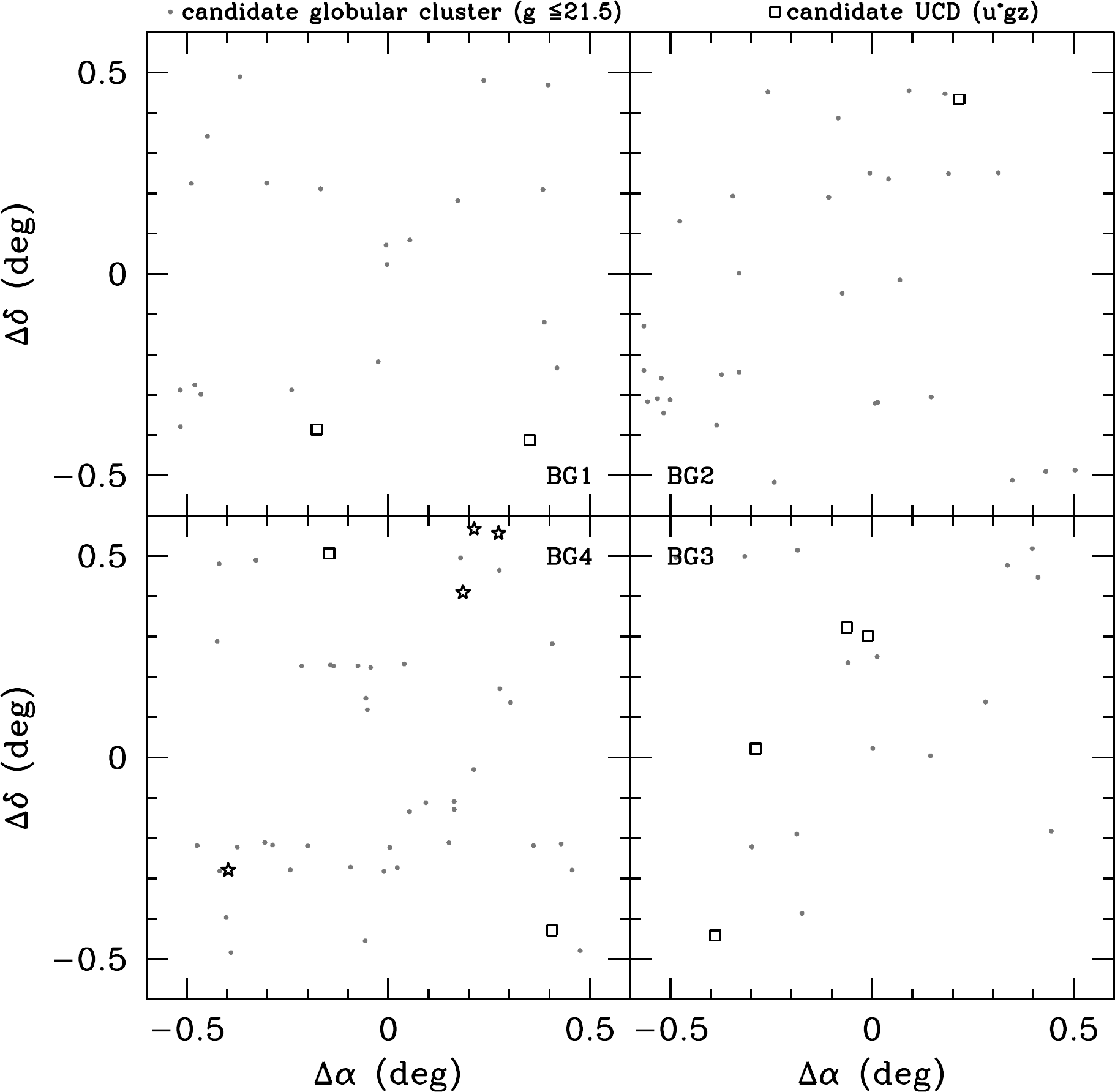}
\caption{Spatial distribution of bright GC- (gray dots) and UCD-like objects (open squares) in the four NGVS background fields. The star symbols plotted in the BG4 field show probable members of the $z \sim 0.02$ galaxy group identified by \citet{2007ApJ_655_790Crook}. They are not identified as GCs or UCDs in our analysis.
}
\label{fig:sd_bg}
\end{figure}

The ellipses in each figure show the best-fit galaxy isophotes at $\mu_g = 20, 21,~...~28$ \magarcsec. In the case of M87, where GCs can be selected with a high level of confidence from the $u^*iK_s$ diagram, we plot all GC candidates brighter than $g = 24$. For M49 and M60, no $K_s$-band imaging is available so we plot only those GCs selected from the $u^*gz$ diagram with $g \le 21.5$ (see \S\ref{sec:gcs}). For comparison, Figure~\ref{fig:sd_bg} shows the spatial distribution of GC- and UCD-like objects in the four background fields (also selected from the $u^*gz$ diagram). This last figure gives an indication of the level of contamination we might expect in the M49 and M60 regions, as well as the field-to-field variance. BG3 contains four UCD-like objects, whereas BG2 contains just one.

The UCD candidates in Figure~\ref{fig:sd_m60} are only weakly concentrated to the center of M60, and at least a few of the candidates may be better associated with M59, a massive early-type galaxy located about 25$^\prime$ ($\sim$ 120~kpc) to the northwest. All in all, the number of UCD candidates in the M60 region appears only slightly larger than what one would expect in a random field (as noted in \S\ref{sec:pop}, we estimate  a total UCD population of $\sim10\pm4$ above background), although the existence of at least some bonafide UCDs in this galaxy is clear: e.g., the high-mass object, M60-UCD1, is projected just $\sim 1\farcm5$ from the center of the galaxy \citep{2013ApJL_775_6Strader}.

\begin{figure}
\epsscale{1.15}
\plotone{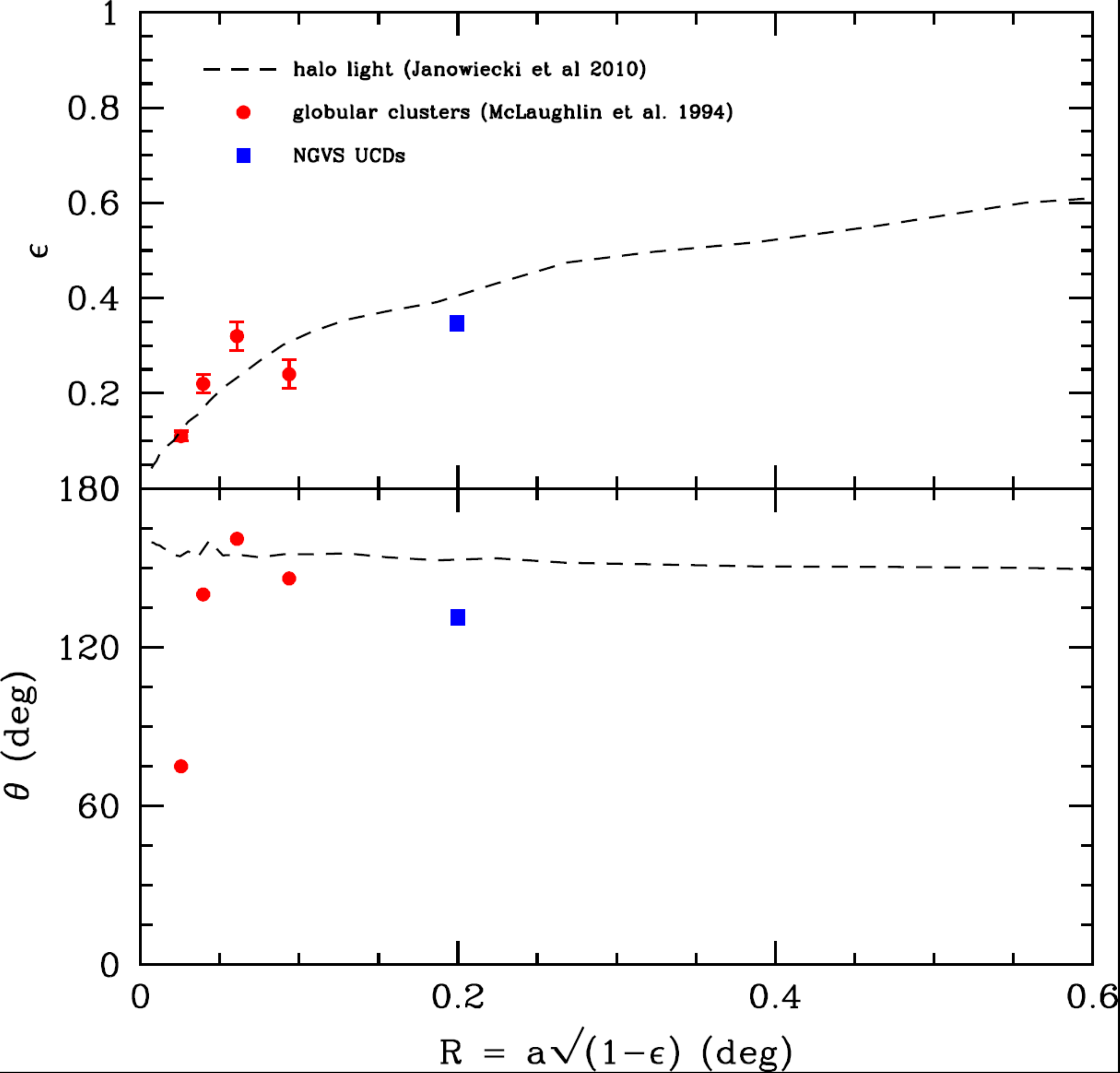}
\caption{Radial variation in ellipticity, $\epsilon$, and position angle, $\theta$, for UCDs surrounding M87 (blue squares). For comparison, we also show profiles for the M87 halo light (dashed curve) and the globular cluster system from \citet{1994ApJ_422_486McLaughlin} (red circles).
}
\label{fig:shape}
\end{figure}

The situation is quite different in M49 and, especially, M87, where there is a clear excess of confirmed or candidate UCDs at the position of the central galaxies. In the latter case, the UCD sample is large enough to permit a more quantitative examination of its flattening and orientation. The dashed lines in Figure~\ref{fig:shape} show the radial variation in ellipticity, $\epsilon$, and position angle, $\theta$, for the isophotes of M87 according to \citet{2010ApJ_715_972Janowiecki}. Although our sample of UCDs is too small to measure reliable {\it profiles} for these parameters, the mean values obtained using the UCDs within a radius of 25$R_e$
\begin{equation}
\begin{array}{rcl}
\langle{\epsilon}\rangle_{\rm UCD} & \simeq &  0.34\\
\langle{\theta}\rangle_{\rm UCD} & \simeq &  129^{\circ}\\
\end{array}
\end{equation}
are in good agreement with those of the galaxy/ICL (intracluster light). Note that the UCD measurements shown in Figure~\ref{fig:shape} (blue squares) are plotted at the mean geometric radius for the sample, $R = 0\fdg20$~($\sim$ 60~kpc). For comparison, we also show  measurements for the GC system in the inner regions of M87 from \citet{1994ApJ_422_486McLaughlin}. For both the GCs and UCDs, the measurements are based on the method of moments formalism described in \S~4.1 of \citet{1994ApJ_422_486McLaughlin}. The GC system in M87, like the UCDs, is also quite similar to that of the galaxy in terms of flattening and position angle, a result which has been noted many times in the past (e.g., \citealt{1994ApJ_422_486McLaughlin, 2012MNRAS_421_635Forte}).

\subsection{Color Distributions}
\label{sec:colors}

For the comparatively rich UCD system surrounding M87, our multi-band imaging --- including $K_s$ data from the NGVS-IR --- makes it possible to carry out a comprehensive study of UCD colors. As is well known, the GC systems of most massive galaxies are known to show multimodal --- usually {\it bimodal} --- color distributions (e.g. \citealt{1992ApJ_384_50Ashman, 1996AJ_111_1529Geisler, 1999AJ_118_1526Gebhardt, 2001AJ_121_2950Kundu, 2007ApJ_660_109Kundu, 2001AJ_121_2974Larsen, 2001MNRAS_327_1116Larsen, 2006ApJ_636_90Harris, 2006ApJ_639_95Peng, 2009ApJ_703_42Peng, 2011ApJ_728_116Liu}).  This bimodality has been interpreted as evidence for a variety of physical processes that may have been involved in the formation of the host galaxy, including hierarchical growth (e.g., \citealt{1998ApJ_501_554Cote, 2000ApJ_533_869Cote, 2002ApJ_567_853Cote, 2002MNRAS_333_383Beasley, 2013ApJ_762_39Tonini}) and cluster formation in gas-rich major mergers \citep{1992ApJ_384_50Ashman, 2010ApJ_718_1266Muratov}. Alternatively, it has been proposed that  a non-linear relationship between metallicity and (some) broadband color indices \citep{2006Sci_311_1129Yoon, 2011ApJ_743_149Yoon, 2013ApJ_768_137Yoon, 2007ApJ_669_982Cantiello, 2012ApJ_746_88Blakeslee, 2012A+A_539_54Chies-Santos} is chiefly responsible for the observed bimodality. Once again, one should keep in mind that these processes need not be mutually exclusive, and it is likely that a combination of effects conspire to produce the ubiquitous bimodality observed for GC systems.

Broadly speaking, existing photometry suggests that UCDs can also have a broad distribution in color, often spanning the full range in GC colors (e.g., \citealt{2005ApJ_627_203Hacsegan, 2006ApJ_653_193Mieske, 2008AJ_136_461Evstigneeva, 2010ApJ_722_1707Madrid, 2011ApJ_737_13Madrid, 2011ApJ_737_86Chiboucas}). There is also some evidence that, at the highest luminosities, this distribution gives way to a trend between luminosity and color, in the sense that the brightest systems have the reddest colors (e.g., \citealt{2004A+A_418_445Mieske, 2008AJ_136_2295Blakeslee}), although there are at least some counterexamples of bright systems that are blue in color \citep{2014MNRAS_439_3808Penny}.

\begin{figure}
\epsscale{1.10}
\plotone{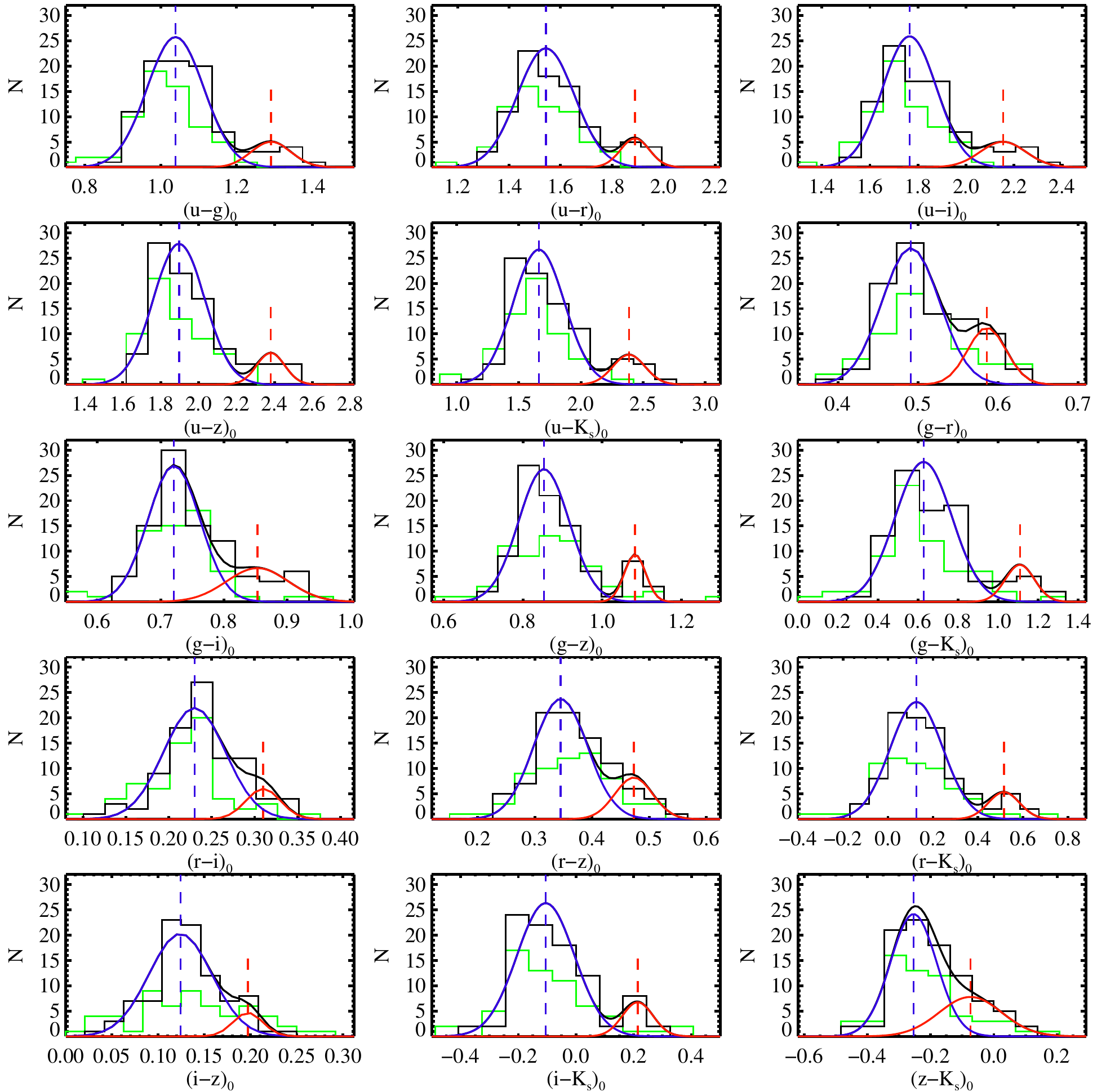}
\caption{Optical and IR ($u^*grizK_s$) color distributions for our sample of 92 probable or confirmed UCDs in the M87 region. The blue and red curves show the individual components of the double Gaussian that best fits each distribution, determined using the GMM code assuming heteroscedasticity (see also Table~\ref{tab:color_ucd_gmm_92}). The dashed vertical lines show the inferred blue and red peaks, while the smooth black curve shows the sum of the individual components. The green histograms present the color distribution for the full sample of 64 dwarf nuclei.}
\label{fig:color_distribution_het}
\end{figure}

Figures~\ref{fig:color_distribution_het} shows optical/IR color histograms for our cleaned sample of 92 likely or confirmed UCDs in the M87 region. The UCD color distributions, plotted as the black histogram in each panel, include many of the possible color combinations, from $(u-g)_0$ to $(z-K_s)_0$. It is obvious from this figure that nearly all of the distributions are bimodal. To quantify this visual impression, we have used the Gaussian Mixture Modeling (GMM) code of \citet{2010ApJ_718_1266Muratov} to evaluate the significance and characteristics of this apparent bimodality. Note that GMM is similar to the KMM code of \citet{1994AJ_108_2348Ashman}, but it implements maximum likelihood parameter estimation. The best-fit GMM models (blue and red curves in Figure~\ref{fig:color_distribution_het}) were calculated assuming heteroscedasticity, i.e., independent dispersions for the two subcomponents. For comparison, the green histogram in each of these two figures shows the color distribution for the sample of 64 dwarf nuclei in this region. While a detailed analysis of the nuclei colors will be presented in a future paper in this series, we note that the nuclei, like the UCDs, are primarily blue. Unlike the UCDs, though, they show no strong evidence for bimodality.

The best-fit, two-component Gaussians from our GMM analysis are shown in Figures~\ref{fig:color_distribution_het} as the blue and red curves in each panel; their sum is shown in black. The two vertical dashed lines in each panel show the locations of the calculated blue and red peaks. The GMM-fitting results for the UCDs are summarized in Table~\ref{tab:color_ucd_gmm_92}. From left to right, this table records for each index, the mean color, $\mu_a$, and dispersion, $\sigma_a$, as well as the mean color and dispersion for the separate blue ($\mu_b$, $\sigma_b$) and red ($\mu_r$, $\sigma_r$) components. The number of blue and red UCDs, $N_b$ and $N_r$, along with the overall fraction of blue objects, $f_b$, are given in the following three columns. The errors on each parameter were determined using a non-parametric bootstrap procedure. The final two columns give the $D$ parameter and p-value for $chi^2$ (see \citealt{2010ApJ_718_1266Muratov} for details). As described in \citealt{2010ApJ_718_1266Muratov}, the $D$ parameter should be larger than 2 and $p$-value should be smaller than 0.05 for a bimodal distribution. As apparent from Figure~\ref{fig:color_distribution_het}, the color distributions are found to be bimodal in all indices with the exceptions of $(r-i)_0$ and $(i-z)_0$, both of which have $p$-values larger than 0.05. The lack of bimodality in these indices is perhaps not surprising given their limited wavelength baseline.

\begin{figure}
\epsscale{1.15}
\plotone{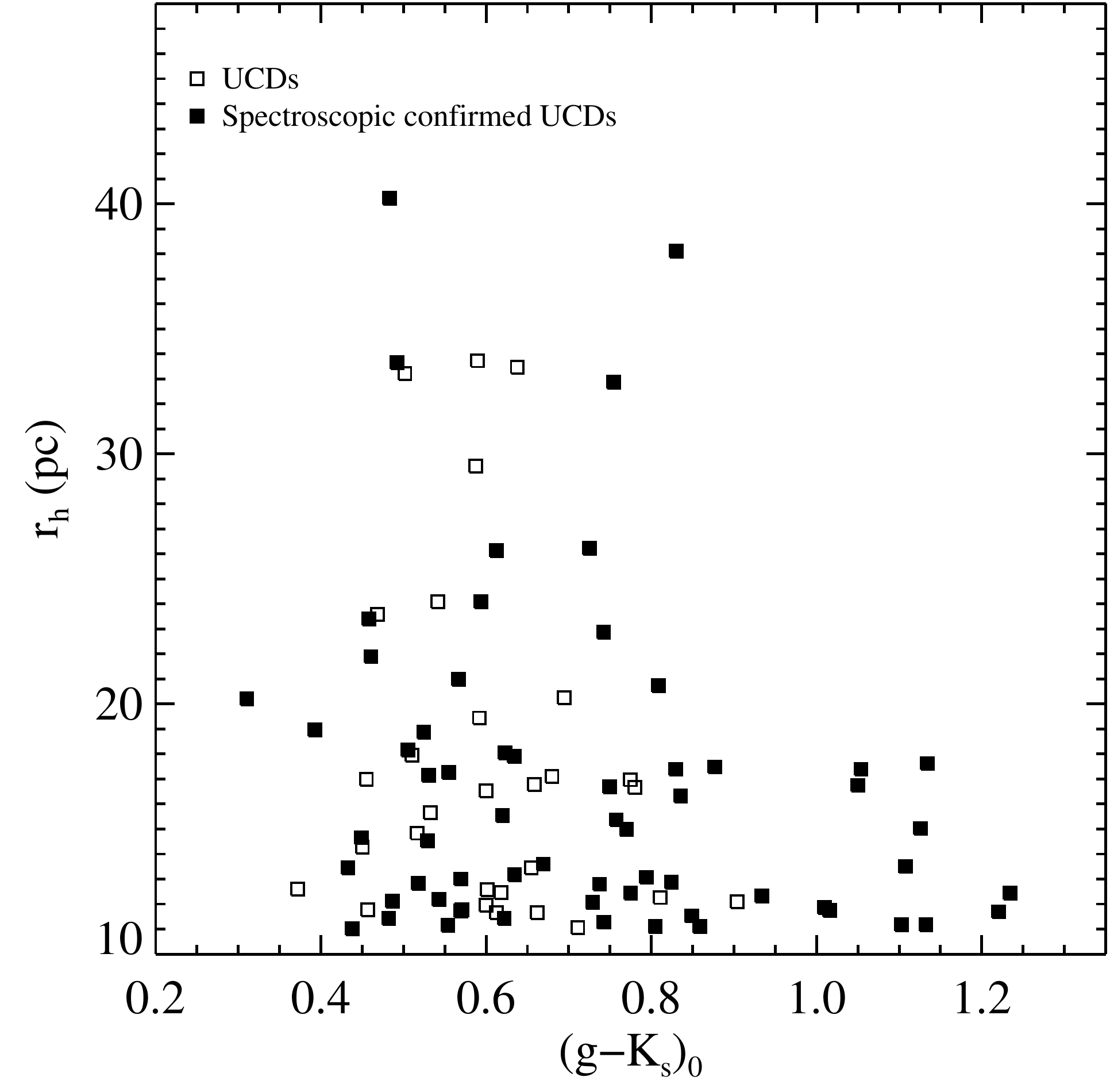}
\caption{The measured half-light radius ($r_h$) versus $(g-K_s)$ color for our probable or confirmed UCDs. Our selection criteria require that a UCD candidate have $r_h\!>\!11$~pc. As shown in Figure~\ref{fig:color_distribution_het}, the vast majority of UCDs are blue in color. The red UCDs are more compact, and all have $r_h\!<\!18$~pc. It is possible that these red UCDs are simply in the tail of the red GC size distribution.}
\label{fig:rhcolor}
\end{figure}

\begin{figure}
\epsscale{1.10}
\plotone{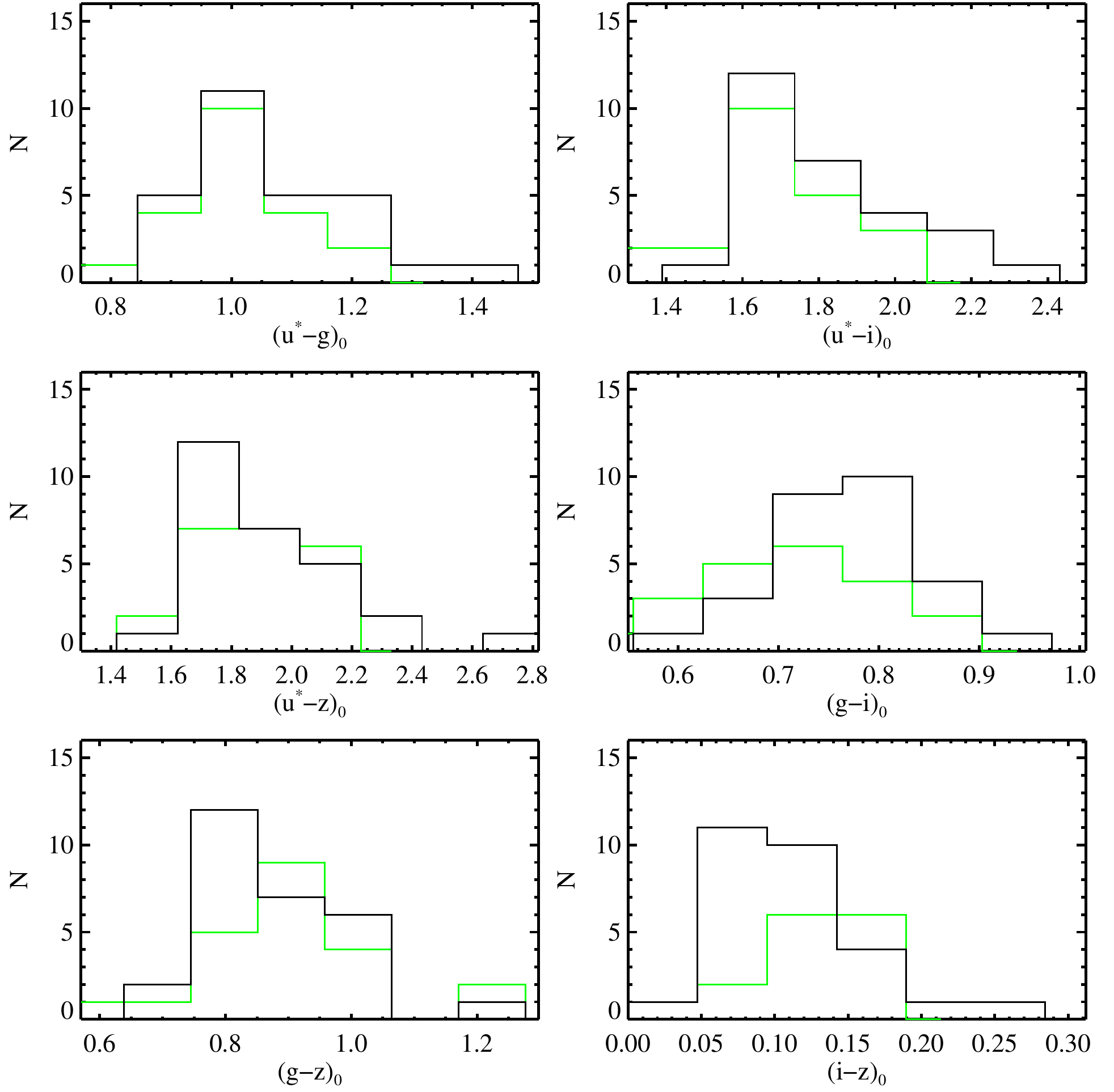}
\caption{Optical ($u^*giz$) color distributions for UCD candidates (black histogram) and dwarf nuclei (green histogram) in the M49 region.}
\label{fig:color_distribution_m49}
\end{figure}

The fraction of blue UCDs is found to be remarkably constant across color index, with a mean value of $f_b = 0.85\pm0.14$. These estimates compare reasonably well with the value of $f_b \simeq 0.71$ measured for GCs using the $(g-z)_0$ index for the 11 ACSVCS galaxies that fall within the M87 region considered here \citep{2006ApJ_639_95Peng, 2009ApJS_180_54Jordan}. In other words, the UCD system in the core of Virgo (sub-cluster A) has a similar, or perhaps slightly larger, fraction of its members belonging to the blue sub-population as does the GC system in the same region.

The interpretation of the UCD color distribution, however, is even less straightforward than it is for GCs. Figure~\ref{fig:rhcolor} plots $r_h$ versus $(g-K_s)$ to show that while the blue UCDs span a large range in sizes---from our lower limit of 11~pc to a maximum of $\sim40$~pc---the red UCDs are all relatively compact. None of the red UCDs have $r_h$ larger than 18~pc. This raises the possibility that these red UCDs are simply the tail of the red GC population to large sizes. This does not mean, however, that there are no red UCDs. Some of the most massive confirmed UCDs in Virgo, such as M60-UCD1 \citep{2013ApJL_775_6Strader, 2014Natur_513_398Seth} and M59c0 \citep{2008MNRAS_385_83Chilingarian}, have colors as red as the reddest GCs. Because of this ambiguity, it is difficult to interpret the color distributions of UCDs. These issues will be explored for a variety of stellar systems (GCs, UCDs, galaxies, nuclei) in a future NGVS paper. For the time being, we simply note that UCDs as a population, defined using our selection criteria, show bimodal color distributions, whereas dwarf nuclei are predominantly blue. The size distributions of the color subpopulations is consistent with a picture where some of the blue UCDs may have origins as stellar nuclei, and the red UCDs in our sample are possibly the largest red GCs.

Figure \ref{fig:color_distribution_m49} shows the optical color distribution of UCD candidates (black lines) and dwarf nuclei (green lines) in the M49 region. The six color indexes shown in this figure are based on the $u^*giz$ wavebands that we have in this region. The UCD colors do not show significant bimodal distributions, but most have ``tails" on the red side, with the exception of the $(g-i)_0$ color index. Due to the limited number of UCD candidates in the M49 region, we did not fit the color distribution using GMM pipeline. The green histograms show the distribution in color of the dwarf nuclei. The nuclei color distribution in the M49 region is similar to that in the M87 region: i.e., similar in color range with the UCDs and showing no obvious bimodality.

\begin{figure}
\epsscale{1.10}
\plotone{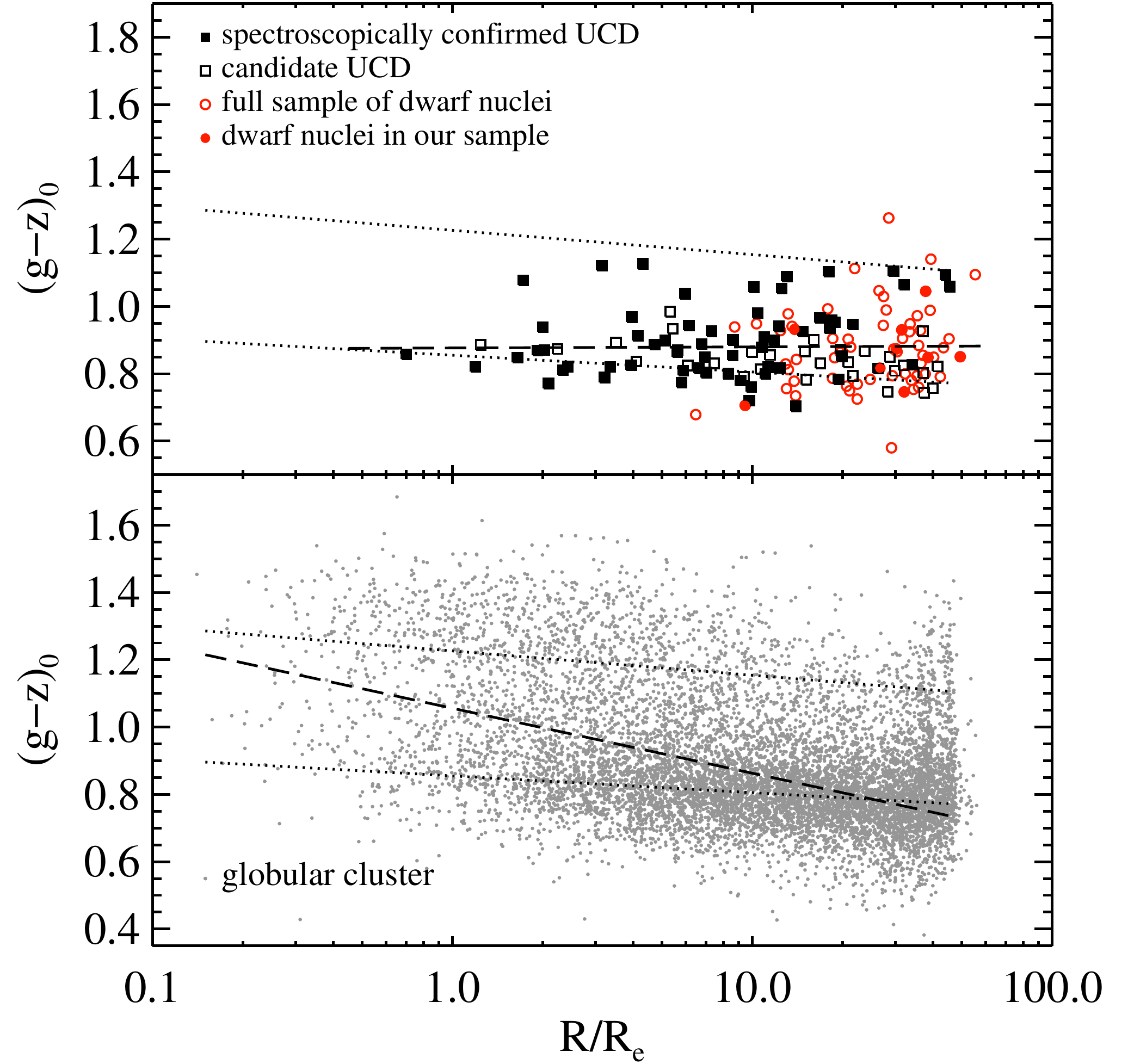}
\caption{Radial variation in color for UCDs and GCs in the M87 region (upper and lower panels, respectively). The GCs show a significant negative color gradient  (dashed line in the lower panel), becoming systematically bluer with increasing radius due mainly to the changing ratio of red-to-blue GCs. The blue and red GCs individually show shallower gradients (dotted lines in the upper and lower panels). The UCDs, on the other hand, show no evidence for a gradient in color (dashed line in the upper panel). The red circles in the upper panel show the dwarf nuclei in this region.
}
\label{fig:color_gradients}
\end{figure}

\begin{figure}
\epsscale{1.10}
\plotone{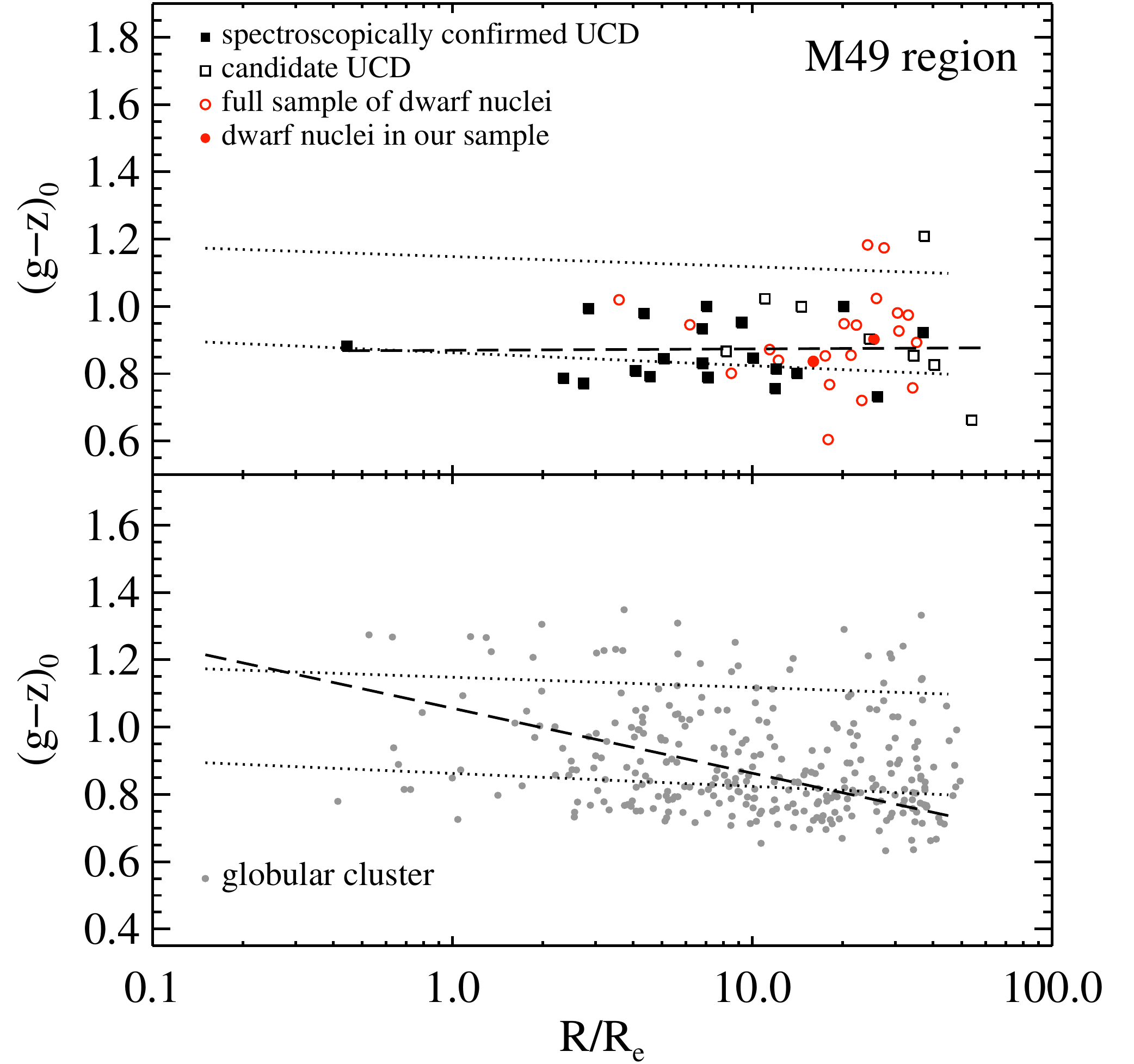}
\caption{Same as figire \ref{fig:color_gradients}, but for the M49 region.
}
\label{fig:color_gradients_m49}
\end{figure}

\subsection{Color Gradients}
\label{sec:grad}

As is well known, massive galaxies usually show radial color gradients, in the sense that their colors become bluer with increasing distance from the center (i.e., a negative gradient). This behavior is found in both early- and late-type galaxies (e.g., \citealt{1983AJ_88_1707Boroson, 1989AJ_98_538Franx, 1990AJ_100_1091Peletier, 1996A+A_313_377DeJong, 2004ApJS_152_175MacArthur, 2009RAA_9_1119Liu, 2011MNRAS_411_1151Gonzalez-Perez, 2012MNRAS_419_2031Peletier}). It is also known that the GC systems associated with massive galaxies show negative color gradients (e.g., \citealt{1981ApJ_245_416Strom, 1996AJ_111_1529Geisler, 2009ApJ_703_939Harris, 2011ApJ_728_116Liu}). In this case, the fairly strong color gradients are known to be driven mainly by the changing ratio of blue-to-red GCs as a function of radius, with the blue clusters being more spatially extended than their red counterparts \citep{1996AJ_111_1529Geisler, 1998AJ_115_947Lee, 1999ApJ_513_733Kundu, 2001AJ_121_2974Larsen, 2006ApJ_639_95Peng, 2008ApJ_681_197Peng}.

The dependence of UCD color on projected distance from M87 is shown in the upper panel of Figure~\ref{fig:color_gradients}. The sample consists of the 92 likely or confirmed UCDs (after the rejection of spectroscopic non-members). UCDs with, and without, measured radial velocities are shown by the filled and open squares, respectively. The dashed line shows the best fit linear relation:
\begin{equation}
\begin{array}{rcl}
(g-z)_0 & = & 0.876(\pm0.026)- 0.003(\pm0.024)\log{R/R_e} \\
\end{array}
\end{equation}
For comparison, we also show the nucleated dwarf galaxies in the M87 region. Neither the dwarf nuclei, which lie in the range $10 \lesssim R/R_e \lesssim 50$, nor the UCDs exhibit any trend with radius.

The lower panel of Figure~\ref{fig:color_gradients} shows the same relation for the GCs of M87. The dashed line in this case
\begin{equation}
\begin{array}{lcl}
(g-z)_0 & = & 1.056(\pm0.003) - 0.193(\pm0.004)\log{R/R_e} \\
\end{array}
\end{equation}
refers to the {\it entire} GC system, with no division on color. Consistent with earlier findings (e.g. \citealt{2004AJ_127_24Jordan, 2007ApJ_668_209Cantiello}), we find that the varying ratio of blue to red GCs is chiefly responsible for the strong negative gradient. The weaker trends observed for the separate red and blue GCs, divided at $(g-z)_0$ = 1.00, are shown by the dotted lines:
\begin{equation}
\begin{array}{lcl}
(g-z)_{red}  & = & 1.215(\pm0.007) - 0.068(\pm0.007)\log{R/R_e} \\
(g-z)_{blue} & = & 0.858(\pm0.003) - 0.046(\pm0.003)\log{R/R_e}
\end{array}
\end{equation}
For comparison, we overlay the fitted relations for the blue and red GCs on the UCDs in the upper panel of this figure.

The existence of a $blue~tilt$ may bias the gradient measured for the luminous blue GCs. To avoid this possibility, it is better to compare the UCDs and blue GCs in the same color and magnitude ranges. We therefore focus on the UCDs and GCs brighter than $g=21.5$ and bluer than $(g-z)_0=1.0$. The best fitted gradients in this case are $\nabla(g-z)_{\rm{blueGC}}=-0.050\pm0.008 ~\rm dex^{-1}$ and $\nabla(g-z)_{\rm{blueUCD}}=-0.018\pm0.018 ~\rm dex^{-1}$. It is clear that luminous blue GC system has a significant gradient (at the $\sim 6 \sigma$ level) while blue UCD system shows no gradient at all.

Figure~\ref{fig:color_gradients_m49} shows color gradients for GCs, UCDs and dwarf nuclei in the M49 region. The GCs in this figure are again only the luminous GCs which are brighter than $g=21.5$. The best fitted lines (dashed lines) of the whole UCD system and whole GC system are:
\begin{equation}
\begin{array}{lcl}
(g-z)_{UCD}  & = & 0.870(\pm0.050) - 0.004(\pm0.046)\log{R/R_e} \\
(g-z)_{GC}    & = & 0.970(\pm0.020) - 0.082(\pm0.019)\log{R/R_e}
\end{array}
\end{equation}
If we fit only the UCDs and GCs that are brighter than $g=21.5$ and bluer than $(g-z)_0=1.0$ , the gradients are $\nabla(g-z)_{\rm{blueGC,M49}}=-0.038\pm0.012 ~\rm dex^{-1}$ and $\nabla(g-z)_{\rm{blueUCD,M49}}=-0.026\pm0.040 ~\rm dex^{-1}$. Again, the luminous blue GC system shows a significant gradient ($\sim$$3 \sigma$) while the blue UCD system shows no evidence of gradient.

From the above analysis, it is clear that the GC systems around M87 and M49 show significant color gradients, while the UCD systems show no gradient. We also notice that the gradients of UCDs have large errors, which makes the gradient values of UCDs and GCs compatible at the $\sim 2 \sigma$ level. A possible difference in the color gradients of GCs and UCDs may point to different formation mechanisms for the two classes of objects.

\begin{figure}
\epsscale{1.06}
\plotone{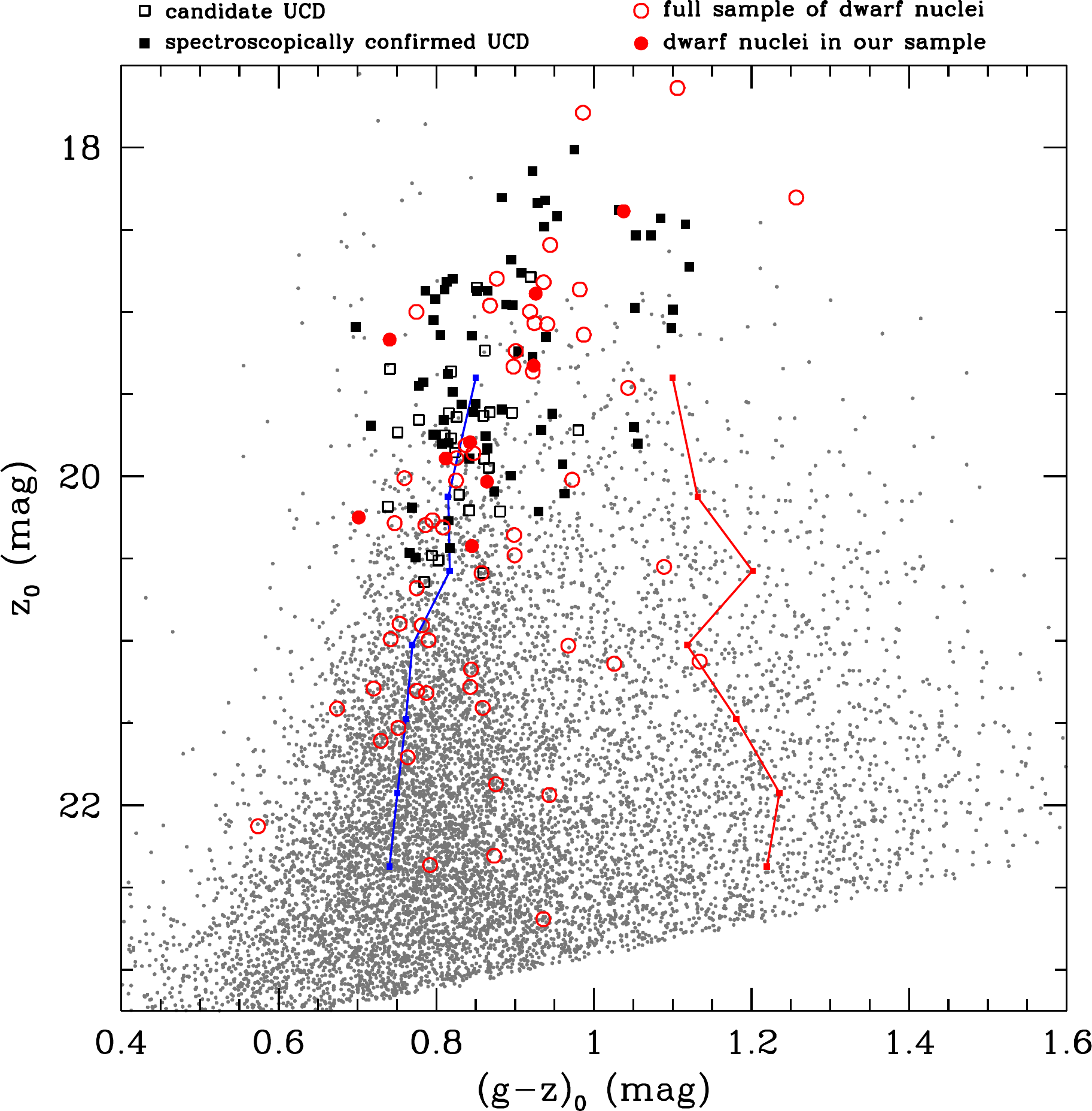}
\caption{Color magnitude diagram for the M87 region showing GCs, UCDs and dwarf nuclei. The blue and red sequences show the variations in the mean color for the blue and red GC sub-populations (see text for details).
}
\label{fig:color_magnitude_faint}
\end{figure}

\begin{figure}
\epsscale{1.06}
\plotone{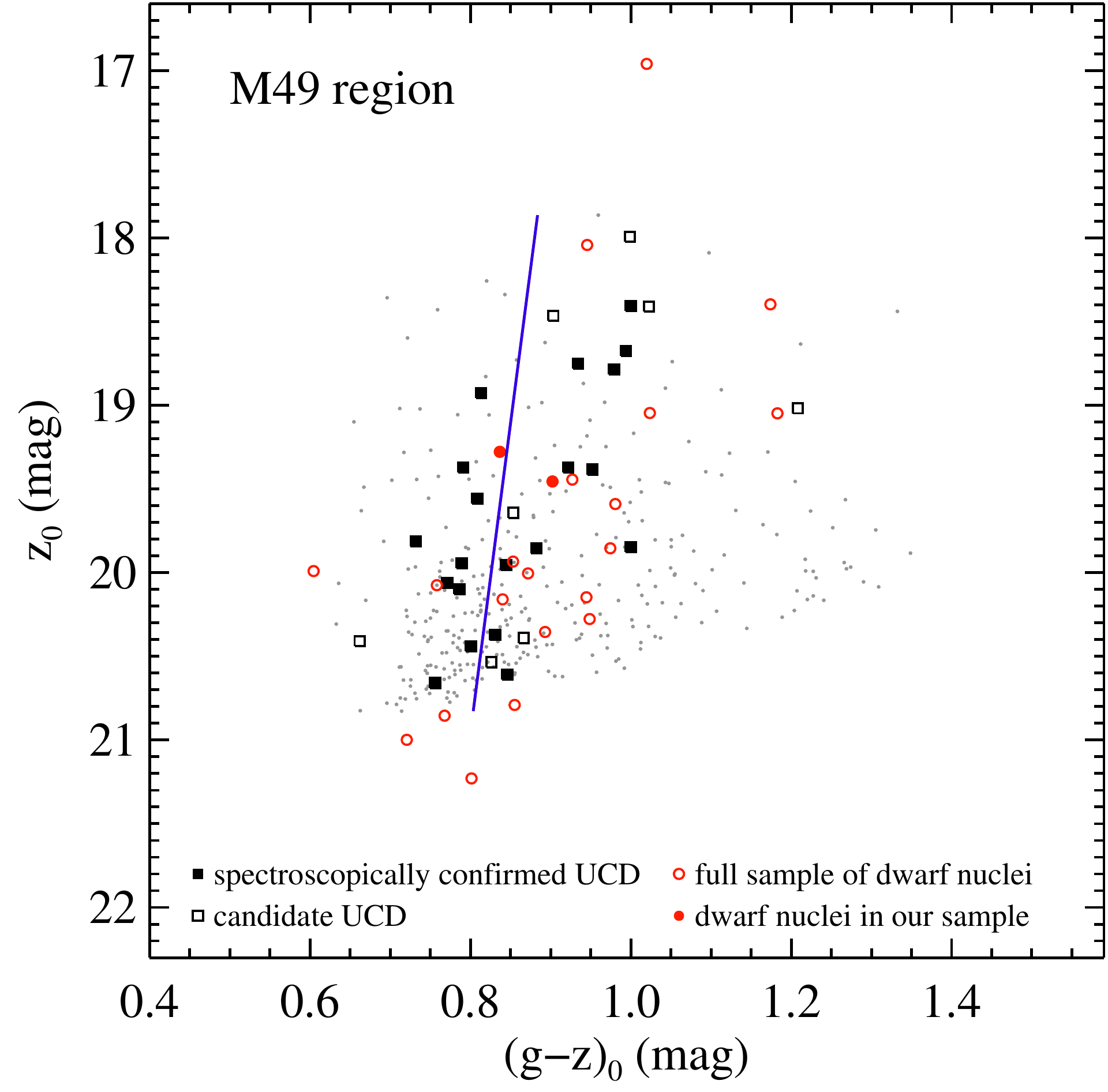}
\caption{Color magnitude diagram for the M49 region. The UCDs (squares), dwarf nuclei (circles) and $u^*gz$-selected bright ($g_0<21.5$) GC candidates (dots) are shown in this figure.
}
\label{fig:color_magnitude_m49}
\end{figure}

\subsection{Color Magnitude Relations}
\label{sec:cmd}

The color-magnitude diagram for sources in the M87 and M49 regions are shown in Figure~\ref{fig:color_magnitude_faint} and \ref{fig:color_magnitude_m49}. GCs are shown as gray dots, spectroscopically confirmed UCDs as filled squares, and candidate UCDs as open squares. Note that the GC sample has been cleaned of any objects that we  have identified here to be UCDs on the basis of size. For reference, the 64 nuclei of dwarf galaxies in the M87 region and 22 nuclei in the M49 region are shown as red circles (open and filled).

As noted previously, the GC system of M87 \citep{2006ApJ_653_193Mieske, 2007MNRAS_382_1947Forte, 2009ApJ_703_42Peng, 2009ApJ_699_254Harris} exhibits a ``blue tilt" \citep{2006ApJ_636_90Harris, 2006AJ_132_2333Strader, 2006ApJ_653_193Mieske}, with the GCs belonging to the blue sub-population becoming progressively bluer with decreasing luminosity. To quantify this behavior, we bin the GCs by $z_0$ magnitude, and fit two-component Gaussians to the data points in each bin, thereby ensuring that all points are independent. The results are shown by the thin blue and red lines, which trace out the central locations of the blue and red components as a function of $z_0$-band magnitude. Our analysis --- which is based on a GC sample that is far larger than that used in any previous study of this galaxy ---  confirms the existence of a significant blue tilt in the M87 GC system. For the M49 region, we only present bright GCs ($g<21.5$) in the Figure~\ref{fig:color_magnitude_m49}. The blue line is the best fit for the blue GCs with $(g-z)_0<1.0$. Similar with M87 GC system, luminous blue GCs in the M49 region show significant blue tilt as well.

Both Figure~\ref{fig:color_magnitude_faint} and Figure~\ref{fig:color_magnitude_m49} show that the UCDs participate in the general trends defined by the GCs. Indeed, at bright magnitudes, a significant fraction of the sources are actually UCD candidates, and so they have been at least partly responsible for driving the color-magnitude trends noted in previous ground-based studies. The abrupt bend to redder colors exhibited by the UCDs is similar to the behavior of UCD candidates in Abell 1689 noted by \citet{2004AJ_128_1529Mieske}, although UCD membership constraints in that case were fairly limited. For both M87 and M49 regions, many of the bright UCDs are confirmed spectroscopic members (filled squares), leaving little doubt over the reality of this color-magnitude relation. Interestingly, the dwarf nuclei seem to follow the trends defined by the UCDs and (blue) GCs. It also noteworthy that all of the 11 UCDs in the M87 region and all of the 3 UCDs in the M49 region with colors redder than $(g-z)_0 \ge 1.0$ are brighter than $z_0 \simeq 20$.

\begin{figure}
\epsscale{1.15}
\plotone{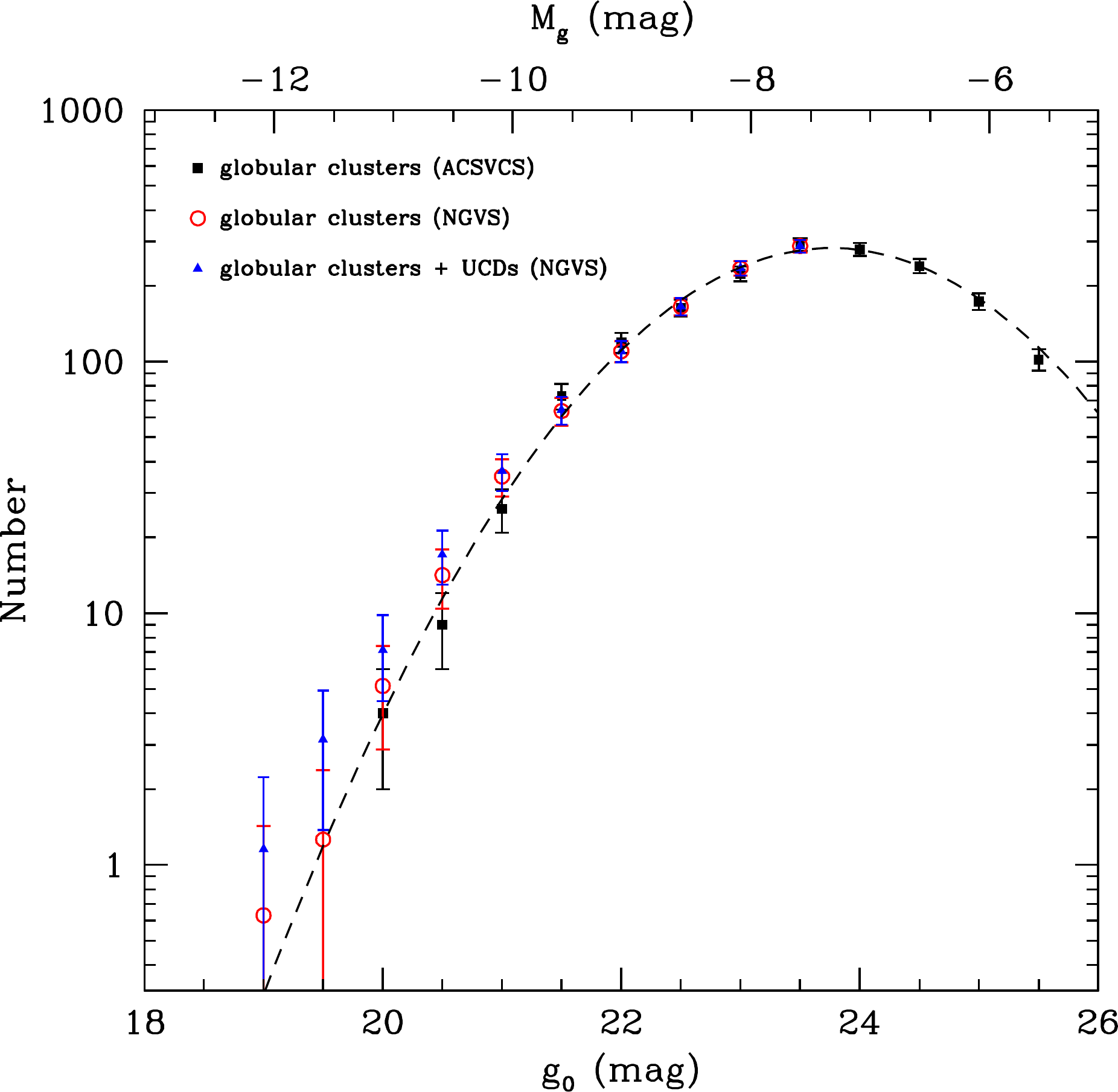}
\caption{The luminosity function of globular clusters in the M87 region from the ACSVCS (black squares), $u^*iK_s$-selected globular clusters from the NGVS ($g<24$ mag, red circles) and the combined sample of NGVS globular clusters and UCDs (blue triangles). The dashed curve shows the gaussian luminosity function fit of \citet{2010ApJ_717_603Villegas} for M87.}
\label{fig:luminosity_function}
\end{figure}

\subsection{Luminosity Functions}
\label{sec:lf}

The M87 GC luminosity function has been studied many times in the past (e.g., \citealt{1994ApJ_422_486McLaughlin, 1999ApJ_513_733Kundu, 2002ApJL_576_113Jordan, 2006ApJ_651_25Jordan, 2007ApJS_171_101Jordan, 2009ApJ_703_42Peng, 2010ApJ_717_603Villegas}), so it is possible to compare these earlier results (many of which are based on high-resolution imaging from HST) with our new measurements from the NGVS. Of particular interest in this context  is a comparison of the UCDs to GCs to see if the two populations are consistent with being drawn from the same parent distributions in luminosity. Although previous studies of this sort have generally failed to provide conclusive evidence for, or against, the hypothesis that the UCDs and GCs are drawn from the same luminosity function (due largely to limitations in UCD sample size and the heterogeneity of the GC and UCD samples used for the comparisons; e.g., \citealt{2004A+A_418_445Mieske, 2007A+A_472_111Mieske, 2009AJ_137_498Gregg}), some recent investigations have argued that the UCD luminosities are consistent with their being the most luminous members of the host galaxy's GC system (e.g., \citealt{0906.0776, 2011MNRAS_414_739Norris, 2012A+A_537_3Mieske}).

Figure~\ref{fig:luminosity_function} shows the GC luminosity function of M87 according to the ACSVCS (black squares). This HST-based measurement has the advantage that typical GCs at the distance of Virgo can be resolved by HST/ACS (e.g., \citealt{2005ApJ_634_1002Jordan}), thus significantly reducing contamination. The data have been taken from the catalog of \citet{2009ApJS_180_54Jordan}, and consist of 1748 GCs identified in a single ACS field located at the center of M87. To allow a direct comparison to the NGVS results, the $g$-band magnitudes from the ACSVCS have been transformed from the SDSS to MegaCam photometric system --- a typical correction of $\sim$ 0.078 mag in the sense that the MegaCam magnitudes are brighter\footnote{For more information, see http://www2.cadc-ccda.hia-iha.nrc-cnrc.gc.ca/megapipe/docs/filters.html}.

The dashed black curve in this figure shows a Gaussian fitted to the transformed ACSVCS data \citep{2010ApJ_717_603Villegas}. The red circles show the corresponding GC luminosity function from the NGVS. The luminosity functions were scaled vertically to match at the peak of the distribution (at $g\sim23$) where the statistical errors are the smallest and both samples should be compete. Any objects in the NGVS GC catalog that have measured radial velocities inconsistent with their being GCs ($v_r \ge 3500$ km~s$^{-1}$) have also been removed, leaving a sample of 9497 objects. Finally, we add to the NGVS GC sample the 92 confirmed or probable UCDs from this study, and plot the combined luminosity function of these 9589 objects as blue open triangles.

There is little to differentiate the two NGVS luminosity functions, although the latter sample (which includes the UCDs) is more noticeably enhanced over the Gaussian curve for $g_0 \lesssim 21$. It is difficult to say with certainty whether either NGVS sample provides a superior match to the ACSVCS results because the UCDs represent only a tiny enhancement  ($\approx$ 1--2\%) to the GC sample, making it is difficult to draw firm conclusions. These difficulties are exacerbated by the differences in the spatial coverage of the HST and NGVS surveys, combined with possible differences in the spatial distribution of GCs and UCDs.

While our ability to isolate large and nearly contaminant-free sample of UCDs and GCs has improved considerably in recent years, thanks to both spectroscopic surveys and imaging programs like the ACSVCS and NGVS, we continue to lack a fully developed and  quantitative theory for the GC luminosity function (i.e., predictive models). Until then, definite conclusions on the correspondence between the UCD and GC luminosity functions will likely remain elusive.

\begin{figure*}
\epsscale{1.10}
\plotone{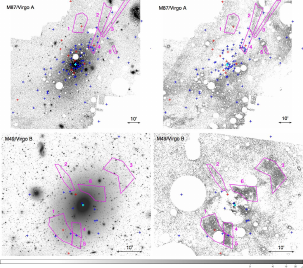}
\caption{(Upper panels). Mosaic $V$-band image of M87 from the study of \citet{2005ApJ_631_41Mihos}. The red and blue UCD candidates identified from the NGVS are shown as the red and blue crosses. The cyan cross shows the M87 center. Regions outlined in magenta are the low surface brightness features identified by \citet{2010ApJ_715_972Janowiecki}. The left and right panels show the original and model galaxy subtracted image, respectively.
(Lower panels). The same as above, except for M49, and showing UCD candidates identified in this region.
}
\label{fig:finder}
\end{figure*}

\subsection{Connections to Features in the Intracluster Light?}

The innermost regions of the principal sub-clusters in Virgo are crowded and dynamic environments in which galaxy interactions have likely played an important role in shaping their present-day appearance. Some UCD formation scenarios invoke processes like satellite stripping, accretion and disruption, so it is natural to ask if there is evidence for a correspondence between the UCDs found in this study and any previously identified features in the diffuse ICL. Given the modest size of the UCD system in M60, we confine the discussion to the UCDs surrounding M87 and M49.

There is evidence for past interactions in both of these regions from several independent studies, including ongoing dwarf galaxy accretion \citep{1994AJ_108_844McNamara, 1997AJ_114_1824Lee, 2012A+A_543_112ArrigoniBattaia}, possible kinematic substructuring among GCs and PNe \citep{2003ApJ_591_850Cote, 2009A+A_502_771Doherty, 2012ApJ_748_29Romanowsky, 2014MNRAS_442_3299Agnello}, dynamical structures (e.g., \citealt{2014A+A_570_69Boselli}), indirect evidence for tidal stripping of certain types of galaxies, including compact ellipticals (e.g., \citealt{2008IAUS_245_395Cote, 2010AIPC_1240_331Cote, 2008ApJ_681_197Peng, 2009A+A_504_347Coenda}), and the discovery of discrete features in the surrounding ICL (e.g., shells, streams and plumes) that almost certainly arose from galaxy interactions and accretions \citep{2005ApJ_631_41Mihos, 2013ApJL_764_20Mihos, 2009ApJ_699_1518Rudick, 2010ApJ_715_972Janowiecki, 2012ApJS_200_4Ferrarese}.

Figure~\ref{fig:finder} shows some of these features for the M87 (upper panels) and M49 (lower panels) regions. The left panels show the original $V$-band images from the study of \citet{2005ApJ_631_41Mihos}, while the images on the right show residuals after subtracting the best-fit galaxy model. The irregular polygons in each panel indicate the features identified by \citet{2010ApJ_715_972Janowiecki}. The blue and red crosses show the UCD candidates in each region after dividing the sample by color at $(g-z)_0 = 1.0$. The UCD samples shown here are the cleaned sample of UCDs in M87 (see \S\ref{sec:cat}), and the $u^*gz$-selected UCD candidates in M49 (see Table~\ref{tab:sample}).

Of the five ICL features in M87 noted by \citet{2010ApJ_715_972Janowiecki}, only eight UCDs fall within their boundaries (a single object in region \#1 and seven more in region \# 4). In the case of M49, two regions contain one or more UCDs: two objects in region \#1, and one more in region \#5. All in all, though, there is little evidence that the UCD populations are preferentially associated with the catalogued features in the ICL of these sub-clusters. Of course, this may simply reflect the fact that most of the known ICL features are found at large radii (where there are relatively few UCDs) presumably because phase-mixing timescales in the inner regions would rapidly erase any such features.

\begin{figure}
\epsscale{1.17}
\plotone{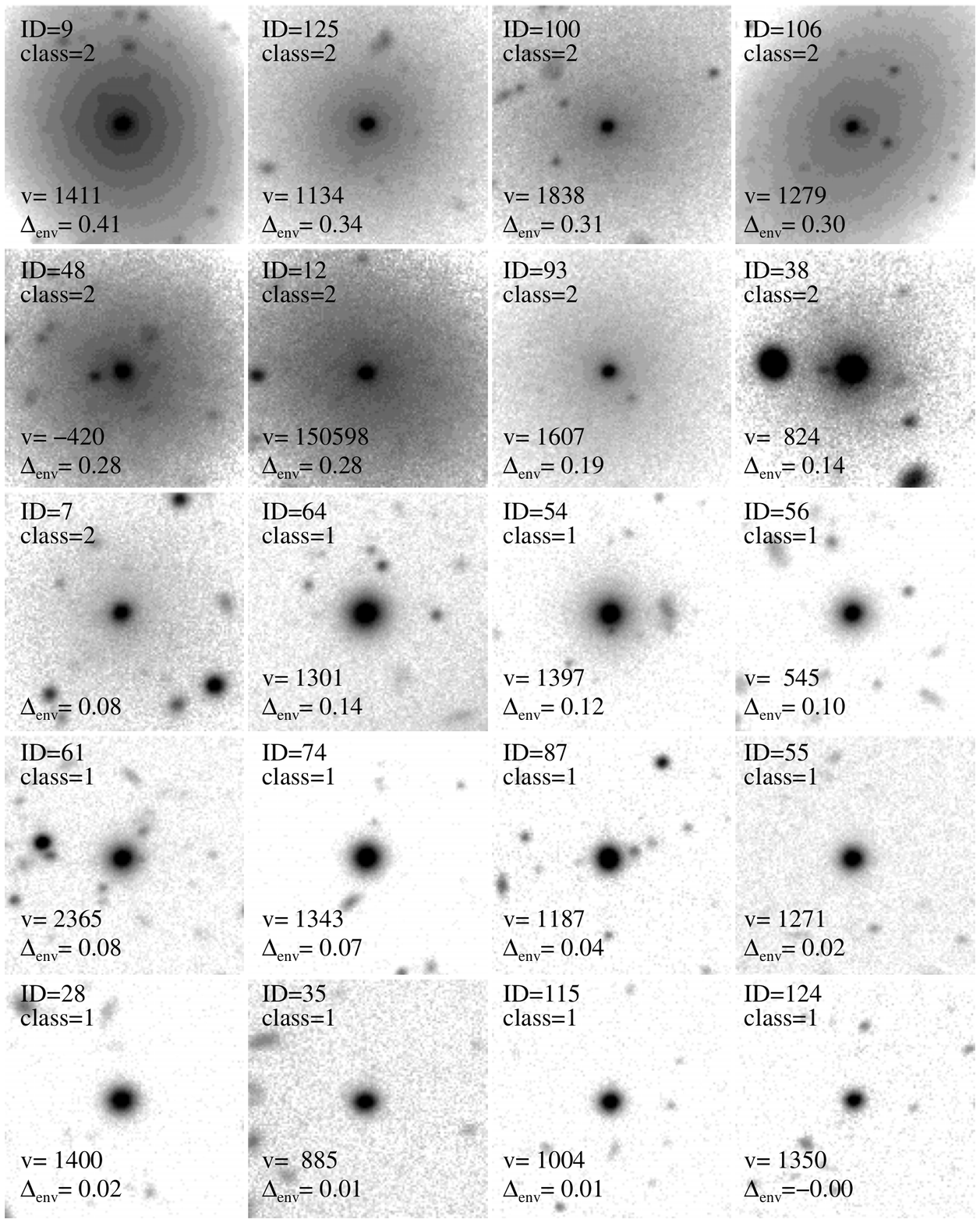}
\caption{Mosaic of images showing 20 of the objectively-selected UCD candidates from Tables~\ref{tab:ucd1} and \ref{tab:ucd2}. The first nine panels show known nucleated dwarf galaxies (\textsf{class}$=$2) from \citet{1985AJ_90_1681Binggeli} or new discoveries from the NGVS. The objects have been loosely arranged according to the visual prominence of an underlying diffuse halo, which is unmistakable in the first objects and non-existent in last ones. Eighteen of these 20 objects are confirmed radial velocity members of the Virgo cluster, the exceptions being two dwarf nuclei (object IDs = 7 and 12). The measured velocities and the envelope parameters $\Delta_{\texttt{env}}$ (see equation \ref{eq:env}) are labeled at the bottom of each panel.
}
\label{fig:seq1}
\end{figure}

\begin{figure}
\epsscale{1.16}
\plotone{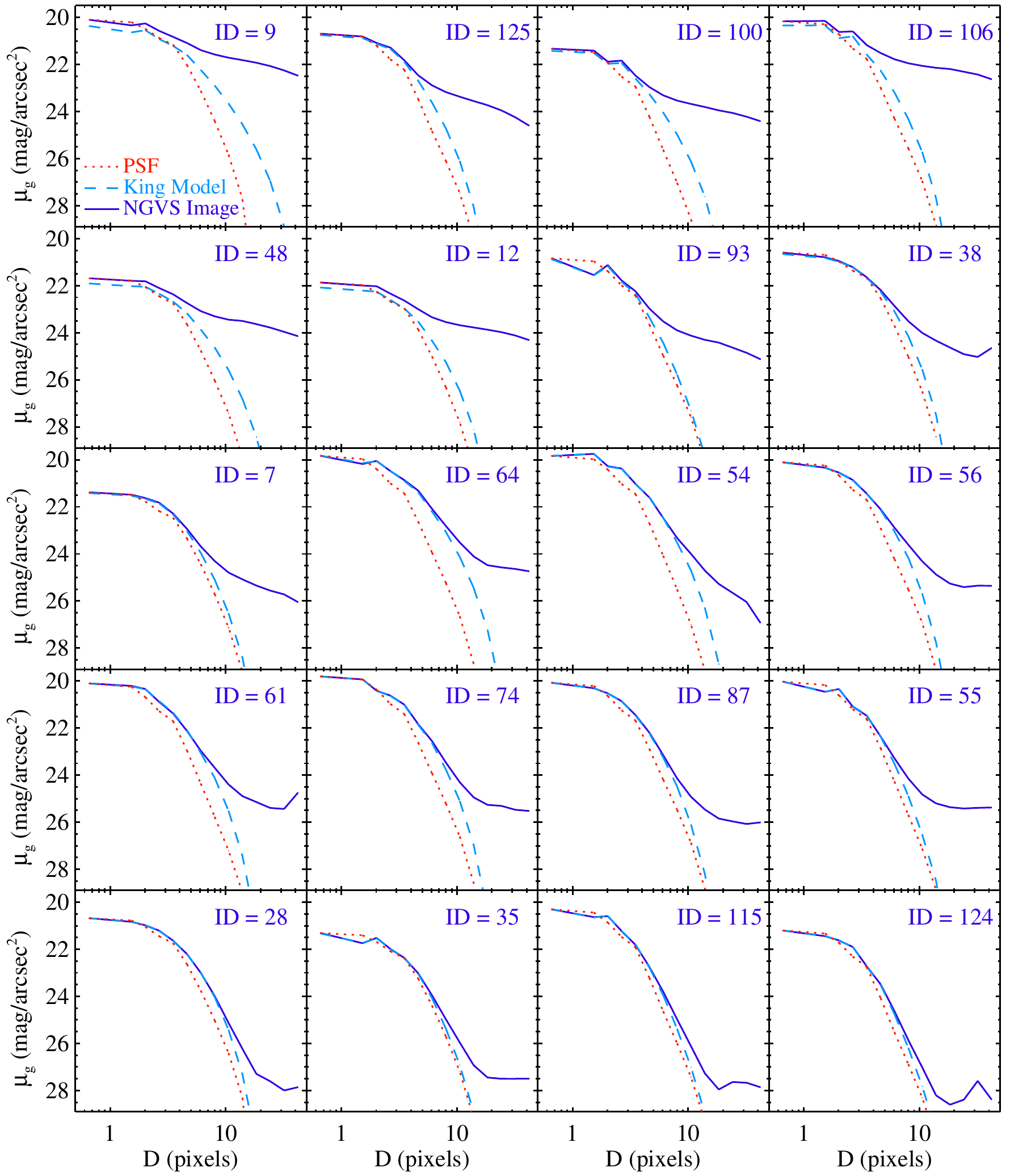}
\caption{The surface brightness profiles (blue solid lines), fitted King models (cyan dashed lines) and PSFs (red dotted lines) of 20 objectively-selected UCD candidates which are shown in Figure \ref{fig:seq1}.
}
\label{fig:seq2}
\end{figure}


\section{Discussion}
\label{sec:discussion}

We showed in \S\ref{sec:obs} that it is possible to measure reliable sizes for UCDs larger than $r_h \sim 10$~pc using NGVS images. By combining these size estimates with cuts in the color-color diagram, visual classifications, and new or published spectroscopic redshifts, we have produced relatively clean samples of UCDs for $\sim$4~deg$^2$ regions centered on each of M87, M49 and M60 --- the three most massive galaxies in the Virgo cluster. In this section, we summarize our key findings and discuss their implications for UCD formation models.

Perhaps most fundamentally, {\it there is clear evidence for large region-to-region differences in the number of UCDs} (\S\ref{sec:pop}). Such variations would {\it not} be expected if $N_{\rm UCD}$ scales with either the luminosity or stellar mass of the central galaxy, which are similar in all three cases. However, these variations would be expected {\it if the number of UCDs scales in proportion to the mass of the surrounding subcluster}; as Figure~\ref{fig:sn} shows, this is precisely what is observed, with $N_{\rm UCD}$ increasing in step with other tracers of sub-cluster mass: X-ray gas mass, dark matter mass, and the number of surrounding galaxies and GCs. Moreover, the flattening and orientation of the UCDs in M87 appear to match that of the central galaxy isophotes and/or ICL in this region.

An increasing number of studies support the view that UCD formation is tied to tidal stripping  (e.g., \citealt{2001ApJL_552_105Bekki, 2003Natur_423_519Drinkwater, 2013MNRAS_433_1997Pfeffer, 2014MNRAS_444_3670Pfeffer, 2014Natur_513_398Seth, 2015MNRAS_449_1716Janz, 1507.02270}). In this scenario, the larger UCD system surrounding M87 ($\sim$ 3.5 and 7.8 times more than M49 and M60) can be understood in terms of the larger reservoir of low-mass galaxies in this more massive environment, while the stronger tidal forces (due to the larger virial mass) would allow low-mass galaxies to be stripped more easily.

Meanwhile, the rather {\it blue colors of most UCDs} (\S\ref{sec:colors}) are consistent with expectations in the tidal stripping scenario if their putative progenitors were low-mass, metal-poor galaxies. Similarly, the fact that both the UCD colors and their {\it location in the color-magnitude relation} closely match the nuclei lends further support to the notion that tidal stripping of low-mass galaxies was a dominant process in UCD formation. Although we find no evidence that the UCDs are preferentially associated with known low-surface brightness features in the ICL, this probably reflects the fact that only those features located at large radii --- beyond the bulk of the UCD populations in each subcluster --- would still be visible at the present time. Clearly, though, this is an area where additional study is needed, preferably using numerical simulations with realistic cosmological initial conditions.

In a companion paper, \citet{2015ApJ_802_30Zhang} have examined the kinematics of UCDs in the M87 region, comparing their properties to GCs and low-mass galaxies. They find the UCDs have a similar velocity dispersion profile with the blue GCs. There is a hint that the velocity anisotropy of the UCDs changes from tangentially biased ($\beta < 0$) to radially biased ($\beta > 0$) with increasing distance from M87, and, to within the uncertainties, the GCs and UCDs are found to have similar anisotropy profiles with possible deviations at large radii. On the other hand, the UCDs show somewhat larger rotation than either the blue or red GCs. Thus, while there is no compelling evidence from kinematics alone that tidal stripping of low-mass galaxies is a primary mechanism for UCD formation, the more radially biased orbits of UCDs compared to blue GCs at large radii are suggestive that stripping plays a role.

Is there evidence from the structural properties of the UCDs and dwarf galaxies themselves that might support a connection between UCDs and low-mass galaxies? Figure~\ref{fig:seq1} uses our sample of 92 likely or confirmed UCDs surrounding M87, along with the nine dwarf nuclei in this same region that were objectively identified in our UCD search, to illustrate a heuristic structural sequence among these $\sim$100 objects. The 20 objects shown in this mosaic have been arranged, from left to right and top to bottom, by the apparent prominence of a diffuse envelope surrounding each compact object. For the first half of the sample, there is clear evidence for the two-component structure (i.e., a nucleus plus an envelope) that is characteristic of dE,N, dS0,N, or dSph,N galaxies. Indeed, all but one of these objects was previously cataloged as a nucleated dwarf galaxy residing in the cluster core \citep{1985AJ_90_1681Binggeli}. The sequence extends to include very compact UCD candidates that show little or no sign of an extended envelope. Figure~\ref{fig:seq2} shows surface brightness profiles, PSFs and fitted King models for each of the objects shown in Figure~\ref{fig:seq1}. The gradual progression from a two- to one-component structure is apparent.

\begin{figure}
\epsscale{1.15}
\plotone{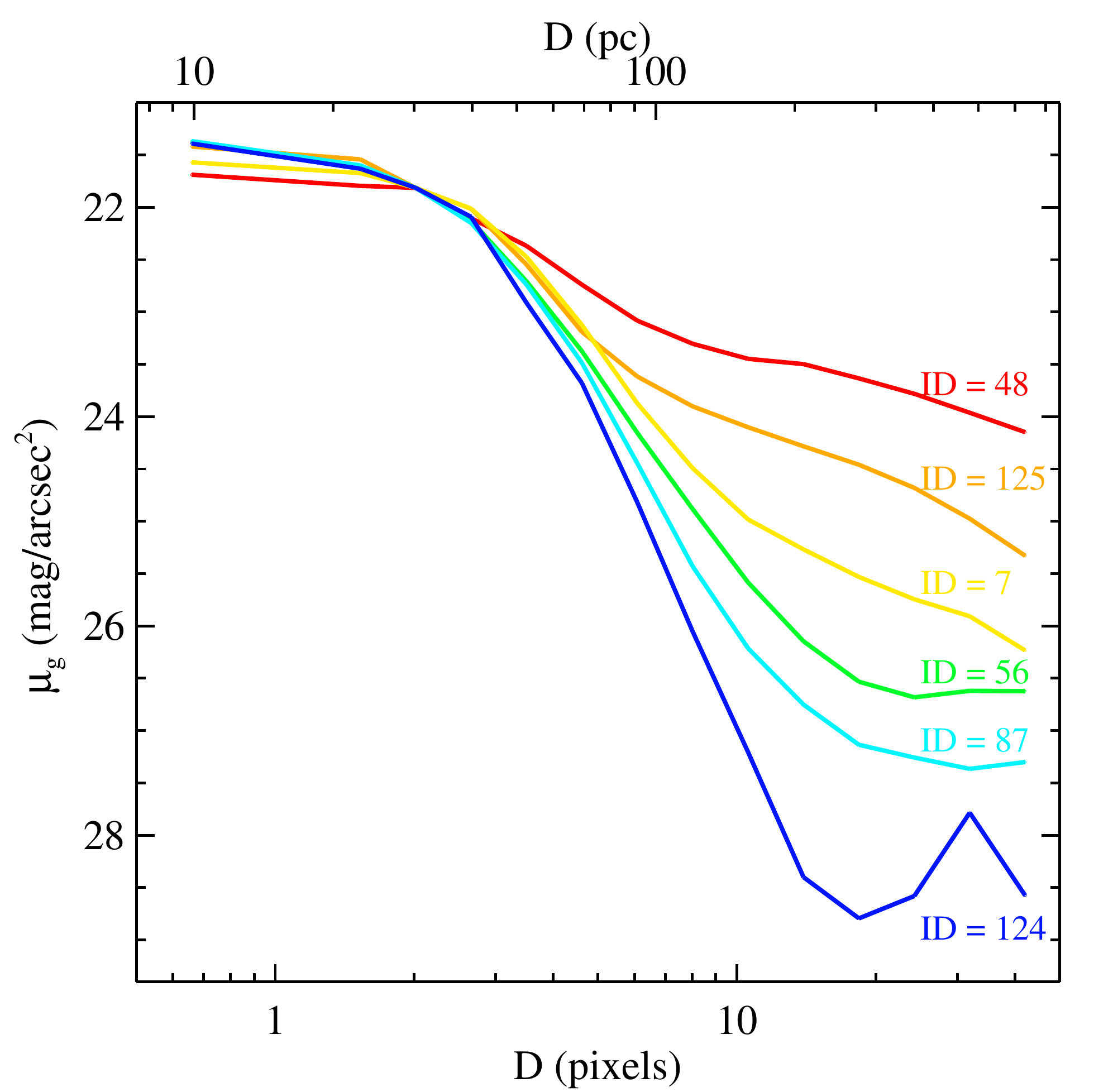}
\caption{Surface brightness profiles of UCDs with a range of envelope fraction. The surface brightness profiles were shifted vertically to match the profile of ID=48 at $D\sim30pc$ (roughly the mean half-light radius of UCDs) to highlight the differences in envelope components.
}
\label{fig:profile}
\end{figure}

In Figure~\ref{fig:profile}, we plot the sequence of UCD and envelope structure on a common axis to highlight the change in envelope fraction relative to the central component. This figure is reminiscent of Figure~5 in \cite{2013MNRAS_433_1997Pfeffer}, which showed the simulated surface brightness profile evolution of a dwarf galaxy being tidally stripped.

\begin{figure}
\epsscale{1.15}
\plotone{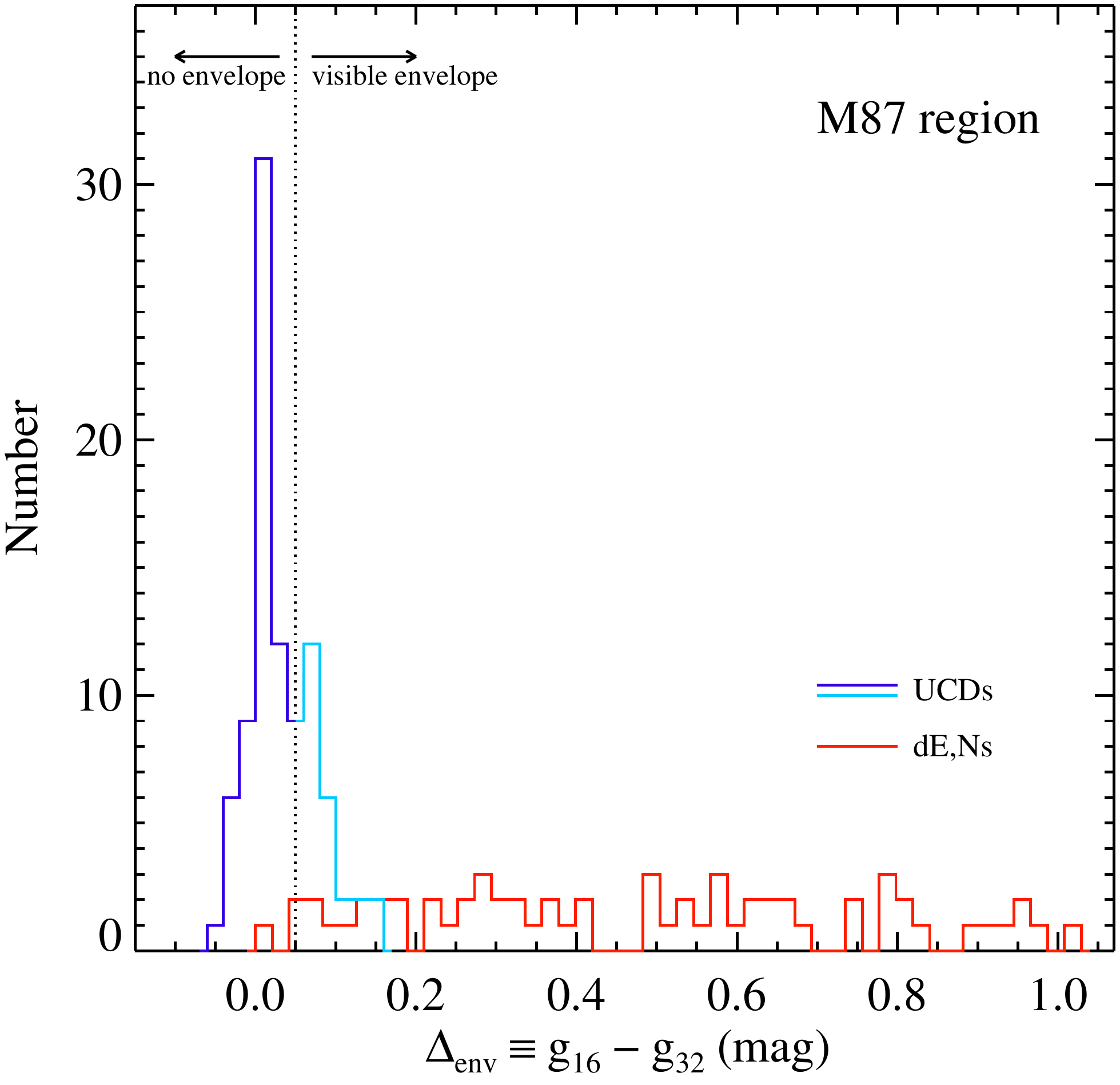}
\caption{Histogram of difference in aperture magnitudes, $\Delta_{\rm env}$, for UCD candidates and dwarf nuclei in the M87 region. Here $\Delta_{\rm env}$ is the magnitude difference measured in apertures of 16- and 32-pixel diameter. A value of $\Delta_{\rm env} = 0.05$~mag (denoted by the dotted vertical line) roughly divides objects with, and without, diffuse extended envelopes.
}
\label{fig:delta1}
\end{figure}

\begin{figure}
\epsscale{1.15}
\plotone{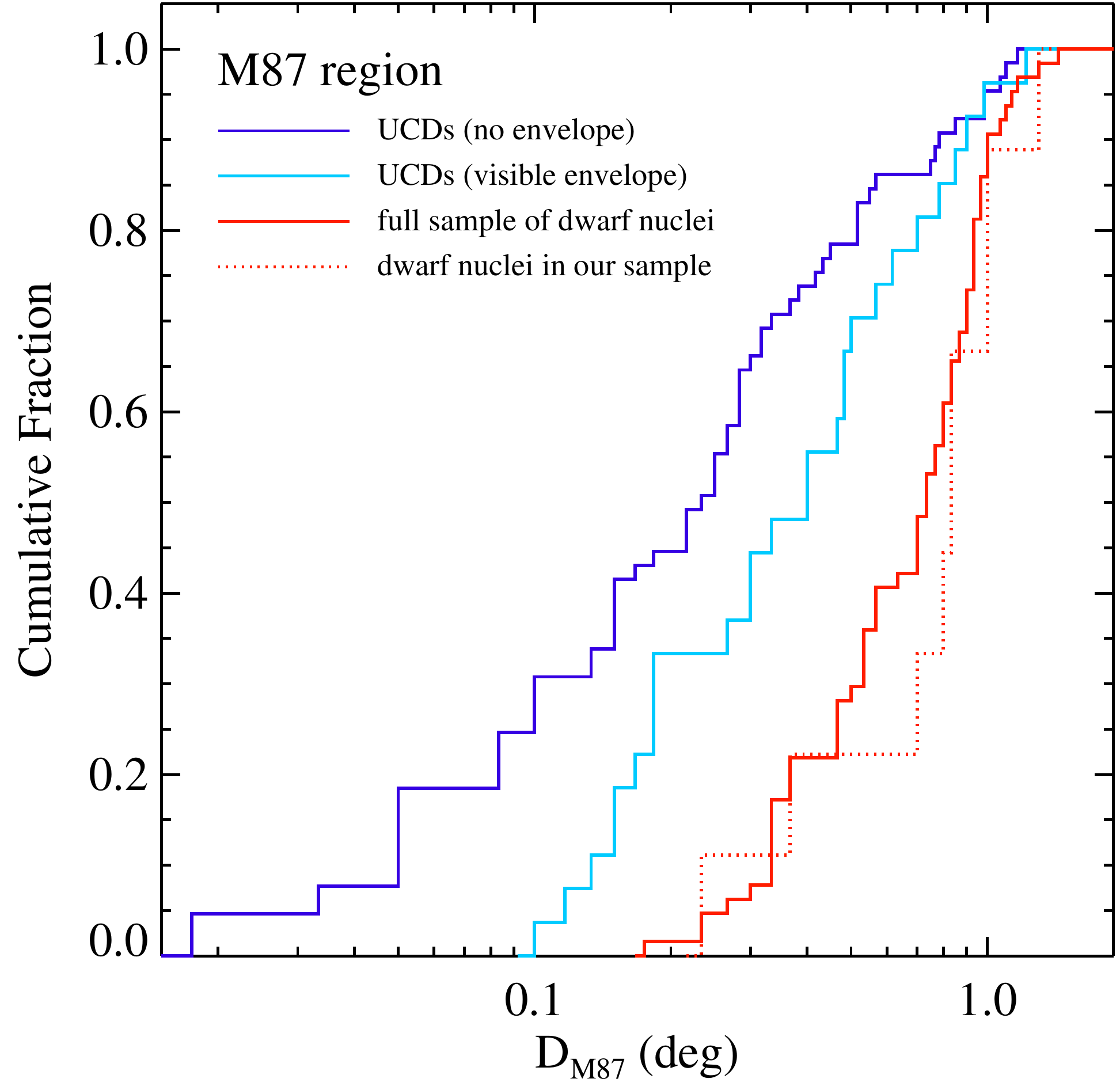}
\caption{Cumulative distributions for D$_{M87}$ (the projected distance from M87) for the objectively identified UCD candidates in the M87 region (blue and cyan lines), the full sample of 64 dwarf nuclei (red solid line), and the nine additional candidates subsequently classified as dwarf nuclei (red dotted line). The cleaned sample of 92 UCD candidates has been divided into two classes by $\Delta_{\rm env}$, which is a measure of the prominence of an outer envelope.
}
\label{fig:delta2}
\end{figure}

\begin{figure}
\epsscale{1.15}
\plotone{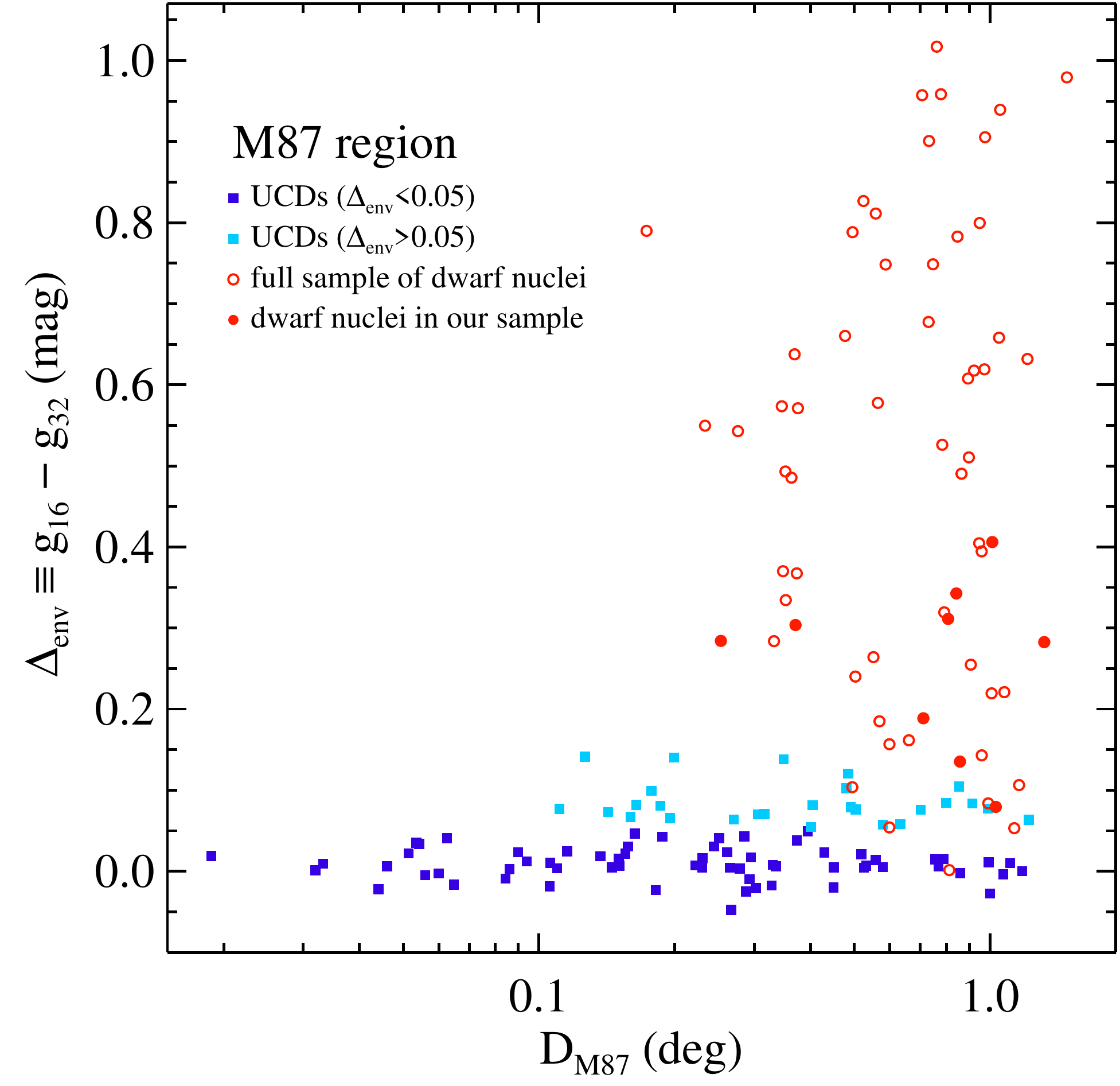}
\caption{Envelope prominence as a function of the projected distance from M87 for UCDs and nucleated dwarf galaxies.
}
\label{fig:delta3}
\end{figure}

\begin{figure}
\epsscale{1.15}
\plotone{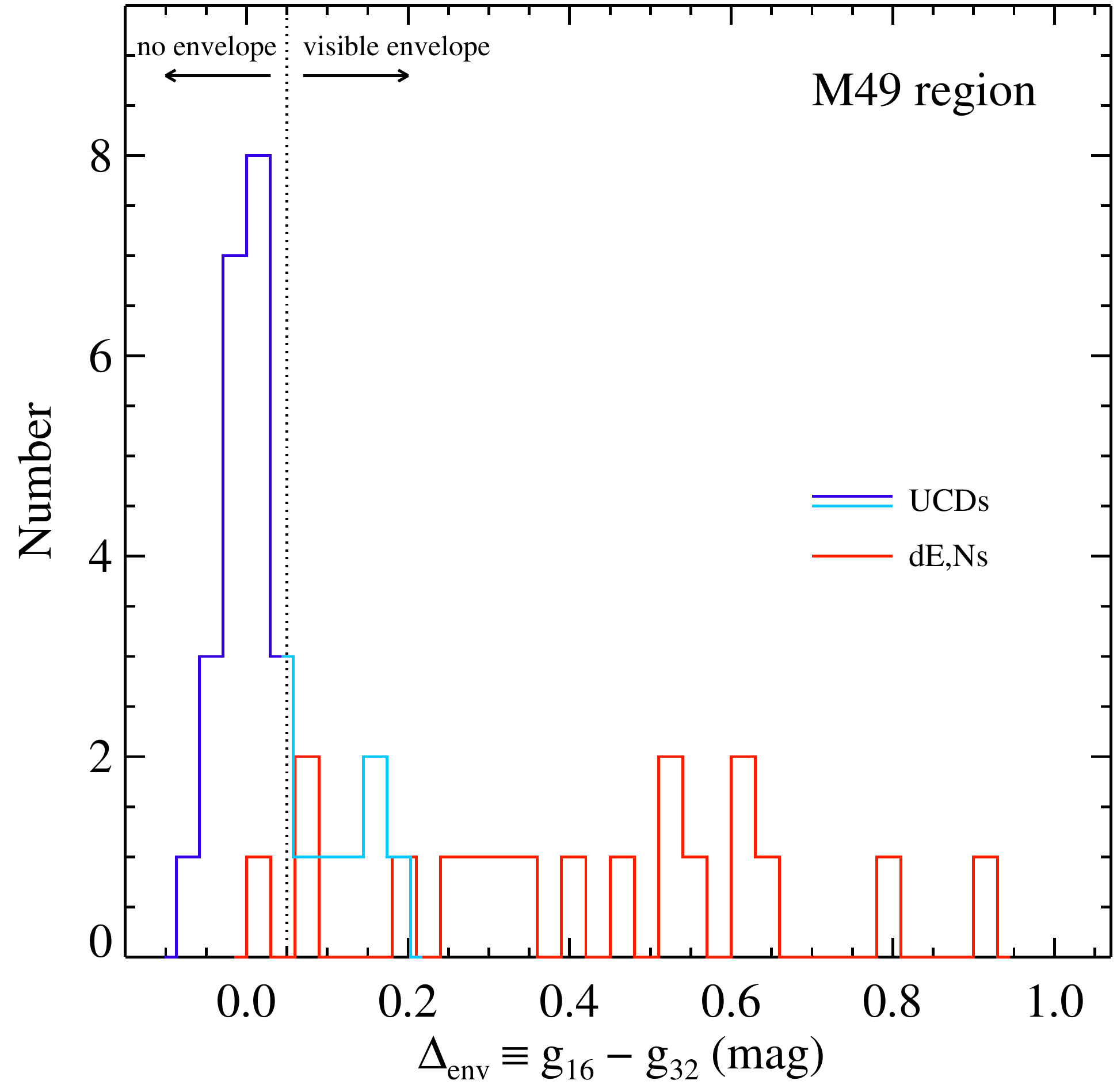}
\caption{The same as Figure \ref{fig:delta1}, except for the M49 region.
}
\label{fig:env1_m49}
\end{figure}

\begin{figure}
\epsscale{1.15}
\plotone{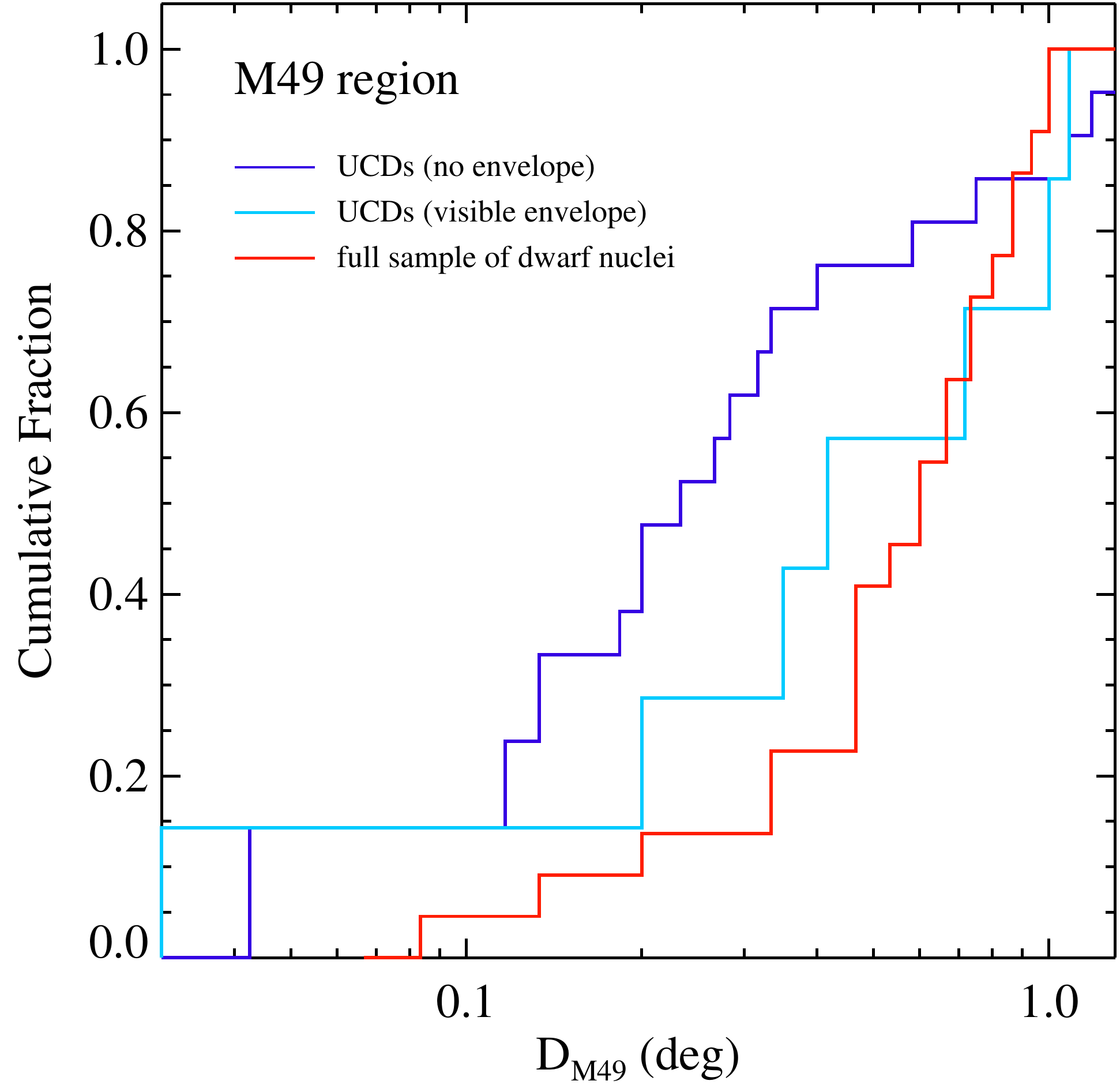}
\caption{The same as Figure \ref{fig:delta2}, except for the M49 region.
}
\label{fig:env2_m49}
\end{figure}

\begin{figure}
\epsscale{1.15}
\plotone{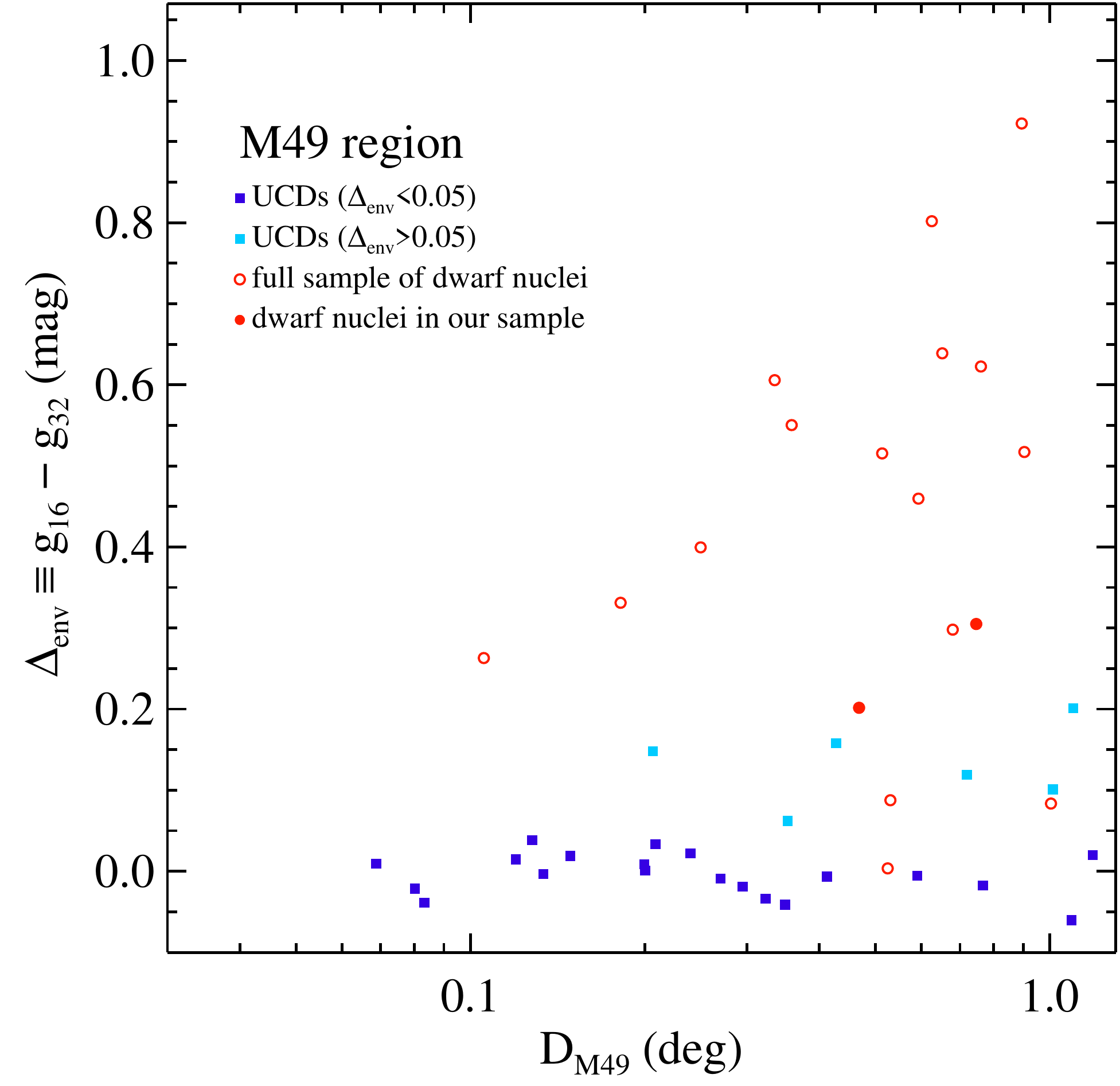}
\caption{The same as Figure \ref{fig:delta3}, except for the M49 region.
}
\label{fig:env3_m49}
\end{figure}

Because this sequence is based purely on visual appearance, a more quantitative approach is desirable. We therefore introduce the parameter
\begin{equation}\label{eq:env}
\begin{array}{rcl}
{\Delta}_{\rm env} & \equiv & g_{\rm 16} - g_{\rm 32} \\
\end{array}
\end{equation}
where $g_{\rm 16}$ and $g_{\rm 32}$ are the $g$-band corrected point-source magnitudes measured in apertures of diameter 16 and 32 pixels (see \S\ref{sec:cata}). These apertures correspond to radii of 1\farcs5 and 3\farcs0, respectively, and their difference should provide a diagnostic of any diffuse light surrounding the candidates.

Figure~\ref{fig:delta1} shows the distribution of $\Delta_{\rm env}$ values for the sample of 92 likely or confirmed UCDs, and 64 dwarf nuclei, in the M87 region. The dwarf nuclei, which are shown in red, are broadly distributed with $0.00 \lesssim \Delta_{\rm env} \lesssim 1.00$~mag. By contrast, the UCDs ({\tt class} = 1 objects), which are shown by the cyan and blue histogram, are strongly peaked at $\Delta_{\rm env} \simeq 0$~mag, with an asymmetry to positive values. We can use the observed dwarf distribution as a guide in separating UCD candidates with, and without, an appreciable envelope by adopting $\Delta_{\rm env} \simeq 0.05$~mag as the dividing point. UCD candidates with $\Delta_{\rm env} < 0.05$~mag and $\Delta_{\rm env} \ge 0.05$~mag are shown in blue and cyan, respectively.

The cumulative distribution of projected distances to M87, D$_{\rm M87}$, for these three samples are compared in Figure~\ref{fig:delta2}. It is obvious that there are dramatic differences in their spatial distribution, with the dwarf nuclei lying preferentially at larger distances than the UCD candidates (see \citealt{2004PASA_21_375Drinkwater} for a similar result in the Fornax cluster). UCDs that show signs of a faint envelope, in turn, have a larger mean distance than those that show no envelope. These same trends are also apparent in Figure~\ref{fig:delta3}, which plots ${\Delta}_{\rm env}$ against D$_{\rm M87}$ for the individual objects. Although the dwarfs have a distribution in $\Delta_{\rm env}$, most are located at large distances (D$_{\rm M87}$ $\ge 0\fdg2$). The UCD candidates with ${\Delta}_{\rm env} < 0.05$~mag are found at all radii, whereas those with ${\Delta}_{\rm env} \ge 0.05$~mag appear to form an intermediate population.

As a check on the significance of the observed trend for compact systems surrounding M87, we show in Figure~\ref{fig:env1_m49}, \ref{fig:env2_m49} and \ref{fig:env3_m49} the analogous distribution for M49, but for its sample of 28 UCDs and 22 nucleated dwarf galaxies. Clearly, the behavior seen in the M87 region is seen here as well.

The most straightforward interpretation of the observed trend is that {\it proximity to the central, massive galaxy in these regions governs the envelope structure of UCDs and dwarf nuclei}. This finding may provide an important new piece of evidence in support of the tidal stripping scenario, since the strong tidal forces near the bottom of the gravitational potential wells in these sub-clusters would act to strip away the envelopes of the nearest UCD progenitors.


\section{Conclusions and Directions for Future Work}

In this paper, we have presented the first homogeneous and complete photometric study of UCDs around the three prominent Virgo sub-clusters, centered on the galaxies M87, M49, and M60. Using deep, multi-color NGVS imaging, we select UCDs with high completeness using their measured colors and half-light radii. Below, we summarize the main findings of our study:

\begin{itemize}
\item Using the excellent image quality in the NGVS data, we can reliably measure the half-light radii of objects in Virgo down to a limiting size of $\sim 10$~pc at $g\approx21.5$~mag, making the data ideally suited for UCD studies.

\item We present samples of 92, 28, and 23 confirmed or probable UCDs associated with M87, M49, and M60 (within $2^{\circ} \times 2^{\circ}$ boxes), respectively, selected to have $18.5<g<21.5$~mag ($-12.7\lesssim M_g\lesssim -9.7$) and half-light radii $11<r_h<100$~pc. These objects were selected using a combination of criteria including luminosity, size, colors, effective surface brightness, and visual inspection.

\item {\it The number of UCDs scales with the total mass of the host system.} M87 contains $\sim3.5$ and 7.8 times more UCDs than M49 and M60 within a radius of 25$R_e$, respectively. However, there are tight correlations between the number of UCDs and the parameters that describe the total mass of each sub-cluster (X-ray gas mass, dark matter mass, number of neighboring galaxies, and number of GCs). Normalized to the parameters such as dark matter mass and the number of neighboring galaxies, M87, M49 and M60 appear to have formed UCDs with comparable efficiency.

\item {\it The spatial distribution of UCDs mirrors the galaxy light.} The M87 UCDs appear to follow the ellipticity and orientation of the low surface brightness stellar light in the halo.

\item {\it Most UCDs have blue colors.} We find that $\sim\!85\%$ of the UCDs in M87 have colors similar to blue GCs or stellar nuclei. The remaining red UCDs are mostly more compact ($11<r_h<18$~pc, and may be consistent with being the tail of the red GC population. However, we note that a few of the most luminous UCDs in the literature are quite red in color, so not all UCDs are blue.

\item {\it M87 and M49 UCDs show no color gradient with galactocentric radius.} The UCDs in the M87 and M49 regions display no significant gradient in color. This lack of a gradient is also found for the nuclei of dwarf galaxies around M87 and M49. Meanwhile, the GCs show a color gradient in both their blue and red sub-populations.

\item {\it M87 and M49 UCDs show a relation between color and magnitude. } Like what has been seen before for blue GCs, the M87 UCDs display a color-magnitude relation, where the more luminous UCDs tend to be redder. Dwarf galaxy nuclei also seem to follow a similar color-magnitude trend. Similar results are found in the M49 region.

\item {\it The M87 UCD system represents a 1--2\% enhancement over a Gaussian GCLF at the bright end.} By combining the GCLF from the ACS Virgo Cluster Survey and the NGVS, we find that inclusion of the UCDs causes the luminosity function at the bright end to be enhanced over a Gaussian extrapolation by a small amount. Lacking a developed theory of the GCLF, however, the interpretation of this result is difficult.

\item {\it There is little evidence that UCDs are preferentially associated with streams or features currently visible in the ICL.} We matched the locations of UCDs with ICL features detected in deep images of the M87 and M49 regions and did not find that UCD numbers were enhanced in these regions.

\item{\it UCDs around M87 follow a morphological sequence defined by the prominence of their outer, low surface brightness envelope, with the envelope fraction correlating with distance from M87.} We quantify the fraction of stellar light in UCDs that is contained in an outer envelope, and find that the M87 UCDs span a range of envelope fraction that merges into the sequence of nucleated dwarf galaxies. The mean UCD envelope fraction rises with distance from M87. The similar result is also found for the UCDs around M49. This sequence suggests that tidal stripping may play an important role in the formation of UCDs.

\end{itemize}

Taken together, our results suggest that many UCDs are distinct from normal GCs, and originate as stellar nuclei that are subsequently subjected to a process of tidal stripping that depends mainly on the total mass of the host halo. Some objects selected as UCDs in our study, however, may be the massive and extended tail of the GC population as these two populations likely overlap.

Several obvious extensions to this work present themselves. First, our study has focused on just three regions covering a combined area of $\sim$ 11 deg$^2$. Since the full NGVS survey spans an area of $\sim$ 104 deg$^2$ \citep{2012ApJS_200_4Ferrarese}, a systematic search for UCDs --- based on both effective radii and location in the $u^*gz$ diagram --- covering a ten-fold larger area is possible and, indeed, will be presented in a future paper in this series (Liu et al., in preparation). Second, the addition of $K_s$-band imaging is obviously invaluable in the identification of both UCDs and GCs (\citealt{2014ApJS_210_4Munoz}; \S\ref{sec:uik}) as it leads directly to a 2--3$\times$ reduction in the level of background contamination (see \S\ref{sec:control_fields}). Programs to acquire additional IR imaging are currently underway (CFHT/WIRCam, VISTA/VIRCam).

Third, a careful examination of the scaling relations for UCDs and GCs requires accurate structural information (e.g, effective radius, mean surface brightness, concentration, etc.) that is beyond the reach of ground-based imaging. While the NGVS image quality is exceptional compared to most ground-based surveys, it is nevertheless suitable only for identifying compact stellar systems with sizes larger than $r_h \sim 10$~pc (see \S\ref{sec:size}). On the other hand, measuring accurate structural parameters for Virgo UCDs and GCs is entirely straightforward with HST, and both new and archival observations could be used for this purpose. Fourth, low-resolution spectroscopy for UCD candidates from the NGVS is within the reach of existing multi-fiber spectrographs on 4m- and 8m-class telescopes, and such spectra would allow the kinematic and chemical properties of NGVS-selected UCDs to be examined in a systematic way (e.g., \citealt{2015ApJ_802_30Zhang}). Finally, high-resolution, integrated-light spectroscopy for the brightest UCDs would, once combined with structural parameters from HST and multi-color photometry from the NGVS and UV/IR surveys, lead to a better understanding of their internal properties, including their initial mass function \citep{2010MNRAS_403_1054Dabringhausen, 2012ApJ_747_72Dabringhausen, 2012MNRAS_422_2246Marks}, dynamical evolution \citep{2011MNRAS_412_1627Chilingarian}, and possible dark matter and/or central black hole content \citep{2005ApJ_627_203Hacsegan, 2008A+A_487_921Mieske, 2013A+A_558_14Mieske, 2011MNRAS_414_70Frank, 2011MNRAS_412_1627Chilingarian,2014Natur_513_398Seth}.

\acknowledgments

The NGVS team owes a debt of gratitude to the director and the staff of the Canada-France-Hawaii Telescope, whose dedication, ingenuity, and expertise have helped make the survey a reality.

We thank Xiaohu Yang, Zhengyi Shao and Shiyin Shen for helpful discussions. This work is supported by the National Key Basic Research Program of China (2015CB857002). CL acknowledges support from the National Natural Science Foundation of China (NSFC, Grant No. 11203017). We also thank the support of a key laboratory grant from the Office of Science and Technology, Shanghai Municipal Government (No. 11DZ2260700). EWP acknowledges support from the National Natural Science Foundation of China under Grant No. 11173003, and from the Strategic Priority Research Program, ¡±The Emergence of Cosmological Structures¡±, of the Chinese Academy of Sciences, Grant No. XDB09000105. JCM is supported by the NSF through grant AST-1108964. HXZ acknowledges support from China Postdoctoral Science Foundation under Grant No. 552101480582. HXZ also acknowledges support from CAS-CONICYT Postdoctoral Fellowship, administered by the Chinese Academy of Sciences South America Center for Astronomy (CASSACA). AJ acknowledges support from BASAL CATA PFB-06. S.M. acknowledges financial support from the Institut Universitaire de France (IUF), of which she is senior member. THP acknowledges support through FONDECYT Regular Project Grant No.\ 1121005 and BASAL Center for Astrophysics and Associated Technologies (PFB-06). PG acknowledges the NSF grant AST-1010039, and support from the LAMOST-PLUS collaboration, a partnership funded by NSF grant AST-09-37523, and NSFC grants 10973015 and 11061120454. HX acknowledges support from the National Natural Science Foundation of China (NSFC, Grant No. 11125313).

This work is based on observations obtained with MegaPrime/MegaCam, a joint project of CFHT and CEA/DAPNIA, at the CanadaFranceHawaii Telescope (CFHT) which is operated by the National Research Council (NRC) of Canada, the Institut National des Sciences de Univers of the Centre National de la Recherche Scientifique (CNRS) of France, and the University of Hawaii. This work is based in part on data products produced at Terapix available at the Canadian Astronomy Data Centre as part of the Canada-France-Hawaii Telescope Legacy Survey, a collaborative project of NRC and CNRS.

This work was supported in part by the Sino-French LIA-Origins joint exchange program, by the French Agence Nationale de la Recherche (ANR) Grant Programme Blanc VIRAGE (ANR10-BLANC-0506-01), and by the Canadian Advanced Network for Astronomical Research (CANFAR) which has been made possible by funding from CANARIE under the Network-Enabled Platforms program. This research used the facilities of the Canadian Astronomy Data Centre operated by the National Research Council of Canada with the support of the Canadian Space Agency. The authors further acknowledge use of the NASA/IPAC Extragalactic Database (NED), which is operated by the Jet Propulsion Laboratory, California Institute of Technology, under contract with the National Aeronautics and Space Administration, and the HyperLeda database (http://leda.univ-lyon1.fr). This publication has made use of data products from the Sloan Digital Sky Survey (SDSS). Funding for SDSS and SDSS-II has been provided by the Alfred P. Sloan Foundation, the Participating Institutions, the National Science Foundation, the U.S. Department of Energy, the National Aeronautics and Space Administration, the Japanese Monbukagakusho, the Max Planck Society and the Higher Education Funding Council for England.

This research uses data obtained through the Telescope Access Program (TAP), which has been funded by the National Astronomical Observatories of China, the Chinese Academy of Sciences (the Strategic Priority Research Program "The Emergence of Cosmological Structures" Grant No. XDB09000000), and the Special Fund for Astronomy from the Ministry of Finance. Observations reported here were obtained at the MMT Observatory, a joint facility of the University of Arizona and the Smithsonian Institution.

{\it Facilities}: CFHT


\clearpage

\tabletypesize{\tiny}
\begin{landscape}
\LongTables
\begin{deluxetable}{lrrrcrrrrrrrc}
 \tablewidth{0pt}
 \tabletypesize{\small}
 \tablecaption{Photometric Properties of UCD Candidates in the M87 Region.
 \label{tab:ucd1}}
 \tablehead{
  \colhead{ID} &
  \colhead{NGVSID} &
  \colhead{$\alpha_{\rm J2000}$} &
  \colhead{$\delta_{\rm J2000}$} &
  \colhead{$E(B-V)$} &
  \colhead{$g_{\rm 16}$} &
  \colhead{$(u^*-g)_{\rm 8}$} &
  \colhead{$(g-r)_{\rm 8}$} &
  \colhead{$(g-i)_{\rm 8}$} &
  \colhead{$(g-z)_{\rm 8}$} &
  \colhead{$(g-K_s)_{\rm 8}$} &
  \colhead{$\Delta_{\rm env}$} &
  \colhead{UCD} \\
  \colhead{ } &
  \colhead{ } &
  \colhead{(deg)} &
  \colhead{(deg)} &
  \colhead{(mag)} &
  \colhead{(mag)} &
  \colhead{(mag)} &
  \colhead{(mag)} &
  \colhead{(mag)} &
  \colhead{(mag)} &
  \colhead{(mag)} &
  \colhead{(mag)} &
  \colhead{} \\
  \colhead{(1)} &
  \colhead{(2)} &
  \colhead{(3)} &
  \colhead{(4)} &
  \colhead{(5)} &
  \colhead{(6)} &
  \colhead{(7)} &
  \colhead{(8)} &
  \colhead{(9)} &
  \colhead{(10)} &
  \colhead{(11)} &
  \colhead{(12)} &
  \colhead{(13)}}
 \startdata
  1 & NGVS-J122623.21+121925.2 & 186.5966882 & 12.3236728 & 0.027 & 19.603$\pm$0.001 & 0.910$\pm$0.003 & 0.435$\pm$0.002 & 0.728$\pm$0.002 & 0.840$\pm$0.003 & 1.231$\pm$0.003 &  0.061 & 0 \\
  2 & NGVS-J122628.69+125642.6 & 186.6195415 & 12.9451739 & 0.029 & 19.695$\pm$0.001 & 1.312$\pm$0.003 & 0.616$\pm$0.002 & 0.920$\pm$0.002 & 1.120$\pm$0.003 & 1.145$\pm$0.004 &  0.064 & 1 \\
  3 & NGVS-J122633.05+123847.6 & 186.6377219 & 12.6465456 & 0.032 & 21.303$\pm$0.003 & 1.009$\pm$0.008 & 0.482$\pm$0.005 & 0.728$\pm$0.006 & 0.808$\pm$0.009 & 0.596$\pm$0.015 &  0.006 & 0 \\
  4 & NGVS-J122636.42+125512.9 & 186.6517616 & 12.9202470 & 0.028 & 18.947$\pm$0.001 & 1.354$\pm$0.002 & 0.622$\pm$0.001 & 0.944$\pm$0.001 & 1.152$\pm$0.002 & 1.226$\pm$0.002 &  0.000 & 1 \\
  5 & NGVS-J122638.52+123702.7 & 186.6604821 & 12.6174078 & 0.032 & 20.575$\pm$0.002 & 1.041$\pm$0.005 & 0.498$\pm$0.003 & 0.753$\pm$0.004 & 0.825$\pm$0.005 & 0.617$\pm$0.009 & -0.004 & 1 \\
  6 & NGVS-J122638.79+124545.8 & 186.6616360 & 12.7627110 & 0.030 & 20.520$\pm$0.002 & 1.029$\pm$0.005 & 0.498$\pm$0.003 & 0.750$\pm$0.004 & 0.886$\pm$0.005 & 0.632$\pm$0.007 &  0.010 & 1 \\
  7 & NGVS-J122643.32+121743.9 & 186.6804595 & 12.2955414 & 0.027 & 20.750$\pm$0.002 & 0.985$\pm$0.007 & 0.517$\pm$0.003 & 0.723$\pm$0.004 & 0.907$\pm$0.007 & 0.490$\pm$0.012 &  0.079 & 0 \\
  8 & NGVS-J122646.48+131732.2 & 186.6936696 & 13.2922923 & 0.027 & 21.395$\pm$0.003 & 0.936$\pm$0.008 & 0.467$\pm$0.005 & 0.727$\pm$0.006 & 0.788$\pm$0.009 & 1.203$\pm$0.013 & -0.006 & 0 \\
\enddata
 \tablenotetext{}{NOTES -- (1) Object ID number; (2) Object name in NGVS; (3) Right Ascension; (4) Declination; (5) Galactic extinction \citep{1998ApJ_500_525Schlegel}; (6) Aperture-corrected $g$ magnitude within a 3-arcsec diameter aperture. Only SExtractor errors are accounted for here; (7-11) Colors measured in a 1.5-arcsec diameter aperture; (12) Envelope parameter defined in \ref{eq:env}; (13) Object flag, 0 = not UCD, 1 = confirmed or possible UCD. }
\end{deluxetable}
\clearpage
\end{landscape}

\tabletypesize{\tiny}
\begin{landscape}
\LongTables
\begin{deluxetable}{llrrrrrrrrrrrrrcl}
 \tablewidth{0pt}
 \tabletypesize{\small}
 \tablecaption{Structural Properties and Redshift Information for UCD Candidates in the M87 Region
 \label{tab:ucd2}}
 \tablehead{
  \colhead{ID} &
  \colhead{NGVSID} &
  \colhead{$g_k$} &
  \colhead{$i_k$} &
  \colhead{$r_{h,g}$} &
  \colhead{$r_{h,i}$} &
  \colhead{$r_h$(ACS)} &
  \colhead{$r_h$(S11)} &
  \colhead{$v_r$(S11)} &
  \colhead{$v_r$(MMT)} &
  \colhead{$v_r$(AAT)} &
  \colhead{$v_r$(SDSS)} &
  \colhead{Class} &
  \colhead{$uiK_s$} &
  \colhead{$ugz$} &
  \colhead{UCD} &
  \colhead{OtherName}  \\
  \colhead{ } &
  \colhead{ } &
  \colhead{(mag)} &
  \colhead{(mag)} &
  \colhead{(pc)} &
  \colhead{(pc)} &
  \colhead{(pc)} &
  \colhead{(pc)} &
  \colhead{(km~s$^{-1}$)} &
  \colhead{(km~s$^{-1}$)} &
  \colhead{(km~s$^{-1}$)} &
  \colhead{(km~s$^{-1}$)} &
  \colhead{ } &
  \colhead{ } &
  \colhead{ } &
  \colhead{ } &
  \colhead{ }  \\
  \colhead{(1)} &
  \colhead{(2)} &
  \colhead{(3)} &
  \colhead{(4)} &
  \colhead{(5)} &
  \colhead{(6)} &
  \colhead{(7)} &
  \colhead{(8)} &
  \colhead{(9)} &
  \colhead{(10)} &
  \colhead{(11)} &
  \colhead{(12)} &
  \colhead{(13)} &
  \colhead{(14)} &
  \colhead{(15)} &
  \colhead{(16)} &
  \colhead{(17)}}
 \startdata
  1 & NGVS-J122623.21+121925.2 & 19.468$\pm$0.022 & 18.742$\pm$0.022 & 36.59$\pm$1.27 & 35.42$\pm$1.08 &         \nodata &   \nodata &   \nodata &   \nodata &     46450 &   \nodata & 1 & 0 & 1 & 0 &                             \\
  2 & NGVS-J122628.69+125642.6 & 19.792$\pm$0.011 & 18.852$\pm$0.013 & 17.38$\pm$0.72 & 15.99$\pm$0.72 &         \nodata &   \nodata &   \nodata &   \nodata &      -186 &   \nodata & 1 & 1 & 1 & 1 &                             \\
  3 & NGVS-J122633.05+123847.6 & 21.391$\pm$0.014 & 20.651$\pm$0.007 & 11.26$\pm$0.57 & 10.99$\pm$0.49 &         \nodata &   \nodata &   \nodata &      5322 &   \nodata &   \nodata & 1 & 1 & 1 & 0 &                             \\
  4 & NGVS-J122636.42+125512.9 & 18.972$\pm$0.006 & 18.021$\pm$0.006 & 18.01$\pm$0.33 & 17.34$\pm$0.27 &         \nodata &   \nodata &   \nodata &   \nodata &       -93 &   \nodata & 1 & 1 & 1 & 1 &                             \\
  5 & NGVS-J122638.52+123702.7 & 20.616$\pm$0.007 & 19.843$\pm$0.008 & 17.86$\pm$0.40 & 18.04$\pm$0.43 &         \nodata &   \nodata &   \nodata &   \nodata &   \nodata &   \nodata & 1 & 1 & 1 & 1 &                             \\
  6 & NGVS-J122638.79+124545.8 & 20.564$\pm$0.007 & 19.823$\pm$0.010 & 15.93$\pm$0.30 & 15.03$\pm$0.42 &         \nodata &   \nodata &   \nodata &   \nodata &   \nodata &   \nodata & 1 & 1 & 1 & 1 &                             \\
  7 & NGVS-J122643.32+121743.9 & 20.830$\pm$0.004 & 20.112$\pm$0.007 & 11.88$\pm$0.39 & 10.28$\pm$0.45 &         \nodata &   \nodata &   \nodata &   \nodata &   \nodata &   \nodata & 2 & 1 & 1 & 0 &                             \\
  8 & NGVS-J122646.48+131732.2 & 21.446$\pm$0.004 & 20.683$\pm$0.013 & 12.97$\pm$0.25 & 14.53$\pm$0.49 &         \nodata &   \nodata &   \nodata &   \nodata &   \nodata &   \nodata & 1 & 0 & 1 & 0 &                             \\
\enddata
\tablenotetext{$\dagger$}{The SDSS redshift measurement has ${\tt ZWARNING}=68$, which means that the spectrum has SMALL\_DELTA\_CHI2 and NEGATIVE\_EMISSION \citep{2012AJ_144_144Bolton}. Although the SDSS velocity indicates this VCC-cataloged galaxy is a background galaxy, its diffuse envelope and prominent nucleus strongly suggest that it is a bonafide cluster member.}
\tablenotetext{}{NOTE-- (1) Object ID number; (2) Object name in NGVS; (3-4) King model magnitudes in $g$ and $i$ bands; (5-6) Half-light radii in $g$ and $i$ bands; (7) Half-light radius measured using HST imaging \citep{2005ApJ_627_203Hacsegan}; (8) Half-light radius from \citet{2011ApJS_197_33Strader}; (9-12) Radial velocity from \citet{2011ApJS_197_33Strader}, MMT/Hectospec and 2dF/AAOmega AAT program \citep{2015ApJ_802_30Zhang}, and SDSS \citep{2000AJ_120_1579York}; (13) Class parameter: 1 = probable UCD, 2 = dwarf nucleus, 3 = background galaxy, 4 = other type of contaminant; (14) Flag for whether the object was selected in the $u^*iK_s$ color-color diagram: 0 = no, 1 = yes; (15) Flag for whether the object selected in the $u^*gz$ color-color diagram: 0 = no, 1 = yes; (16) Object flag, 0 = not UCD, 1 = confirmed or possible UCD; (17) Alternative names, if available.}
\end{deluxetable}
\clearpage
\end{landscape}

\tabletypesize{\tiny}
\LongTables
\begin{deluxetable}{lrrrrrrrrrrrc}
 \tablewidth{0pt}
 \tabletypesize{\small}
 \tablecaption{Photometric Properties of UCD Candidates in the M49 Region
 \label{tab:ucd_m49_1}}
 \tablehead{
  \colhead{ID} &
  \colhead{NGVSID} &
  \colhead{$\alpha_{\rm J2000}$} &
  \colhead{$\delta_{\rm J2000}$} &
  \colhead{$E(B-V)$} &
  \colhead{$g_{\rm 16}$} &
  \colhead{$(u^*-g)_{\rm 8}$} &
  \colhead{$(g-i)_{\rm 8}$} &
  \colhead{$(g-z)_{\rm 8}$} &
  \colhead{$\Delta_{\rm env}$} &
  \colhead{UCD} \\
  \colhead{ } &
  \colhead{ } &
  \colhead{(deg)} &
  \colhead{(deg)} &
  \colhead{(mag)} &
  \colhead{(mag)} &
  \colhead{(mag)} &
  \colhead{(mag)} &
  \colhead{(mag)} &
  \colhead{(mag)} &
  \colhead{} \\
  \colhead{(1)} &
  \colhead{(2)} &
  \colhead{(3)} &
  \colhead{(4)} &
  \colhead{(5)} &
  \colhead{(6)} &
  \colhead{(7)} &
  \colhead{(8)} &
  \colhead{(9)} &
  \colhead{(10)} &
  \colhead{(11)}}
 \startdata
  1 & NGVS-J122708.94+074228.3 & 186.7872607 &  7.7078639 & 0.024 & 19.429$\pm$0.001 & 1.025$\pm$0.003 & 0.813$\pm$0.002 & 0.955$\pm$0.002 &  0.239 & 1 \\
  2 & NGVS-J122718.40+081154.2 & 186.8266527 &  8.1983810 & 0.023 & 20.966$\pm$0.002 & 1.076$\pm$0.008 & 0.794$\pm$0.004 & 0.900$\pm$0.007 & -0.140 & 0 \\
  3 & NGVS-J122735.11+072518.7 & 186.8962982 &  7.4218483 & 0.022 & 20.992$\pm$0.003 & 0.923$\pm$0.008 & 0.678$\pm$0.005 & 0.790$\pm$0.008 &  0.085 & 0 \\
  4 & NGVS-J122741.97+075613.3 & 186.9248642 &  7.9370367 & 0.025 & 21.193$\pm$0.003 & 1.146$\pm$0.012 & 0.747$\pm$0.006 & 0.862$\pm$0.010 &  0.075 & 0 \\
  5 & NGVS-J122744.26+072542.5 & 186.9344272 &  7.4284640 & 0.022 & 20.645$\pm$0.002 & 0.968$\pm$0.006 & 0.683$\pm$0.004 & 0.778$\pm$0.006 &  0.103 & 1 \\
  6 & NGVS-J122803.46+065343.1 & 187.0144136 &  6.8953085 & 0.020 & 21.446$\pm$0.004 & 0.987$\pm$0.011 & 0.737$\pm$0.007 & 0.868$\pm$0.011 &  0.140 & 1 \\
  7 & NGVS-J122806.14+065909.0 & 187.0255686 &  6.9858243 & 0.020 & 20.272$\pm$0.002 & 1.485$\pm$0.008 & 0.984$\pm$0.004 & 1.250$\pm$0.005 &  0.321 & 1 \\
  8 & NGVS-J122806.31+080342.8 & 187.0262901 &  8.0618891 & 0.022 & 21.475$\pm$0.003 & 0.923$\pm$0.013 & 0.697$\pm$0.007 & 0.795$\pm$0.013 &  0.092 & 0 \\
\enddata
 \tablenotetext{}{NOTES -- (1) Object ID number; (2) Object name in NGVS; (3) Right Ascension; (4) Declination; (5) Galactic extinction \citep{1998ApJ_500_525Schlegel}; (6) Aperture-corrected $g$ magnitude within a 3-arcsec diameter aperture. Only SExtractor errors are accounted for here; (7-9) Colors measured in a 1.5-arcsec diameter aperture; (10) Envelope parameter defined in \ref{eq:env}; (11) Object flag, 0 = not UCD, 1 = confirmed or possible UCD. }
\end{deluxetable}
\clearpage

\LongTables
\begin{deluxetable}{llrrrrrrrrrrrrrcl}
 \tablewidth{0pt}
 \tabletypesize{\small}
 \tablecaption{Structural Properties and Redshift Information for UCD Candidates in the M49 Region
 \label{tab:ucd_m49_2}}
 \tablehead{
  \colhead{ID} &
  \colhead{NGVSID} &
  \colhead{$g_k$} &
  \colhead{$i_k$} &
  \colhead{$r_{h,g}$} &
  \colhead{$r_{h,i}$} &
  \colhead{$v_r$(MMT)} &
  \colhead{Class} &
  \colhead{$u^*gz$} &
  \colhead{UCD}  \\
  \colhead{ } &
  \colhead{ } &
  \colhead{(mag)} &
  \colhead{(mag)} &
  \colhead{(pc)} &
  \colhead{(pc)} &
  \colhead{(km~s$^{-1}$)} &
  \colhead{ } &
  \colhead{ } &
  \colhead{ }  \\
  \colhead{(1)} &
  \colhead{(2)} &
  \colhead{(3)} &
  \colhead{(4)} &
  \colhead{(5)} &
  \colhead{(6)} &
  \colhead{(7)} &
  \colhead{(8)} &
  \colhead{(9)} &
  \colhead{(10)}}
 \startdata
  1 & NGVS-J122708.94+074228.3 & 24.228$\pm$0.010 & 23.403$\pm$0.010 & 45.60$\pm$0.70 & 44.22$\pm$0.65 & \nodata & 1 & 1 & 1 & \\
  2 & NGVS-J122718.40+081154.2 & 25.947$\pm$0.026 & 25.062$\pm$0.047 & 13.48$\pm$0.82 & 18.41$\pm$1.46 & \nodata & 3 & 1 & 0 & \\
  3 & NGVS-J122735.11+072518.7 & 26.010$\pm$0.025 & 25.270$\pm$0.014 & 15.53$\pm$0.89 & 18.11$\pm$0.64 &   38059 & 1 & 1 & 0 & \\
  4 & NGVS-J122741.97+075613.3 & 26.139$\pm$0.019 & 25.337$\pm$0.024 & 23.11$\pm$0.74 & 23.14$\pm$0.85 &   31816 & 1 & 1 & 0 & \\
  5 & NGVS-J122744.26+072542.5 & 25.688$\pm$0.006 & 24.948$\pm$0.007 & 12.77$\pm$0.37 & 14.70$\pm$0.40 &    1107 & 1 & 1 & 1 & \\
  6 & NGVS-J122803.46+065343.1 & 26.373$\pm$0.014 & 25.524$\pm$0.037 & 21.72$\pm$0.70 & 28.08$\pm$1.46 & \nodata & 1 & 1 & 1 & \\
  7 & NGVS-J122806.14+065909.0 & 25.353$\pm$0.014 & 24.326$\pm$0.015 & 15.98$\pm$1.12 & 20.12$\pm$0.58 & \nodata & 1 & 1 & 1 & \\
  8 & NGVS-J122806.31+080342.8 & 26.529$\pm$0.007 & 25.861$\pm$0.012 & 13.06$\pm$0.36 &  7.84$\pm$0.48 &   15078 & 1 & 1 & 0 & \\
\enddata
\tablenotetext{}{NOTE-- (1) Object ID number; (2) Object name in NGVS; (3-4) King model magnitudes in $g$ and $i$ bands; (5-6) Half-light radii in $g$ and $i$ bands; (7) Radial velocity from MMT/Hectospec; (8) Class parameter: 1 = probable UCD, 2 = dwarf nucleus, 3 = background galaxy, 4 = other type of contaminant; (9) Flag for whether the object selected in the $u^*gz$ color-color diagram: 0 = no, 1 = yes; (10) Object flag, 0 = not UCD, 1 = confirmed or possible UCD.}
\end{deluxetable}
\clearpage

\LongTables
\begin{deluxetable}{lrrrrrrrrrrrc}
 \tablewidth{0pt}
 \tabletypesize{\small}
 \tablecaption{Photometric Properties of UCD Candidates in the M60 Region
 \label{tab:ucd_m60_1}}
 \tablehead{
  \colhead{ID} &
  \colhead{NGVSID} &
  \colhead{$\alpha_{\rm J2000}$} &
  \colhead{$\delta_{\rm J2000}$} &
  \colhead{$E(B-V)$} &
  \colhead{$g_{\rm 16}$} &
  \colhead{$(u^*-g)_{\rm 8}$} &
  \colhead{$(g-i)_{\rm 8}$} &
  \colhead{$(g-z)_{\rm 8}$} &
  \colhead{$\Delta_{\rm env}$} &
  \colhead{UCD} \\
  \colhead{ } &
  \colhead{ } &
  \colhead{(deg)} &
  \colhead{(deg)} &
  \colhead{(mag)} &
  \colhead{(mag)} &
  \colhead{(mag)} &
  \colhead{(mag)} &
  \colhead{(mag)} &
  \colhead{(mag)} &
  \colhead{} \\
  \colhead{(1)} &
  \colhead{(2)} &
  \colhead{(3)} &
  \colhead{(4)} &
  \colhead{(5)} &
  \colhead{(6)} &
  \colhead{(7)} &
  \colhead{(8)} &
  \colhead{(9)} &
  \colhead{(10)} &
  \colhead{(11)}}
 \startdata
  1 & NGVS-J123909.32+112147.9 & 189.7888329 & 11.3633175 & 0.033 & 21.159$\pm$0.003 & 1.415$\pm$0.010 & 1.004$\pm$0.005 & 1.175$\pm$0.007 &  0.057 & 1 \\
  2 & NGVS-J123948.33+105157.0 & 189.9513674 & 10.8658447 & 0.022 & 20.858$\pm$0.002 & 1.244$\pm$0.009 & 0.907$\pm$0.004 & 1.043$\pm$0.006 &  0.130 & 0 \\
  3 & NGVS-J123949.21+110135.3 & 189.9550398 & 11.0264758 & 0.024 & 21.167$\pm$0.003 & 1.338$\pm$0.011 & 0.981$\pm$0.005 & 1.173$\pm$0.007 &  0.078 & 1 \\
  4 & NGVS-J123950.99+112805.1 & 189.9624727 & 11.4680886 & 0.039 & 20.763$\pm$0.002 & 1.208$\pm$0.007 & 0.934$\pm$0.004 & 1.162$\pm$0.005 &  0.114 & 0 \\
  5 & NGVS-J124002.08+105517.2 & 190.0086686 & 10.9214418 & 0.022 & 21.420$\pm$0.003 & 0.994$\pm$0.012 & 0.664$\pm$0.007 & 0.735$\pm$0.012 &  0.261 & 1 \\
  6 & NGVS-J124049.54+121138.1 & 190.2064072 & 12.1939068 & 0.039 & 21.348$\pm$0.003 & 1.219$\pm$0.011 & 0.981$\pm$0.005 & 1.127$\pm$0.008 &  0.083 & 1 \\
  7 & NGVS-J124115.49+103811.9 & 190.3145415 & 10.6366475 & 0.024 & 21.350$\pm$0.003 & 1.119$\pm$0.011 & 0.865$\pm$0.006 & 1.037$\pm$0.009 &  0.069 & 1 \\
  8 & NGVS-J124151.11+105311.7 & 190.4629499 & 10.8865742 & 0.021 & 19.805$\pm$0.001 & 1.405$\pm$0.005 & 1.014$\pm$0.002 & 1.242$\pm$0.003 &  0.306 & 0 \\
\enddata
  \tablenotetext{}{NOTES -- (1) Object ID number; (2) Object name in NGVS; (3) Right Ascension; (4) Declination; (5) Galactic extinction \citep{1998ApJ_500_525Schlegel}; (6) Aperture-corrected $g$ magnitude within a 3-arcsec diameter aperture. Only SExtractor errors are accounted for here; (7-9) Colors measured in a 1.5-arcsec diameter aperture; (10) Envelope parameter defined in \ref{eq:env}; (11) Object flag, 0 = not UCD, 1 = confirmed or possible UCD. }
\end{deluxetable}

\LongTables
\begin{deluxetable}{llrrrrrrrrrrrrrcl}
 \tablewidth{0pt}
 \tabletypesize{\small}
 \tablecaption{Structural Properties and Redshift Information for UCD Candidates in the M60 Region
 \label{tab:ucd_m60_2}}
 \tablehead{
  \colhead{ID} &
  \colhead{NGVSID} &
  \colhead{$g_k$} &
  \colhead{$i_k$} &
  \colhead{$r_{h,g}$} &
  \colhead{$r_{h,i}$} &
  \colhead{$v_r$(AAT)} &
  \colhead{Class} &
  \colhead{$u^*gz$} &
  \colhead{UCD}  \\
  \colhead{ } &
  \colhead{ } &
  \colhead{(mag)} &
  \colhead{(mag)} &
  \colhead{(pc)} &
  \colhead{(pc)} &
  \colhead{(km~s$^{-1}$)} &
  \colhead{ } &
  \colhead{ } &
  \colhead{ }  \\
  \colhead{(1)} &
  \colhead{(2)} &
  \colhead{(3)} &
  \colhead{(4)} &
  \colhead{(5)} &
  \colhead{(6)} &
  \colhead{(7)} &
  \colhead{(8)} &
  \colhead{(9)} &
  \colhead{(10)}}
 \startdata
  1 & NGVS-J123909.32+112147.9 & 26.190$\pm$0.009 & 25.224$\pm$0.011 & 12.39$\pm$0.68 & 12.60$\pm$0.58 & \nodata & 1 & 1 & 1 & \\
  2 & NGVS-J123948.33+105157.0 & 25.805$\pm$0.029 & 24.874$\pm$0.039 & 21.08$\pm$0.94 & 23.58$\pm$1.43 & \nodata & 3 & 1 & 0 & \\
  3 & NGVS-J123949.21+110135.3 & 26.214$\pm$0.009 & 25.268$\pm$0.009 & 11.72$\pm$0.43 & 11.56$\pm$0.30 & \nodata & 1 & 1 & 1 & \\
  4 & NGVS-J123950.99+112805.1 & 25.740$\pm$0.024 & 24.742$\pm$0.012 & 21.20$\pm$1.06 & 27.76$\pm$0.80 & \nodata & 3 & 1 & 0 & \\
  5 & NGVS-J124002.08+105517.2 & 26.523$\pm$0.020 & 25.875$\pm$0.025 & 12.72$\pm$0.57 & 14.98$\pm$0.77 & \nodata & 1 & 1 & 1 & \\
  6 & NGVS-J124049.54+121138.1 & 26.376$\pm$0.008 & 25.463$\pm$0.021 & 13.93$\pm$0.45 & 13.44$\pm$1.13 & \nodata & 1 & 1 & 1 & \\
  7 & NGVS-J124115.49+103811.9 & 26.359$\pm$0.013 & 25.518$\pm$0.008 & 14.31$\pm$0.64 & 13.39$\pm$0.51 & \nodata & 1 & 1 & 1 & \\
  8 & NGVS-J124151.11+105311.7 & 24.887$\pm$0.017 & 23.810$\pm$0.013 & 20.24$\pm$0.82 & 25.90$\pm$0.65 & \nodata & 3 & 1 & 0 & \\
\enddata
\tablenotetext{}{NOTE-- (1) Object ID number; (2) Object name in NGVS; (3-4) King model magnitudes in $g$ and $i$ bands; (5-6) Half-light radii in $g$ and $i$ bands; (7) Radial velocity from 2dF/AAOmega AAT program \citep{2015ApJ_802_30Zhang}; (8) Class parameter: 1 = probable UCD, 2 = dwarf nucleus, 3 = background galaxy, 4 = other type of contaminant; (9) Flag for whether the object selected in the $u^*gz$ color-color diagram: 0 = no, 1 = yes; (10) Object flag, 0 = not UCD, 1 = confirmed or possible UCD.}
\end{deluxetable}
\clearpage

\tabletypesize{\small}
\LongTables
\begin{deluxetable*}{lrrrrrrr}
 \tablewidth{0pt}
 \tabletypesize{\small}
 \tablecaption{UCD Samples in Program Regions
 \label{tab:sample}}
 \tablehead{
  \colhead{Region} &
  \colhead{$\langle{\alpha_{\rm J2000}}\rangle$} &
  \colhead{$\langle{\delta_{\rm J2000}}\rangle$} &
  \colhead{$\Omega$} &
  \colhead{$N_{\rm UCD,ugz}$} &
  \colhead{$\Sigma_{\rm UCD,ugz}$} &
  \colhead{$N_{\rm UCD,uiK_s}$} &
  \colhead{$\Sigma_{\rm UCD,uiK_s}$} \\
  \colhead{ } &
  \colhead{(deg)} &
  \colhead{(deg)} &
  \colhead{(deg$^2$)} &
  \colhead{} &
  \colhead{(deg$^2$)} &
  \colhead{} &
  \colhead{(deg$^2$)} \\
  \colhead{(1)} &
  \colhead{(2)} &
  \colhead{(3)} &
  \colhead{(4)} &
  \colhead{(5)} &
  \colhead{(6)} &
  \colhead{(7)} &
  \colhead{(8)}}
 \startdata
    M87                & 187.60 & 12.44 & 3.61872 & 110 & 30.4$\pm$2.9 & 96 & 26.5$\pm$2.7 \\
    NGVS$+$0$+$0 & 187.94 & 12.05 & 0.90390 &  50 & 55.3$\pm$7.8 & 47 & 52.0$\pm$7.6 \\
    NGVS$+$0$+$1 & 188.03 & 12.94 & 0.90314 &  17 & 18.8$\pm$4.6 & 17 & 18.8$\pm$4.6 \\
    NGVS$-$1$+$0 & 187.15 & 12.01 & 0.90656 &  14 & 15.4$\pm$4.1 &  6 &  6.6$\pm$2.7 \\
    NGVS$-$1$+$1 & 187.09 & 12.94 & 0.90512 &  29 & 32.0$\pm$5.9 & 26 & 28.7$\pm$5.6 \\
 & & & & & & & \\
    M49                    & 187.57  & 07.75   & 3.61230 & 40 & 11.1$\pm$1.8 & \nodata & \nodata \\
    NGVS$-$1$-$4      & 187.14  & 08.20   & 0.90095 & 19 & 21.1$\pm$4.8 & \nodata & \nodata \\
\enddata
 \tablenotetext{}{NOTE -- (1) Region designation (2-3) Mean Right Ascension and Declination of the region; (4) Field of view; (5-6) Number and number density of UCDs selected based on the $u^*gz$ color-color diagram; (7-8) Number and number density of UCDs selected based on the $u^*iK_s$ color-color diagram.}
\end{deluxetable*}
\clearpage

\tabletypesize{\small}
\begin{landscape}
\LongTables
\begin{deluxetable}{lrrrrrrrrrrrr}
 \tablewidth{0pt}
 \tabletypesize{\small}
 \tablecaption{UCD Color Distributions in the M87 region
 \label{tab:color_ucd_gmm_92}}
 \tablehead{
  \colhead{color$_8$} &
  \colhead{$\mu_a$} &
  \colhead{$\sigma_a$} &
  \colhead{$\mu_{b}$} &
  \colhead{$\mu_{r}$} &
  \colhead{$\sigma_{b}$} &
  \colhead{$\sigma_{r}$} &
  \colhead{$N_{b}$} &
  \colhead{$N_{r}$} &
  \colhead{$f_{b}$} &
  \colhead{$D$} &
  \colhead{$p(\chi^2)$} \\
  \colhead{(mag)} &
  \colhead{(mag)} &
  \colhead{(mag)} &
  \colhead{(mag)} &
  \colhead{(mag)} &
  \colhead{(mag)} &
  \colhead{(mag)} &
  \colhead{} &
  \colhead{} &
  \colhead{} &
  \colhead{} &
  \colhead{} \\
  \colhead{(1)} &
  \colhead{(2)} &
  \colhead{(3)} &
  \colhead{(4)} &
  \colhead{(5)} &
  \colhead{(6)} &
  \colhead{(7)} &
  \colhead{(8)} &
  \colhead{(9)} &
  \colhead{(10)} &
  \colhead{(11)} &
  \colhead{(12)}}
 \startdata
 ($u^*-g$) & $ 1.068\pm0.011$ &  $0.109\pm0.008$ & $ 1.039\pm0.019$ & $ 1.291\pm0.057$ &  $0.075\pm0.010$ &  $0.052\pm0.026$ & $81$ & $11$ &  $0.880\pm0.120$ & $3.90$ & $0.001$ \\
 ($u^*-r$) & $ 1.580\pm0.015$ &  $0.151\pm0.011$ & $ 1.543\pm0.038$ & $ 1.889\pm0.092$ &  $0.111\pm0.020$ &  $0.056\pm0.038$ & $82$ & $10$ &  $0.891\pm0.185$ & $3.94$ & $0.004$ \\
 ($u^*-i$) & $ 1.820\pm0.017$ &  $0.174\pm0.014$ & $ 1.765\pm0.036$ & $ 2.154\pm0.099$ &  $0.112\pm0.023$ &  $0.095\pm0.042$ & $79$ & $13$ &  $0.859\pm0.163$ & $3.74$ & $0.001$ \\
 ($u^*-z$) & $ 1.951\pm0.019$ &  $0.200\pm0.016$ & $ 1.897\pm0.034$ & $ 2.380\pm0.103$ &  $0.136\pm0.021$ &  $0.074\pm0.047$ & $82$ & $10$ &  $0.891\pm0.130$ & $4.40$ & $0.001$ \\
 ($u^*-K_s$) & $ 1.750\pm0.033$ &  $0.311\pm0.023$ & $ 1.661\pm0.047$ & $ 2.386\pm0.144$ &  $0.207\pm0.035$ &  $0.129\pm0.064$ & $80$ & $11$ &  $0.879\pm0.121$ & $4.21$ & $0.001$ \\
   ($g-r$) & $ 0.512\pm0.005$ &  $0.052\pm0.003$ & $ 0.491\pm0.008$ & $ 0.586\pm0.012$ &  $0.036\pm0.007$ &  $0.024\pm0.007$ & $72$ & $20$ &  $0.783\pm0.098$ & $3.10$ & $0.038$ \\
   ($g-i$) & $ 0.751\pm0.007$ &  $0.070\pm0.005$ & $ 0.721\pm0.012$ & $ 0.853\pm0.034$ &  $0.041\pm0.007$ &  $0.049\pm0.016$ & $71$ & $21$ &  $0.772\pm0.130$ & $2.93$ & $0.001$ \\
   ($g-z$) & $ 0.882\pm0.009$ &  $0.098\pm0.007$ & $ 0.853\pm0.013$ & $ 1.083\pm0.036$ &  $0.064\pm0.009$ &  $0.027\pm0.017$ & $80$ & $12$ &  $0.870\pm0.098$ & $4.67$ & $0.001$ \\
\enddata
\tablenotetext{}{NOTE -- (1) Color index; (2-3) Mean color of all the objects and corresponding sigma; (4-5) Mean color of blue and red UCD subpopulations; (6-7) Sigma of blue and red UCD subpopulations; (8-9) Number of  blue and red UCDs; (10) Blue UCD fraction; (11) $D$ parameter \citep{2010ApJ_718_1266Muratov} (12) $p$-value for chi$^2$ \citep{2010ApJ_718_1266Muratov}.}
\end{deluxetable}
\clearpage
\end{landscape}

\end{document}